 \documentclass[final,3p,12pt,times]{elsarticle}

\usepackage{graphicx}
\usepackage{pgf}  
\usepackage{subfig}
\usepackage{amsmath}
\usepackage{longtable}    
\usepackage{morefloats}  
\usepackage{graphicx}
\usepackage{xcolor}
\usepackage{url}
\usepackage{caption}
\usepackage{multicol}
\usepackage{multirow}
\usepackage{array}
\usepackage{lscape}
\usepackage{boxedminipage}
\usepackage{multicol}
\usepackage{multirow}
\usepackage{booktabs}
\usepackage[hidelinks]{hyperref}
\usepackage{color,soul}
\usepackage[pagewise]{lineno}
\usepackage[titletoc,title]{appendix}
\usepackage{amssymb}
\usepackage[Symbol]{upgreek}

\usepackage[mathscr]{euscript}  
\DeclareMathAlphabet{\mathscrbf}{OMS}{mdugm}{b}{n}  

\usepackage[nodots,nocompress]{numcompress}

\biboptions{numbers, square}

\newcommand{\eref}[1]{Eq.~\ref{#1}}
\newcommand{\fref}[1]{Fig.~\ref{#1}}
\newcommand{\tref}[1]{Table~\ref{#1}}
\newcommand{\sref}[1]{Section~\ref{#1}}

\journal{}

\begin{document}

\begin{frontmatter}



\title{A discontinuous Galerkin method based on a hierarchical orthogonal basis for Lagrangian hydrodynamics on curvilinear grids }

\author[label1]{Xiaodong Liu \corref{cor1}}
\ead{liufield@gmail.com}
\author[label2]{Nathaniel R. Morgan}
\ead{nmorgan@lanl.gov}
\author[label2]{Evan J. Lieberman}
\ead{elieberman3@lanl.gov}
\author[label2]{Donald E. Burton}
\ead{burton@lanl.gov}
\cortext[cor1]{Corresponding author}

 \address[label1]{Remcom Inc; State College, PA, USA}
 \address[label2]{X-Computational Physics Division; Los Alamos National Laboratory; P.O. Box 1663, Los Alamos, NM, USA}

\begin{abstract}
We present a new high-order accurate Lagrangian discontinuous Galerkin (DG) hydrodynamic method to simulate material dynamics (for \textit{e.g.}, gasses, fluids, and solids) with up to fourth-order accuracy on cubic meshes. 
The variables, such as specific volume, velocity, specific total energy, and deformation gradient fields within a cell, are represented with a polynomial constructed from a novel hierarchical orthogonal basis about the center of mass, 
which decouples the moments of the solution because the mass matrix is diagonal.
The discontinuity in the polynomials at the cell boundary is addressed by solving a multi-directional Riemann problem at the vertices of the cell and a 1D Riemann problem at additional non-vertex quadrature points along the edges so that the surface integral is exact for the polynomial order. The uniqueness lies in that the vertices of the curvilinear grid work as the quadrature points for the surface integral of DG methods. 
To ensure robust mesh motion, the pressure for the Riemann problem accounts for the difference between the density variation over the cell and a density field from subcell mesh stabilization (SMS).  
The accuracy and robustness of the new high-order accurate Lagrangian DG hydrodynamic method is demonstrated by simulating a diverse suite of challenging test problems covering gas and solid dynamic problems on curvilinear grids.  
\end{abstract}


\end{frontmatter}


\section{Introduction}  

Lagrangian hydrodynamic methods solve the governing physics equations for material dynamics on a 
mesh that moves with the flow and historically are based on linear cells (also called elements or zones) that have edges defined by a polynomial of degree 1.  One of the first Lagrangian hydrodynamic methods was the finite volume (FV) staggered-grid hydrodynamic (SGH) method by von Neumann and Richtmyer \cite{VNR}.  A range of FV SGH methods  \cite{WilkinsSGH,BurtonSGH,Caramana,MorganMARS,Maire3} have been proposed since the work of von Neumann and Richtmyer.  The FV Lagrangian cell-centered hydrodynamic (CCH) approach was proposed by Godunov \cite{Godunov1, Godunov2} and has garnered significant interest following the work of Despr{\'e}s and Mazeran \cite{Despres} who developed a nodal Riemann solver that greatly improved the robustness of CCH methods over using a Riemann solver at the cell faces \cite{CAVEAT}.   A range of robust nodal Riemann solvers have been proposed for use in Lagrangian CCH methods \cite{Maire1,Maire2,BurtonCCH,MorganContact}.  Discontinuous Galerkin (DG) \cite{ShuDG5jcp1998, luodgtaylor2008, luodghweno2007, luodgpmulti2006, xiaIRK2015,LiurDGcaf2017, Liuirkcaf2016,Wangdgaleaiaa2018,corriganMDG2019,corriganMDGviscous2020, Chengddgcicp2017,ChengGRPDGcaf2019,Lidgadaptive2020, ChengrDGcaf2016,Liudgturbulence2016, ChengrDGvr2019, ChengrDGcls2019,nourgalievfrontdg2017, nourgalievorthdg2016, bassiorthodg2012,KimlimiterDG2009, Lourdghyperbolic2019, LouhyperbolicrDGadvec-diff2017, LouhyperbolicrDGnonlinear-diff2018, LouhyperbolicrDGexplicit2018, LingfvVR2017, LingrdgVR2017, Linghyperbolicfv2018, LinghyperbolicrdgNS2019, Linghyperbolicrdgunsteady2019, LinghyperbolicrdgJCP2021} hydrodynamic methods have been developed for Lagrangian motion with linear meshes in Cartesian coordinates \cite{JiaDGlagjcp2011,VilarDGlagcaf2012, VilarDGlagjcp2014, JiaDGlagcaf2014, LiuDGlagcaf2017,LiebermanStrength1D2017,WangDGlagcaf2020} and 2D R-Z coordinates \cite{Liudgrzaiaa2018,LiuDGlagRZ2018}.  The DG method evolves a polynomial of order Pn forward in time, which is denoted as DG(Pn), whereas FV CCH methods fit the neighboring cells to build a polynomial.  Lagrangian hydrodynamic methods with linear cells deliver up to second-order accuracy on multi-dimensional problems.

Recently, some authors have proposed third-order accurate 2D Lagrangian hydrodynamic methods for quadratic cells, which have edges defined by a polynomial of degree 2.  Cheng and Shu \cite{ChengCurv2008} developed a weighted essentially non-oscillatory (WENO) FV CCH method for quadratic cells. Vilar \textit{et al.} \cite {VilarDGlagjcp2014} proposed a third-order accurate Lagrangian DG(P2) method  for solving the 2D gas dynamics on polygonal quadratic cells using a hierarchical slope limiter for shock capturing \cite{kuzminbarth2010, kuzminslopelim2014}.  Spurious mesh motion can still arise on some strong shock problems with quadratic cells.  Cheng and Shu addressed the spurious mesh motion by controlling the curvature of the cell surface.  In a context of the second-order FV and DG method with a Barth slope limiter \cite{BJlimiter1989}, Morgan \textit{et al.} \cite{MorganVeloFilter2017} proposed a velocity filter to dissipate spurious mesh motion by modifying the velocity reconstructions.   Liu \textit{et al.} \cite{LiuDGlagSMS2019, LiuDGlagWENO2020} developed a novel subcell mesh stabilization (SMS) scheme for ensuring robust mesh motion with a third-order Lagrangian DG method on quadratic cells. \cite{LiuDGlagSMS2019} limited the second-order DG(P1) solution truncated from the high-order polynomial near the shock and \cite{LiuDGlagWENO2020} proposed symmetry-preserving Hermite WENO for shock capturing.  Later, Liu \textit{et al.} \cite{LiuDGlagSMStria2019} also explored the Lagrangian DG(P1) method on quadratic triangles \cite{Morgandgtriaaiaa2018} with SMS for improved accuracy on large deformation flows.  Lieberman \textit{et al.}\cite{LiebermanStrength2D2019} extended the third-order accurate Lagrangian DG method with SMS \cite{LiuDGlagSMS2019} to support hyper and hypo-elastic plastic models for simulating solid dynamic problems using quadratic cells and to simulate multi-phase high-explosives detonation physics \cite{LiebermanHE2020}.

The objective in this paper is to create a new high-order accurate Lagrangian DG hydrodynamic method to simulate material dynamics (for gases, fluids, and solids) up to fourth-order accuracy with cubic cells, which have edges defined by a polynomial of degree 3. A cell with edges of more degrees of freedom is essential to accurately simulate large deformation problems because it is less likely to tangle compared with quadratic- or linear-order cells. To-date, there are no fourth-order accurate compact stencil Lagrangian hydrodynamic methods with cubic cells. High-order Lagrangian methods that use a compact numerical stencil are desirable for achieving better scaling and performance on modern computer architectures. The SMS method was first proposed by Liu \textit{et al.} \cite{LiuDGlagSMS2019} to ensure accurate and robust mesh motion for quadratic cells, and in this work, we extend SMS to cubic cells in the context of this new high-order accurate Lagrangian DG hydrodynamic method.
The polynomials in this work are constructed using a novel hierarchical orthogonal  basis about the center of mass, which differs from prior Lagrangian DG research efforts that used Taylor basis functions.  Using a Taylor basis will create a mass matrix that must be inverted at the start of a calculation, and couples together the equations for the high-order terms, which is problematic when limiting the second-order (DG(P1)) solution truncated from the high-order polynomial for preserving monotonicity near a shock.
We show that the hierarchical orthogonal basis combined with SMS is essential for robust mesh motion with cubic cells on strong shock problems using the aforementioned limiting strategy.  The orthogonal basis is also constant in time for Lagrangian motion, so they only need to be calculated at the start of a simulation, which is key for a computationally efficient scheme.  An explicit multi-step Runge-Kutta (RK) method is employed for time-marching. The new high-order accurate Lagrangian DG hydrodynamic method conserves mass, momentum, and total energy, produces robust solutions on strong shock problems and yields high-order accurate solutions on smooth flows using curvilinear grids. 

The layout of the paper is as follows. The governing equations are described in \sref{GovEqn}. 
The constitutive models are described in \sref{models}.  
\sref{Discretization} discusses 
the new hierarchical orthogonal  basis,
the Lagrangian DG method, 
the evaluation of the right hand side ($rhs$) of the discretized equations,
the SMS scheme,
and the limiting strategy for shock capturing.
 The results from a large suite of test problems are reported in \sref{TestCases} covering solid and gas dynamics. Concluding remarks are given in \sref{Conclusion}.

\section{Governing equations}
\label{GovEqn}
In Lagrangian hydrodynamics, the differential  equations governing the evolution of the specific volume ($v$), velocity (${\textbf {\textit{u}}}$), specific total energy ($\tau$), and deformation gradient ${\text F}$ are given by,

\begin{equation}
\label{dspecVol}
\rho \frac{{dv}} {dt}  = \nabla \cdot  {\textbf {\textit{u}}} ,
\end{equation}

\begin{equation}
\label{dMomentum}
\rho \frac{d{\textbf {\textit{u}}} }{dt}  = \nabla \cdot {{\upsigma}} ,
\end{equation}

\begin{equation}
\label{dTotalEnergy}
\rho \frac{d \tau}{dt}  = \nabla  \cdot ( {\upsigma}  {\textbf {\textit{u}}} ),
\end{equation}

\begin{equation}
\label{dF}
\frac{d \text F}{dt}  = \nabla_{\textbf {\textit X}}{\textbf {\textit{u}}} ,
\end{equation}

\noindent here the subscript $\textbf {\textit X}$ in  \eref{dF} denotes the gradient with respect to the initial coordinates $\textbf {\textit X}$ instead of the current physical coordinates $\textbf{\textit x}$, and
\begin{equation}
\label{Jacobian}
\nabla_{\textbf {\textit X}}{\textbf {\textit{u}}} =
\begin{bmatrix}
 \frac{\partial u}{\partial X}  &  \frac{\partial v}{\partial X} \\
 \frac{\partial u}{\partial Y}  &  \frac{\partial v}{\partial Y} 
\end{bmatrix}.
\end{equation}
The Cauchy stress tensor, pressure, specific internal energy, and specific kinetic energy are denoted as ${\upsigma}$, $p$, $e$, and $k$ respectively. The specific total energy is $\tau = e+k$. For gas dynamics, the stress is given by ${\upsigma} = -p \text I$, where the pressure is calculated using an equation of state (EOS) for the material, $p=\text{EOS}(\rho, e)$; for solid dynamics, the stress is given by ${\upsigma} = -p \text I + {\upsigma}^{\prime}$, where either a hypo- or hyper-elastic-plastic model is used to calculate the deviatoric stress ${\upsigma}'$.  The derivations in this work are valid for both gas and solid dynamics, where gas dynamics can be regarded as a special case. 
The time derivatives are total derivatives that move with the flow.  The rate of change of the position is,

\begin{equation}
\label{dx}
\frac{d {\textbf {\textit{x}}} }{dt}  ={{\textbf {\textit{u}}}}.
\end{equation}

\noindent  The equations presented in this paper will adhere to the following conventions. A slanted character is used to denote a scalar such as density $\rho$ and specific total energy $\tau$. A slanted bold character is used to denote a vector such as velocity $\textbf {\textit{u}}$ and Riemann forces ${\textbf {\textit{F}}}_{i}^*$.  
 An upright character denotes a tensor such as stress $\upsigma$, deformation gradient $\text F$, and Jacobian matrix $\text J$. A blackboard bold character $\mathbb{U}$  is used to denote an arbitrary unknown. The fluxes in a generalized evolution equation will be denoted as $\mathbb{H}$, which can be either a vector or tensor.   
Since the following sections involve both the  inner product of two vectors and the matrix product,
in order to describe these operations in the paper more easily and clearly, 
unless stated, the vectors (\textit{e.g.}, ${\textbf {\textit{u}}}$ and ${\textbf {\textit{n}}}$) are column vectors. 
Therefore, ${\textbf {\textit{n}}} \cdot {\textbf {\textit{u}}}$ denotes the inner product and ${\upsigma}{\textbf {\textit{u}}}$ (or ${\upsigma}{\textbf {\textit{n}}}$) means the matrix product.
In this work we follow the convention of matrix manipulation if the multiplication or product involves one matrix, and the dot product is used to represent the inner product of two vectors.

\subsection{Constitutive models}
\label{models}

We will begin with describing the EOS's for calculating the pressure with gases and solids followed by a discussion on two approaches to calculate the deviatoric stress with a solid material.
The gases are modeled using a gamma-law EOS model given by  

\begin{equation}
\label{gammalaw}
p = \rho e (\gamma -1),
\end{equation}

\noindent where $\gamma$ is the adiabatic index (ratio of specific heats).  
The pressure in solids is modeled using a Gr{\"u}neisen EOS model that is given by

\begin{equation}
\label{gruneisen}
\begin{aligned}
P_H (\eta)   &= \frac{\rho^0 c_0^2 \eta}{(1-s\eta )^2} , \\
E_H (\eta) &= \frac{\eta P_H}{2 \rho^0} , \\
p (\eta,e)  &= P_H+\Gamma\rho (e-E_H),  \\
\end{aligned}
\end{equation}

\noindent Here $\eta=1-\rho^0/\rho$, $\rho^0$ and $\rho$ are the initial and current density; $e$ is the specific internal energy; $P_H$ and $E_H$ are the Hugoniot pressure and energy;
$ \Gamma$ is the Gr{\"u}neisen coefficient; and $c_0$ and $s$ are parameters that relate the shock speed and the particle velocity according to $u_s = c_0+s u_p$. 


We consider strength formulations based on infinitesimal strain and also finite strain.  We will start by describing the infinitesimal strain formulation followed by the finite strain formulation.  The infinitesimal strain is $\upvarepsilon = sym({\text F} - {\text I})$ and the deviatoric infinitesimal strain is $\upvarepsilon^{\prime} = \upvarepsilon - \frac{1}{3} \text{tr}(\upvarepsilon)$.  The deviatoric stress is

\begin{equation}
\label{linearElastic}
{\upsigma}^{\prime}=2G{\upvarepsilon}^{\prime}_e,
\end{equation}

\noindent where $G$ is the material shear modulus and ${\upvarepsilon}^{\prime}_e$ is the deviatoric elastic strain considering the effect of the yield criterion, that is based on a $J_2$ radial return method. 
Assuming no new plastic deformation has occurred, a trial deviatoric stress $\hat {\upsigma}^{\prime}$ can be calculated by $\hat {\upsigma}^{\prime} = 2G({\upvarepsilon}^{\prime}-{\upvarepsilon}^{\prime}_p)$. The equivalent form of the Cauchy stress tensor is defined as,

\begin{equation}
\label{J2eq}
{\sigma}_{eq}=\sqrt{3J_2} = \sqrt{\frac{3}{2}{\upsigma}^{\prime}:{\upsigma}^{\prime}}
\end{equation}
The material yield stress is $Y$. If ${\sigma}_{eq}<Y$, the assumption that no new plastic occurs holds. This means the deviatoric plastic strain ${\upvarepsilon}^{\prime}_p$ is kept the same as the one of the previous time step. Therefore, 
\begin{equation}
\label{nonewplas}
\begin{split}
&{\upsigma}^{\prime} = \hat {\upsigma}^{\prime},\\
&{\upvarepsilon}^{\prime}_e={\upvarepsilon}^{\prime}-{\upvarepsilon}^{\prime}_p.
\end{split}
\end{equation}
If ${\sigma}_{eq}>Y$, then the $J_2$ radial return method is imposed to capture the plastic deformation. 
\begin{equation}
\label{newplas}
\begin{split}
&{\upsigma}^{\prime} = \hat {\upsigma}^{\prime}\frac{Y}{{\sigma}_{eq}},\\
& {\upvarepsilon}^{\prime}_p = {\upvarepsilon}^{\prime}-\frac{{\upsigma}^{\prime} }{2G}.
\end{split}
\end{equation}
\noindent
We will now shift the discussion to focus on the finite strain approach.

The finite strain model uses the Green-Lagrange strain ${\text E} = \frac{1}{2}( {\text C}- {\text I})$, where the  right Cauchy-Green strain ${\text C}={\text F}^{T}{\text F}$.
Accordingly, the above strain definitions can also be extended for the elastic (\textit{i.e.}, ${\text C}_e$ and ${\text E}_e$) or deviatoric strain (\textit{i.e.}, ${\text C}^\prime$ and ${\text E}^\prime$) if we use the elastic or deviatoric part (\textit{i.e.}, ${\text F}_e$ or ${\text F}^\prime$) of the deformation gradient. The deviatoric elastic Green-Lagrange strain ${\text E}^{\prime}_e$ is calculated using the deviatoric elastic deformation gradient ${\text F}^{\prime}_e = (\text{det}{\text F}_e)^{-\frac{1}{3}}{\text F}_e$. The elastic deformation gradient is calculated from the multiplicative decomposition of the deformation gradient ${\text F}_e = {\text F}{\text F}_p^{-1}$.  The second Piola-Kirchoff (PK2) stress $\tilde{\text S}$ is calculated by

\begin{equation}
\label{hElastic1}
\tilde{\text S}=2G{\text E}^{\prime}_e.
\end{equation}

\noindent The deviatoric PK2 stress in the intermediate configuration is found by

\begin{equation}
\label{hElastic2}
\tilde{\text S}^{\prime} = \tilde{\text S} - \frac{1}{3}\left(  \tilde{\text S}: ({\text F}_e^T {\text F}_e) \right) \left({\text F}_e^T {\text F}_e\right)^{-1}.
\end{equation}

\noindent The deviatoric Cauchy stress in the current configuration is found from the deviatoric PK2 stress using,

\begin{equation}
\label{hElastic3}
{ \upsigma}^{\prime} = \frac{1}{\text{det} {\text F}} {\text F}_e \cdot \tilde{\text S}^{\prime} \cdot {\text F}_e^T.
\end{equation}

\noindent Further details on the finite strain elastic-plastic model with the DG(P2) method can be found in Lieberman et al. \cite{LiebermanStrength2D2019}.

\section{Discretization}
\label{Discretization}
\begin{figure} 
	\centering
	\includegraphics[trim=0 0 0 0.1in, width=3.75in,height=!,clip=true]{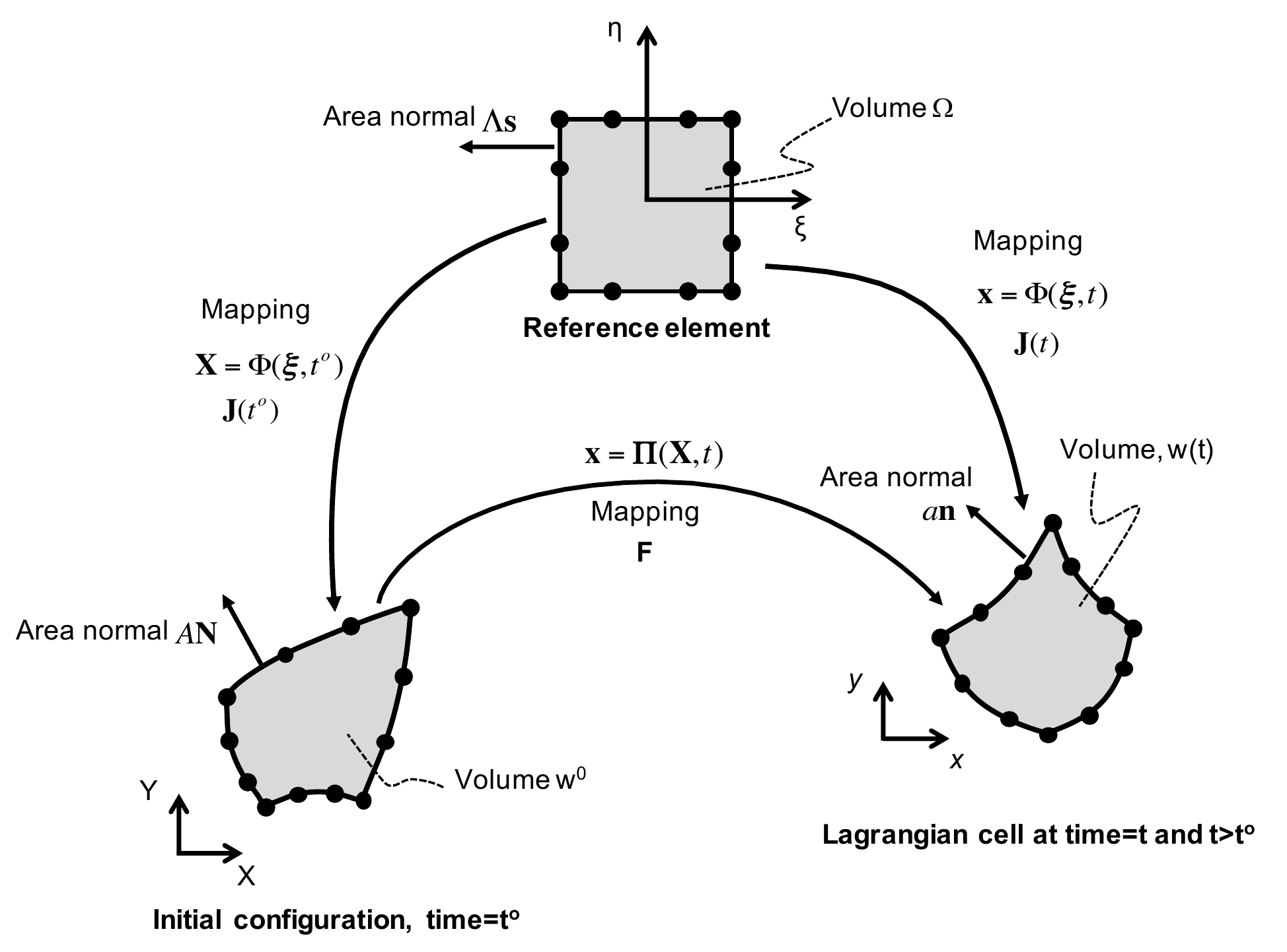}
	\caption{\label{NomenclatureFig} {The maps are illustrated for a cubic quadrilateral cell.  The cell map from the reference coordinates to the physical coordinates is created using the cubic serendipity Lagrange interpolation basis functions.  A fourth-order accurate Lagrangian hydrodynamic method requires a cubic cell.} }
   \end{figure}

The computational domain is decomposed into non-overlapping cells (also called elements or zones) that connect together to create a mesh.  
The initial cell volume is $w^0$, where the superscript $0$ denotes the initial time $t=t^0$.  Each cell has a constant mass, $\frac{dm}{dt}=0$. 
The cells will deform with the flow and the volume at a later time will be $w(t)$ (\fref{NomenclatureFig}). 
The deformation gradient $\text F$ and Jacobain matrix $\text J$  is, 

\begin{equation}
\label{Jacobian2}
{\text F} = \frac{\partial {\textbf {\textit{x}}}}{\partial {\textbf {\textit{X}}}}=
\begin{bmatrix}
 \frac{\partial x}{\partial X}  &  \frac{\partial x}{\partial Y} \\
 \frac{\partial y}{\partial X}  &  \frac{\partial y}{\partial Y} 
\end{bmatrix}, \quad
{\text J} = \frac{\partial {\textbf {\textit{x}}}}{\partial {\boldsymbol \xi}}=
\begin{bmatrix}
 \frac{\partial x}{\partial \xi}  &  \frac{\partial x}{\partial \eta} \\
 \frac{\partial y}{\partial \xi}  &  \frac{\partial y}{\partial \eta} 
\end{bmatrix}. 
\end{equation}

\noindent Here ${\textbf {\textit{X}}}, \textit{i.e.}, (X, Y)$ and ${\boldsymbol \xi}, \textit{i.e.}, (\xi, \eta)$ denote the coordinate system for the initial and reference configuration.  Both $\text F$ and $\text J$ vary in time.

\subsection{Basis functions for DG methods}
\label{polynomial}
In DG methods,  numerical polynomial solutions ${\mathbb U}_h$ in each cell can be represented using a range of basis functions including the Taylor basis \cite{ShuDG5jcp1998, luodgtaylor2008}, standard Lagrange finite element basis \cite{bassicurved1997}, or the Legendre-Dubiner basis \cite{dubinerDGbasis1991,Lidgadaptive2020}. Taylor basis functions have been used in many Lagrangian DG hydrodynamic methods; however, in this work we create a scheme based on new hierarchical orthogonal basis functions\cite{bassiorthodg2012}.  
The polynomial in a cell for $v$, $\textbf {\textit{u}}$, and $\tau$ can be written as

\begin{equation}
 {\mathbb U}_h({\boldsymbol \xi},t) = \sum \limits_k {\phi}_k \left( {\boldsymbol \xi} \right) \cdot {\mathbb U}_k \left( t \right)
 \quad \quad k = 1,...,N(P) 
\end{equation}

\noindent where $\mathbb U_h$ denotes the respective field, ${\phi}_k$ denotes the hierarchical orthogonal basis functions, ${\mathbb U}_k$ are the unknown coefficients and  $N(P)$ denotes the total number of terms in the numerical solution expansions for the polynomial with degree $P$ (  $N(P)$ is equal to 10 for DG(P3) in 2D).  The basis functions ${\phi}_k$ are calculated by applying the Gram-Schmidt orthogonalization to the Taylor-basis functions ${\psi}_k$ described in the following equation.  

 \begin{equation}
 \label{GramSchmidt}
             \begin{array}{l}
              \phi_1 = 1, \\
              \phi_2 = \psi_2 - \frac{ \langle \psi_2 , \phi_1 \rangle }{ \langle \phi_1 , \phi_1 \rangle } \phi_1, \\
              \phi_3 = \psi_3 - \frac{ \langle \psi_3 , \phi_2 \rangle }{ \langle \phi_2 , \phi_2 \rangle } \phi_2 - \frac{ \langle \psi_3 , \phi_1 \rangle }{ \langle \phi_1 , \phi_1 \rangle } \phi_1, \\
              \phi_4 = \psi_4 - \frac{ \langle \psi_4 , \phi_3 \rangle }{ \langle \phi_3 , \phi_3 \rangle } \phi_3 - \frac{ \langle \psi_4 , \phi_2 \rangle }{ \langle \phi_2 , \phi_2 \rangle } \phi_2 - \frac{ \langle \psi_4 , \phi_1 \rangle }{ \langle \phi_1 , \phi_1 \rangle } \phi_1, \\
              \vdots \\
               \phi_n = \psi_n - \frac{ \langle \psi_{n} , \phi_{n-1} \rangle }{ \langle \phi_{n-1} , \phi_{n-1} \rangle } \phi_{n-1} - \dots -  \frac{ \langle \psi_n , \phi_0 \rangle }{ \langle \phi_0 , \phi_0 \rangle } \phi_0,
            \end{array}   
 \end{equation}

\noindent where the inner product of two functions $g(\boldsymbol \xi)$ and $h(\boldsymbol \xi)$ is defined as,

\begin{equation}
\label{Orthog}
               \langle g , h \rangle = \int \limits_{\Omega_h} \rho g h j^0d\Omega.
\end{equation}

\noindent The Taylor basis functions up to P3 for $v_h$, $\textbf {\textit{u}}_h$, and $\tau_h$ are about the center of mass ($cm$) and defined as

\begin{equation}
\label{TaylorBasis-details}
\begin{array}{l}
 \psi_1 = 1, \\
 \psi_2 = \xi - \xi_{cm} ,  \\
 \psi_3 = \eta - \eta_{cm}, \\ 
 \psi_4 = ( \xi - \xi_{cm})^2 ,\\
 \psi_5 = ( \eta - \eta_{cm})^2,\\
 \psi_6 = ( \xi - \xi_{cm})( \eta - \eta_{cm}) ,\\
 \psi_7 =  (\xi - \eta_{cm})^3, \\
 \psi_8 =   (\eta - \eta_{cm})^3, \\
 \psi_9 =  (\xi - \xi_{cm})^2(\eta - \eta_{cm}), \\
 \psi_{10} =   (\xi - \xi_{cm})(\eta - \eta_{cm})^2.
\end{array}  
\end{equation}

\noindent Here, the subscript $cm$ denotes the center of mass, that is defined by 
 ${\boldsymbol{\xi}_{cm}}= \frac{1}{m} {\int \limits_{w_h} \rho {\boldsymbol{\xi}} dw}$, $m$ denotes the mass of the cell, namely $m={\int \limits_{w_h} \rho dw}$.  
In \ref{BFproof}, we show that the coefficients(\textit{i.e.},  $ \frac{ \langle \psi_{2} , \phi_{1} \rangle }{ \langle \phi_{1} , \phi_{1} \rangle } $,  
$ \frac{ \langle \psi_{3} , \phi_{2} \rangle }{ \langle \phi_{2} , \phi_{2} \rangle } $,$ \frac{ \langle \psi_{3} , \phi_{1} \rangle }{ \langle \phi_{1} , \phi_{1} \rangle } $, etc.)  are temporally constant, meaning this set of coefficients are only calculated once at the beginning of the calculation. Temporally constant basis functions, $\frac{d{\phi}_k}{dt} =0$, are an essential ingredient in the presented Lagrangian hydrodynamic method.

The discussion will now address the polynomial for the deformation gradient.  The basis functions for $\text F_h$ are about the center of the reference cell $\boldsymbol{\xi}_c = \frac{1}{v} {\int \limits_{w^0} {\boldsymbol{\xi}} dw}$ because it is a per volume quantity. Likewise, the Taylor basis functions are,

\begin{equation}
\label{TaylorBasis-details2}
\begin{array}{l}
\hat \psi_1 = 1,  \\
\hat  \psi_2 = \xi - \xi_{c} ,\\
\hat \psi_3 = \eta - \eta_{c}, \\ 
\hat  \psi_4 = ( \xi - \xi_{c})^2 , \\
 \hat  \psi_5 = ( \eta - \eta_{c})^2,\\ 
 \hat  \psi_6 = ( \xi - \xi_{c})( \eta - \eta_{c}) ,\\
 \hat \psi_7 =  (\xi - \eta_{c})^3,\\
 \hat \psi_8 =   (\eta - \eta_{c})^3,\\ 
 \hat  \psi_9 =  (\xi - \xi_{c})^2(\eta - \eta_{c}),\\
 \hat  \psi_{10} =   (\xi - \xi_{c})(\eta - \eta_{c})^2.
\end{array}
\end{equation}

\noindent The orthogonal basis function for $\text F_h$ are denoted as $\hat {\boldsymbol{\phi}}$ and are calculated using Eq. $\ref{Orthog}$ without the density.  The basis functions are constant in the calculation.  
If the initial density is constant in the cell, then the Taylor basis functions are identical for the per mass quantities and per volume quantities, namely ${\boldsymbol {\psi}}=\hat {\boldsymbol {\psi}}$.

\subsection{Discontinuous Galerkin method}
\label{DGsection}
In this section, we present a DG discretization of the governing evolution equations. 
The DG approach creates a system of equations for evolving the polynomial coefficients for $v$, $\textbf {\textit{u}}_h$, $\tau_h$, and $\text F_h$.  The evolution equations for specific volume, velocity, and specific total energy have the general form of

\begin{equation*}
\label{GeneralForm}
\rho \frac{ d \mathbb{U}_h }{dt}  = \nabla \cdot \mathbb{H}
\end{equation*}

\noindent where $\mathbb{U}_h$ is the respective field and $\mathbb{H}$ is the associated flux.  This equation is multiplied by a test function and integrated over the cell in the physical coordinates,

\begin{equation*}
\int \limits_{w_h} \phi_m \left( \rho \frac{d \mathbb{U}_h }{dt} - \nabla \cdot \mathbb{H} \right) dw = 0, \quad m =1,..., N(P)
\end{equation*}

\noindent Substituting the polynomial expansion for the field gives,

\begin{equation*}
\label{SomeEqn1}
\sum \limits_n \int \limits_{w_h} \rho  {\phi_m  \phi_n} dw  \frac{d\mathbb{U}_n}{dt} = \int \limits_{\Omega_h} {\phi_m} (\nabla \cdot \mathbb{H}) dw,  \quad \quad m = 1,...,N(P)
\end{equation*}

\noindent The first term is the mass matrix for the cell, ${M}_{mn} = \sum \limits_k \int \limits_{\Omega_h} \rho  {\phi_m  \phi_n} dw$, and it is constant in time.  
Since the basis functions are orthogonal, the mass matrix is diagonal. Using integration by parts for the $rhs$, the DG equations become, 

\begin{equation*}
\label{DGEquation0}
 {\text M}_{mm} \frac{d\mathbb{U}_m}{dt} = 
\int \limits_{w(t)} 
 \nabla  \cdot ( {\phi}_m \mathbb {H} )dw
- \int \limits_{w(t)}  
 \mathbb {H} \cdot (\nabla {\phi}_m) dw  , \qquad m = 1,...,N(P),
\end{equation*}
\noindent where $\nabla {\phi}_m = [\frac{\partial {\phi}_m}{\partial x} \quad \frac{\partial {\phi}_m}{\partial y}]^T$.
The DG equations require solving a Riemann problem on the surface of the reference cell to resolve the discontinuity inside the polynomials.  The flux from the Riemann solver is denoted by a superscript $*$.  Using the Riemann solution and transforming the volume integral  terms  to be in the reference coordinate system,
the resulting evolution equations for the unknown basis coefficients for specific volume $v_n$, velocity ${\textbf {\textit{u}}}_n$, and specific total energy $\tau_n$ are,

\begin{equation}
\label{DGEquation-SpecV}
 {\text M}_{mm} \frac{d{v}_n}{dt} = \oint \limits_{\partial w(t)} {\phi}_m ({\textbf {\textit{n}}}\cdot{\textbf {\textit{u}}}^*) da -
\int \limits_{\mathit{\Omega}} {\textbf {\textit{u}}} \cdot \left(({\text J}^{-1})^T\nabla_\xi {\phi}_m \right) j d{\mathit{\Omega}},
\end{equation}

\begin{equation}
\label{DGEquation-Momen}
 {\text M}_{mm} \frac{d{\textbf {\textit{u}}}_n}{dt} = \oint \limits_{\partial w(t)} {\phi}_m ({\upsigma}^*{\textbf {\textit{n}}}) da -
	\int \limits_{\mathit{\Omega}}  {\upsigma} \left(({\text J}^{-1})^T\nabla_\xi {\phi}_m \right)  j d{\mathit{\Omega}},
\end{equation}

\begin{equation}
\label{DGEquation-Energy}
 {\text M}_{mm} \frac{d{\tau}_n}{dt} = \oint \limits_{\partial w(t)}  {\phi}_m {\textbf {\textit{n}}} \cdot ({\upsigma}^* \textbf {\textit{u}}^*)  da -
	\int \limits_{\mathit{\Omega}}  ({\upsigma}\textbf {\textit{u}}) \cdot \left(({\text J}^{-1})^T \nabla_\xi {\phi}_m \right)  j d{\mathit{\Omega}}.
\end{equation}

\noindent  For the $rhs$, the first (\textit{i.e.}, the surface integral) and second (\textit{i.e.}, the volume integral) terms are evaluated by Gauss
quadrature formulas. This will be discussed in detail in \sref{RHS-cal}. All the volume integrals can be calculated on the reference cell by 

\begin{equation}
\label{dwdomega}
 dw = jd\Omega,
\end{equation}
where $j$ is the determinant of the Jacobian matrix \text{J}. 

The discussion will now focus on the DG approach applied to the deformation gradient.  Multiplying the evolution equation by the test function and integrating over the initial cell gives,

\begin{equation*}
\int \limits_{w^0_h} \hat\phi_m \left( \frac{d {\text F}_h  }{dt} - \nabla_{\textbf {\textit{X}}} {\textbf {\textit{u}}}\right) dw^0 = 0, \qquad m=1,...,N(P).
\end{equation*}
\noindent By introducing the volume matrix for the cell, ${V}^0_{mn} = \int \limits_{\Omega_h}  {\hat\phi_m  \hat\phi_n} dw^0$,
the previous system of equations reduces to

\begin{equation}
\label{DGEquation2}
 {V}^0_{mm} \frac{d\text {F}_m}{dt} = \oint \limits_{\partial w_h^0} \hat{\phi}_m (\textbf {\textit{n}} \cdot  \textbf {\textit{u}}^*) dA -
\int \limits_{\mathit{\Omega}} \textbf {\textit{u}} \cdot (({\text J}^{-1})^T {\nabla_\xi \, \hat{\phi}_m })  j^0 d{\mathit{\Omega}} , \quad \quad m = 1,...,N(P)
\end{equation}
\noindent where ${V}^0_{mm}$ is the $m^{th}$ row and $m^{th}$ column of the mass matrix.  There is no matrix to invert.  The details on the derivation of the DG evolution equations with Taylor basis functions for $v$, $\textbf {\textit{u}}_h$, $\tau_h$, and $\text F_h$ can be found in papers \cite{LiuDGlagSMS2019, LiebermanStrength2D2019, LiebermanHE2020}.

\subsection{Temporal integration}
In this work, explicit strong-stability-preserving (SSP) Runge-Kutta (RK) schemes \cite{ShuTVD1988,ShuOsherENO1988,SpiteriRK5,ShuDG5jcp1998} are used to evolve the semi-discrete equations for each polynomial coefficient from time level $n$ to $n+1$.  
We consider the three-stage third-order accurate TVD RK method (denoted as SSPRK(3,3) in the paper) and the five-stage fourth-order accurate SSP RK method \cite{SpiteriRK5} (denoted as SSPRK(5,4) in the paper).  The SSPRK(3,3) time integration has been applied to the third-order accurate Lagrangian DG(P2) method and is discussed by a range of papers, see \cite{VilarDGlagjcp2014, LiuDGlagSMS2019, LiebermanStrength2D2019}. The equation to evolve a polynomial coefficient ${\mathbb U}_k$ has the general form of

\begin{equation}
\frac{d {\mathbb U}_k}{dt} = \frac{1}{{\text A}_{kk}} \mathbb{R}_k
\end{equation}

\noindent where ${\text A}_{kk} = \text M_{kk}$ when evolving the coefficients for $v$, $\textbf {\textit{u}}$, and $\tau$; likewise, $\text A_{kk} = \text V_{kk}^0$ when evolving the coefficients for $\text F$.  RK methods for advancing ${\mathbb U}_k$ forward in time from time level $n$ to $n+1$ can be written as

\begin{equation}
\label{RKmethod}
\begin{array}{lll}
{\mathbb U}^{(0)}_k  &=&   {\mathbb U}^{n}_k  \\
{\mathbb U}^{(i)}_k   &=&  \sum \limits_{j=0}^{i-1} \alpha^{i,j} {\mathbb U}^{(j)}_k   +  \frac{\Delta t}{{\text A}_{kk}} \sum \limits_{j=0}^{i-1} \beta^{i,j} {\mathbb{R}_k^{(j)}}, \quad i=1,...,s \\
{\mathbb U}^{n+1}_k  &=&   {\mathbb U}^{(s)}_k. 
\end{array}
\end{equation}

\noindent The superscripts $0$, $i$, and $j$ denote the solution at a RK step, where $s$ is the number of stages.  Table \ref{rkcoefs} presents the coefficients $\alpha^{i,j}$ and  $\beta^{i,j}$.  The mesh position is temporally advanced using the same RK method as used to advance the polynomial coefficients.

\begin{equation}
\label{RKmethod2}
\begin{array}{lll}
{  \textbf{\textit{x}} }^{(0)}_V  &=&   {  \textbf{\textit{x}} }^{n}_V  \\
{  \textbf{\textit{x}} }^{(i)}_V   &=&  \sum \limits_{j=0}^{i-1} \alpha^{i, j} { \textbf{\textit{x}} }^{(j)}_V   +  \Delta t \sum \limits_{j=0}^{i-1} \beta^{i,j} {  \textbf{\textit{u}} }_V^{*(j)}, \\
{  \textbf{\textit{x}} }^{n+1}_V  &=&  {  \textbf{\textit{x}} }^{(s)}_V. 
\end{array}
\end{equation}

\noindent where ${ \textbf{\textit{x}} }_V$ is the position of a vertex and ${  \textbf{\textit{u}} }_V^{*}$ is the velocity of the vertex that comes from the Riemann solver.

In the context of DG(P3) methods, we found no obvious advantage for the SSPRK(5,4) over SSPRK(3,3) in terms of the numerical accuracy on test cases. For instance, the accuracy of the calculations of the Taylor-Green vortex test problem (\sref{TGV}) was the same with both time integration methods for the mesh resolutions and time step sizes considered. As such, robust, accurate, and computationally efficient solutions are possible using SSPRK(3,3) with this DG(P3) method on cubic cells.

\begin{table}
\scriptsize
\centering
\caption{The coefficients for two RK time integration methods are shown.  SSPRK(3,3) has three steps and is third-order accurate.  SSPRK(5,4) has five steps and is fourth-order accurate. }
 \begin{tabular}{c c c c c c} 
 \hline
 SSPRK(3,3),    Stages s=3\\
 \hline
 \multirow{3}{1em}{$\alpha^{i,j}$}	& 1 \\
					   		& 3/4 & 1/4 \\
					   		& 1/3 & 0 & 2/3   \\
\hline
 \multirow{3}{1em}{$\beta^{i,j}$}	& 1 \\
							& 0 & 1/4 \\
							& 0 & 0 & 2/3   \\
\hline
\\
\\
 \hline
SSPRK(5,4),   Stages s=5 \\
 \hline
 \multirow{5}{1em}{$\alpha^{i,j}$}	& 1 \\
							& 0.444370493651235  & 0.555629506348765 \\
							& 0.620101851488403 & 0 & 0.379898148511597   \\
							& 0.178079954393132 & 0 & 0 & 0.821920045606868 \\
							& 0 & 0 & 0.517231671970585 & 0.096059710526147 & 0.386708617503269 \\
\hline
 \multirow{5}{1em}{$\beta^{i,j}$}	& 0.391752226571890 \\
							& 0 & 0.368410593050371 \\
							& 0 & 0 & 0.251891774271694  \\
							& 0 & 0 & 0 & 0.544974750228521 \\
							& 0 & 0 & 0 & 0.063692468666290 & 0.226007483236906   \\
\hline
\end{tabular}
\label{rkcoefs}
\end{table}

\subsection{Quadrature rules}
\label{RHS-cal}

Gauss quadrature formulas are used to evaluate the surface and volume integrals in the DG equations. 
The quadrature rules are chosen to exactly integrate the integrals, where 
the number of quadrature points depends on both the polynomial of degree $P$ and the order of the cell. 
On a linear cell, the number of quadrature points is selected to integrate exactly polynomials of order $2P+1$ on the reference cell.
For quadratic and cubic meshes, the number of the quadrature points increases by 1 and 2 respectively in theory. 
The number of quadrature points used to solve the surface and volume integrals is shown in \tref{tab:quadrature-no}. 

For Lagrangian hydrodynamics, it is essential to use a nodal Riemann solver to calculate a vertex velocity ${\textbf {\textit{u}}}_V^*$ that is then used to spatially move the vertex (\textit{i.e.}, for the vertices at the corner and along the edge). 
Therefore, Gauss-Lobatto quadrature rules are used to calculate the surface integrals as they will spatially overlap some or all of the vertices on the cell surface. For Lagrangian hydrodynamics, the mesh is updated every time step.
So another issue is how to guarantee the vertices always overlap the quadrature points especially for the edge vertices. 
Due to a favorable property in Lagrangian hydrodynamics,  

\begin{equation}
\label{dxidteq0}
\frac{d \boldsymbol \xi}{dt} =0,
\end{equation}

\noindent
so the reference coordinates ${\boldsymbol \xi}$ for any quadrature point is the same as the initial one. 
In other words, as long as the initial vertex overlaps the Gauss-Lobatto quadrature point, then the vertex can always be used as the quadrature point (\tref{tab:vertex-quadrature}). Table \ref{tab:Lobatto-rule} in the Appendix presents the Gauss-Lobatto quadrature point locations and values. The volume integrals are calculated using the Gauss-Legendre quadrature rules because they are more accurate than the Gauss-Lobatto rules using the same number of quadrature points. 

For higher-order DG methods, it is necessary to solve 1D Riemann problems at multiple locations along the edge in addition to that at the vertices.
As such, the reference coordinate locations for the surface vertices will coincide with some of the Gauss-Lobatto quadrature point locations, see \tref{tab:vertex-quadrature}. For cubic meshes, the number of required Lobatto quadrature points to exactly integrate the $rhs$ varies, meaning that
the initial cubic meshes are slightly different for DG methods with different polynomial degrees. 
The details on the Riemann solvers will be discussed in the following subsections. 
 
\begin{table*}[t]
\begin{center}
\caption
{The information for the Gauss quadrature point number is shown in the context of DG methods. Gauss-Lobatto quadrature points are used for the surface integral 
because they spatially overlap the vertices where a nodal Riemann solver is used. The Gauss-Legendre quadrature points are adopted for the volume integral due to their high accuracy. For the surface integral, the number in the parentheses is not used since we want to use the Riemann information at the vertices.}
\label{tab:quadrature-no}
\mbox{
\begin{tabular}{l c c c c c c c c}
\hline
 & DG(P1) &  DG(P2) &  DG(P3)\\
 \hline
Surface integral \\
\hline
Linear edge         & 3  &    4  &   5\\
Quadratic edge    & 5 (4) &    5  &   6\\
Cubic edge          & 6  (5)  &   6  &   7\\
\hline
Volume integral\\
\hline
Linear cell        & 2   & 3 & 4\\
Quadratic cell   & 3   & 4  & 5\\
Cubic cell         & 4   & 5  & 6\\
\hline
\end{tabular}}
\end{center}
\end{table*}

\begin{table*}[t]
\begin{center}
\caption
{The position in the reference coordinates for each surface vertex position is shown for differing orders.  The reference coordinates are over interval $[-1, 1]$. The vertices overlap all or only some of the surface quadrature points.
 For a cubic edge, different quadrature rules are used for DG methods with differing polynomial orders, which leads to using different initial cubic meshes.}
\label{tab:vertex-quadrature}
\mbox{
\begin{tabular}{l c c c c c c c c}
\hline
 & \textcircled{1} &  \textcircled{2} &  \textcircled{3} & \textcircled{4}\\
\hline
Linear edge \\
\hline
        & -1    & 1 \\
\hline
Quadratic edge \\
\hline
    & -1   &  0   &   1\\
\hline
Cubic edge \\
\hline
6-point quadrature rule  & -1   &  $-\sqrt{\frac{1}{21}\left( 7 - 2\sqrt{7} \right) }$   &  $\sqrt{\frac{1}{21}\left( 7 - 2\sqrt{7} \right) }$  & 1\\
7-point quadrature rule  & -1   &  $-\sqrt{\frac{5}{11}-\frac{2}{11}\sqrt{\frac{5}{3}} }$   &  $\sqrt{\frac{5}{11}-\frac{2}{11}\sqrt{\frac{5}{3}} }$  & 1\\
\hline
\end{tabular}}
\end{center}
\end{table*}

\subsubsection{Riemann solvers}
\label{RiemannProblem}

The surface integral requires solving Riemann problems at the quadrature points along the cell surface. 
In this work, a multi-directional approximate Riemann solver (MARS) is used at the vertices of the cell and a 1D Riemann solver is used at additional locations along the cell edge.
 MARS are widely used in the Lagrangian finite volume CCH where the first one was proposed by B. Despr{\'e}s and C. Mazeran \cite{Despres} for gas dynamics problems.  Maire and various authors \cite{Maire1, Maire2} extended the work in \cite{Despres} and proposed a new MARS that had improved mesh robustness properties.  Burton \textit{et al.} \cite{BurtonCCH} proposed another robust MARS that was suitable for materials with strength.  We use the MARS by Morgan \textit{et al.} \cite{MorganVeloFilter2017} and the discussion that follows will briefly describe that MARS.  

The force acting on a surface segment connected to the vertex is given by

\begin{equation}
\label{RiemannForce}
{\textbf {\textit{F}}}_{i}^* = a_{i}{\upsigma}_{i}^* {\textbf {\textit{n}}}_{i}= a_{i}{\upsigma}_{c} {\textbf {\textit{n}}}_{i} + \mu_{i}  a_{i} \left( {  {\textbf {\textit{u}}}_V^* - {\textbf {\textit{u}}}_{c}  }\right).
\end{equation}

\noindent The subscript $c$ denotes the value in the cell at the vertex, the subscript $i$ denotes the surface segment, the subscript $V$ is a vertex value, and the superscript $*$ denotes a variable that comes from solving a Riemann problem. The shock impedance \cite{MorganSGHdisp} is 

\begin{equation}
\label{RF-impedance}
\mu_{i} = \rho_z(c_z + \frac{\gamma+1}{2} | ( {  {\textbf {\textit{u}}}_V^* - {\textbf {\textit{u}}}_{c}  })\cdot {\textbf {\textit{n}}}_{i}|),
\end{equation}

\noindent where the subscript $z$ denotes the mass average value.
Momentum conservation at the vertex $V$ requires that the summation of all forces around the vertex to be equal to zero. 
A single Riemann velocity is found at the vertex that guarantees momentum conservation.

\begin{equation}
\label{RiemannVelocity}
{\textbf {\textit{u}}}_V^* = \frac{{\sum\limits_{i\in V} { \left( { \mu_{i}  a_i  {\textbf {\textit{u}}}_{c} - a_i {\boldsymbol{\upsigma}_{c}} } {\textbf {\textit{n}}}_i \right)}  }}{{\sum\limits_{i \in V} { \mu_{i} a_i} }}.
\end{equation}

\noindent Analysis is presented in \cite{LiuDGlagcaf2017} to show that the Lagrangian DG method, combined with this Riemann solver, satisfies the second-law of thermodynamics. We also use the subcell surfaces (Section \ref{SMSsec}) with the MARS so that there is consistency in solving for the Riemann velocity (and associated forces) between the corner and edge vertices (\fref{Nomenclature3Fig}).

\begin{figure} 
	\centering
	\subfloat[The corners for the vertex quadrature points]{
	\includegraphics[trim=0 0 0 0.0in, width=3.5in,height=!,clip=true]{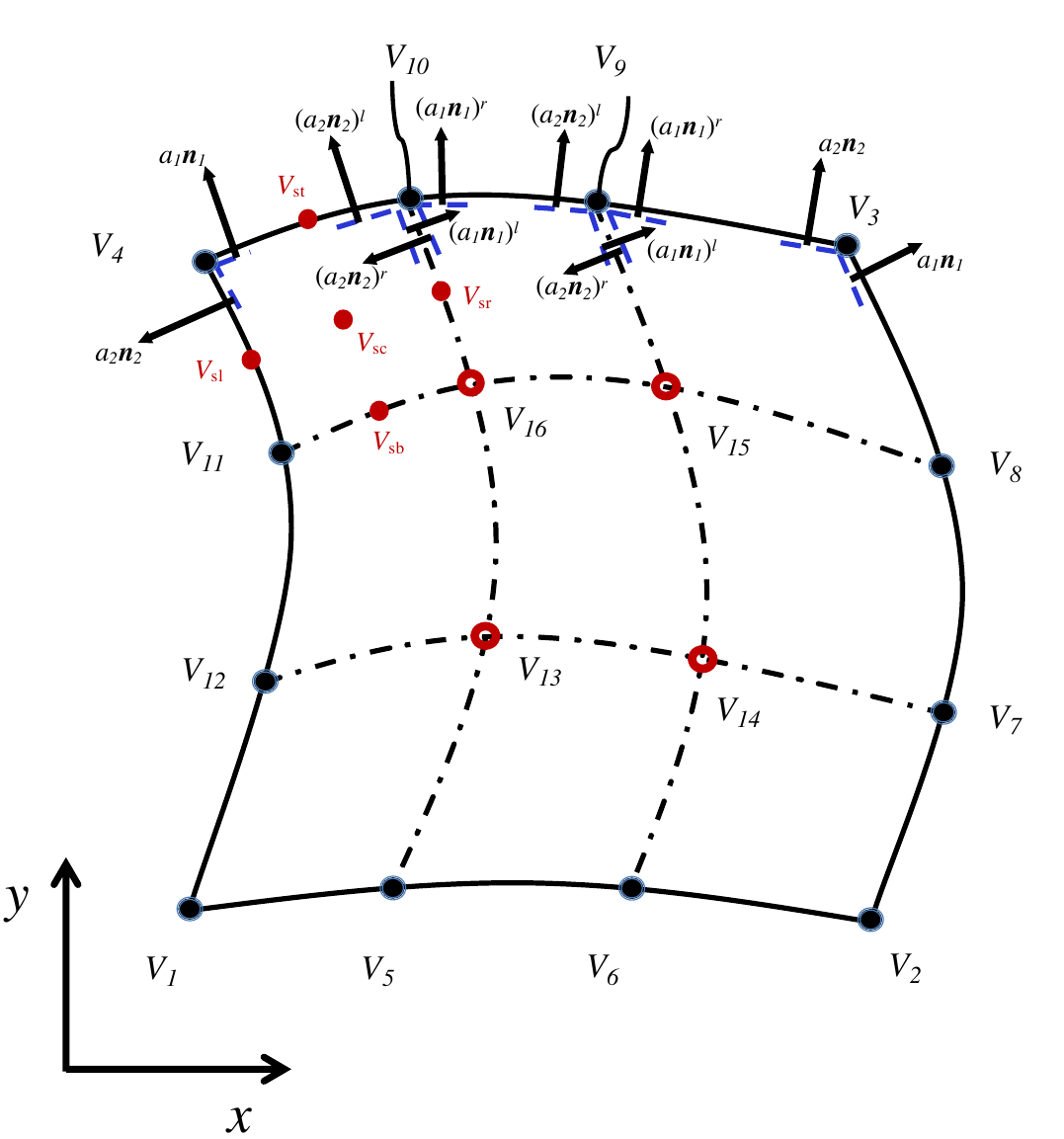}}
	\subfloat[The corners for the non-vertex quadrature points]{
	\includegraphics[trim=0 0 0 0.0in, width=3.5in,height=!,clip=true]{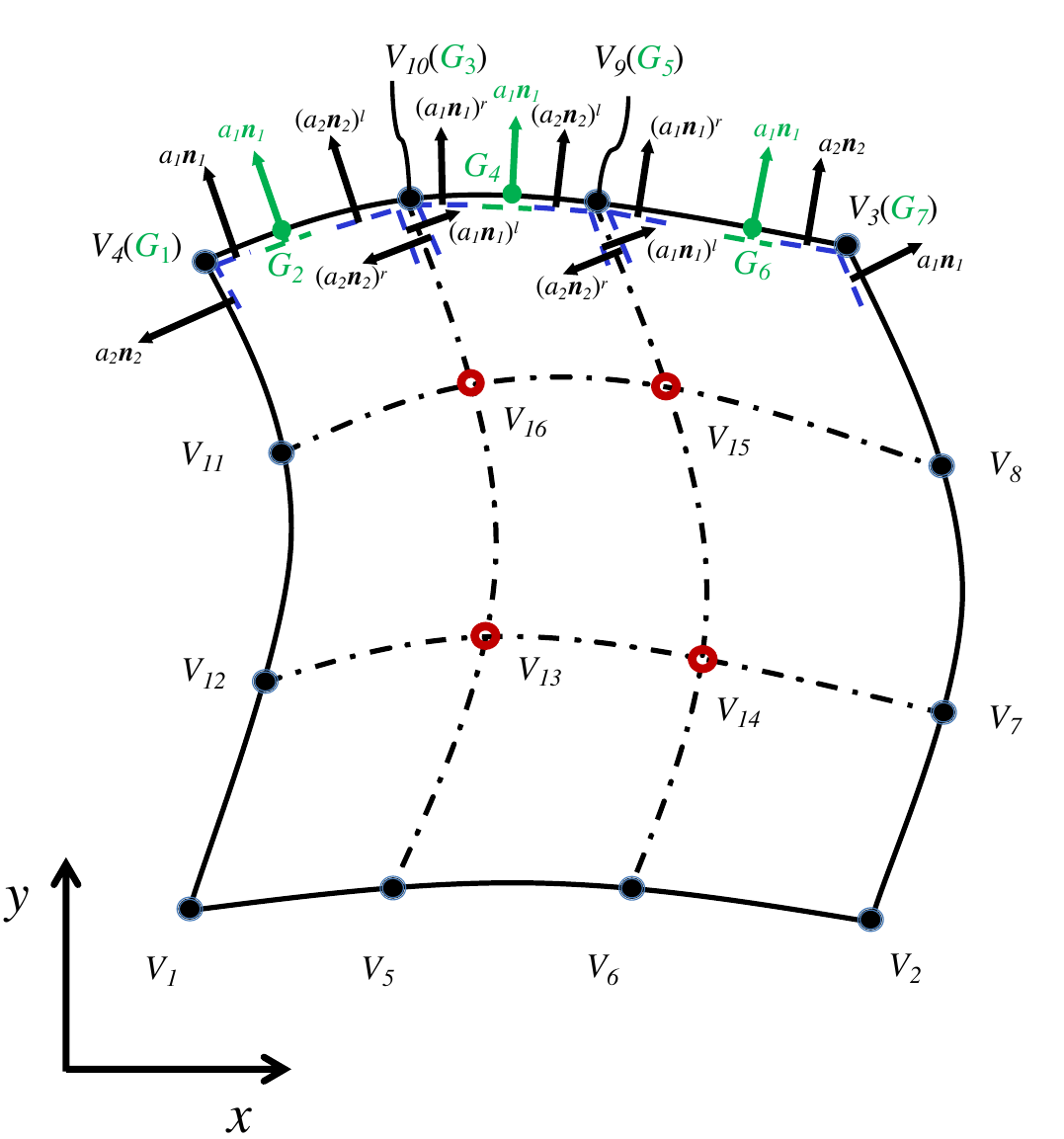}}
	\caption{\label{Quadraturept} The nomenclatures along the surface $V_3V_4$  in the current coordinate system. (a) In cell 1, there are two segments associated with every corner. And they are ordered in an anticlockwise way. For the edge vertex $V_9$ and $V_{10}$, there exist two corners each. The superscript $l$ and $r$ denote the left and right corner associated with the edge vertex. (b) The quadrature points are marked by green.
	 In addition to the quadrature points overlapping the vertices (\textit{i.e.}, $G_1$, $G_3$, $G_5$, $G_7$),  the non-vertex quadrature points (\textit{i.e.}, $G_2$, $G_4$ and $G_6$)  are also required. Different from the vertex quadrature point, in cell 1, only one segment $a_{1}  {\textbf {\textit{n}}}_{1}$ contributed to the non-vertex quadrature point.}
      \end{figure}

The accuracy of the surface integral for the DG(P3) method is increased by using more quadrature points along the edge.  The goal is to increase the quadrature order without adding more kinematic degrees of freedom to the cell. The resulting Riemann solutions correspond to the Gauss-Lobatto quadrature rule where the integration weights and locations in the reference coordinate system are provided in Table \ref{tab:Lobatto-rule}.  The 6-point quadrature rule is exact for integrating a polynomial of order $9$, and is substantially more accurate than using the 4 points that coincide with the vertices along the cell edge.  
The discussion that follows will focus on the 1D Riemann solutions at the non-vertex quadrature points that are illustrated in \fref{Quadraturept}.  

For the 1D Riemann problem, there will be two forces acting on a surface segment of the cell.
\begin{equation}
\label{RiemannForce-G5}
\begin{split}
{\textbf {\textit{F}}}_{1}^* = a_{1} {\upsigma}_{1}^* {\textbf {\textit{n}}}_{1}= a_{1}{\upsigma}_{c1} {\textbf {\textit{n}}}_{1} + \mu_{1}  a_{1} \left( {  {\textbf {\textit{u}}}_G^* - {\textbf {\textit{u}}}_{c1}  }\right)\\
{\textbf {\textit{F}}}_{2}^* = a_{2} {\upsigma}_{2}^* {\textbf {\textit{n}}}_{2} = a_{2}{\upsigma}_{c2} {\textbf {\textit{n}}}_{2}+ \mu_{2}  a_{2} \left( {  {\textbf {\textit{u}}}_G^* - {\textbf {\textit{u}}}_{c2}  }\right)
\end{split}
\end{equation}

\noindent Here, the subscripts $1$ and $2$ denote the respective inputs from the two cells that share this surface segment,
 and ${\textbf {\textit{u}}}_G^*$ denotes the Riemann velocity at the non-vertex Gauss quadrature point.  
The surface weighted area normals for the segment are equal and opposite, so $a_{1}{\textbf {\textit{n}}}_{1} = -a_{2}{\textbf {\textit{n}}}_{2}$. 
A single Riemann velocity ${\textbf {\textit{u}}}_G^*$ can be calculated by enforcing momentum conservation $\sum \limits_{i} {\textbf {\textit{F}}}_{i}^* = 0$.  
For a non-vertex Gauss quadrature point, the calculated Riemann velocity ${\textbf {\textit{u}}}_G^*$ is not necessarily equal to the velocity of the edge ${\textbf {\textit{u}}_G}$.
Along an edge, the velocity ${\textbf {\textit{u}}_G}$ at the non-vertex Gauss quadrature point is calculated using Lagrangian interpolation where the coefficients are the Riemann velocities at the vertices of this edge ${\textbf {\textit{u}}}_V^*$.  
The fluxes in the specific volume evolution equation at the non-vertex quadrature points are calculated using the edge velocity ${\textbf {\textit{u}}_G}$.
The fluxes in the specific total energy evolution equation at the non-vertex quadrature points are calculated by multiplying the forces (\eref{RiemannForce-G5}) by the edge velocity ${\textbf {\textit{u}}_G}$,   

\begin{equation}
\label{RiemannForceVel-G5}
\begin{split}
{\textbf {\textit{F}}}_{1}^* \cdot {\textbf {\textit{u}}_G} 
 = a_{1} {\upsigma}_{1}^* {\textbf {\textit{n}}}_{1} \cdot {\textbf {\textit{u}}_G}  \\
 {\textbf {\textit{F}}}_{2}^* \cdot {\textbf {\textit{u}}_G} 
 =  a_{2} {\upsigma}_{2}^* {\textbf {\textit{n}}}_{2} \cdot {\textbf {\textit{u}}_G} 
\end{split}
\end{equation}

\noindent These total energy fluxes guarantee the conservation of total energy.

\subsection{Subcell Mesh Stabilization (SMS)}
\label{SMSsec}
\begin{figure} 
      \centering
       \includegraphics[trim=0 0 0 0.0in, width=3.5in,height=!,clip=true]{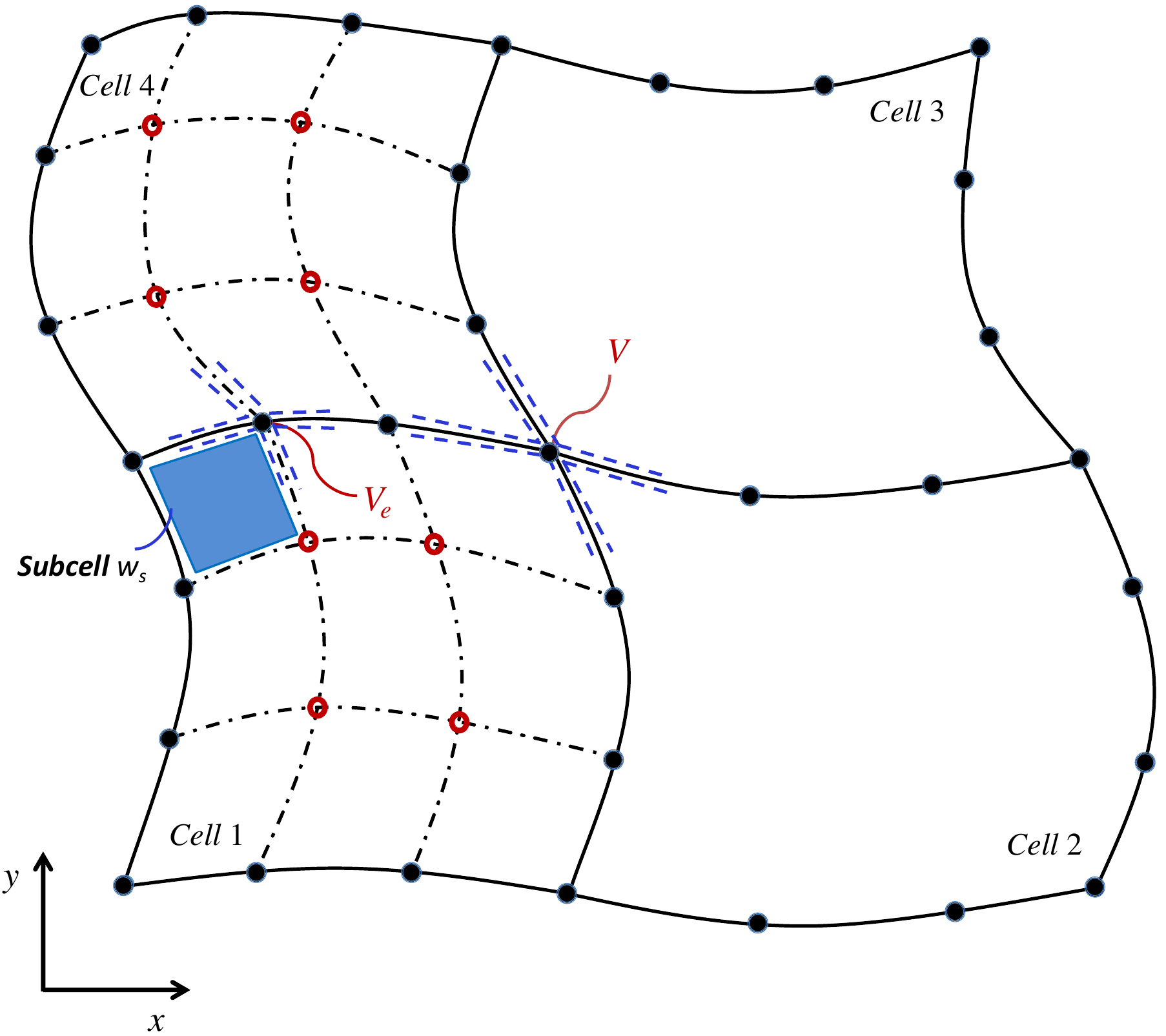} 
      \caption{\label{Nomenclature3Fig} 
      A patch of four cells is shown in the current coordinate system.
      In the context of SMS, a multi-directional approximate Riemann problem is solved at the vertex $V$ in the corner of the cell and the edge vertex $V_e$ of the cell. 
      The control surface for the Riemann solver is constructed from all segments connected to the corner vertex $V$ (or the edge vertex $V_e$) and is denoted as $i \in V$ (or $i \in V_e$). 
      The control surface for the Riemann solver is highlighted in the figure with a dashed-line. }
      \end{figure}

In this work, the SMS scheme \cite{LiuDGlagSMS2019} is extended to work with cubic cells.  The goal of SMS is to remove spurious vorticity that can arise on flows with strong shocks and to obtain a more consistent Riemann solution between the corner and edge vertices.
For a corner vertex $V$ shown in \fref{Nomenclature3Fig}, there are eight segments contributing to the Riemann problem. 
This motivates us to use more segments to construct the Riemann problem for an edge vertex by creating 
subcells that surround the edge vertex. A cubic cell is decomposed into nine subcells (\fref{Nomenclature3Fig}).  The subcell surfaces are used in the MARS at the edge vertices. 
The subcell decomposition also enables us to introduce corrective pressures in the MARS to remove spurious vorticity.
The concept in SMS is to calculate a density for each subcell and then correct the pressure inputs to the MARS at every exterior vertex if the subcell density deviates from the modal density field for the cell.  The subscells move in a Lagrangian manner so the average density of a subcell is

\begin{equation}
\label{ModalDensityEqn1}
 \bar { \rho}_{s} =  \frac{ m_s }{ w_s(t)} = \frac{\int \limits_{\Omega_s} \rho({\boldsymbol \xi}, t^0)  j_s^0 d\mathit{\Omega} }{ \int \limits_{\Omega_s} j_s d\mathit{\Omega}}.
\end{equation}

\noindent  The subscript $s$ denotes the variables for the subcell, the determinant of the Jacobian matrix for the subcell is denoted as $j_s$ and the superscript $0$ denotes the initial configuration.  Inconsistencies between the cell density field and the subcell density may arise.  The average density in the subcell using the specific volume field $v_h$ that is evolved with the DG approach is

\begin{equation}
\label{DensitySg}
\bar \rho_{vs} = \frac{\int \limits_{w_s(t)} \frac{1}{v_h} dw}{w_s(t)} = \frac{\int \limits_{\Omega_s} \frac{1}{v_h} j_s d{\Omega}}{w_s(t)},
\end{equation}

\noindent SMS adds a correction term to the density used to calculate the pressure used in the MARS,

\begin{equation}
\label{DensityCorect}
\hat {\rho}_c = \rho_c + \chi \left( \bar \rho_{s} -   \bar{\rho}_{vs} \right),
\end{equation}

\noindent where $\hat {\rho}_c$ is the corrected density for all exterior corners of the subcell (\textit{i.e.,} in the corner at an edge vertex and in the cell corner), $\bar \rho_{s}$ is the average density of a subcell based on mass conservation (\eref{ModalDensityEqn1}), $\bar \rho_{vs}$ is the average density of the subcell based on the specific volume $v_h$ distribution for the cell, and $\chi$ is a user settable coefficient in the range of zero to one.   If the cell deforms in a manner consistent with the specific volume field then the density correction term is zero, and higher-order accuracy can be achieved with the Lagrangian DG method, \textit{i.e.}, fourth-order accuracy with DG(P3).
  
\subsubsection{Surface integral}
\label{FaceIntegral}
This section focuses on the evaluation of the surface integral in \eref {DGEquation-SpecV}, \ref{DGEquation-Momen}, and \ref{DGEquation-Energy}.
Taking  \eref {DGEquation-SpecV} as an example, for the 7-point Lobatto quadrature rule on cubic meshes, 

\begin{equation}
\label{Face-integral}
\begin{array}{l}
\int_{V_3V_4}  ({\textbf {\textit{n}}}\cdot{\textbf {\textit{u}}}^*) da 
=  (a_{1}  {\textbf {\textit{n}}}_{1})_{G_1}  \cdot {\textbf {\textit{u}}}_{G_1} + (a_{1}  {\textbf {\textit{n}}}_{1})_{G_2}  \cdot {\textbf {\textit{u}}}_{G_2} 
+ \sum_{i=1}^2  (a_{i}  {\textbf {\textit{n}}}_{i})_{G_3^r}  \cdot {\textbf {\textit{u}}}_{G_3}+ \sum_{i=1}^2  (a_{i}  {\textbf {\textit{n}}}_{i})_{G_3^l}   \cdot {\textbf {\textit{u}}}_{G_3}\\
+ (a_{1}  {\textbf {\textit{n}}}_{1})_{G_4}  \cdot {\textbf {\textit{u}}}_{G_4} 
+ \sum_{i=1}^2  (a_{i}  {\textbf {\textit{n}}}_{i})_{G_5^r}  \cdot {\textbf {\textit{u}}}_{G_5}+ \sum_{i=1}^2  (a_{i}  {\textbf {\textit{n}}}_{i})_{G_5^l}   \cdot {\textbf {\textit{u}}}_{G_5}
+ (a_{1}  {\textbf {\textit{n}}}_{1})_{G_6}  \cdot {\textbf {\textit{u}}}_{G_6} + (a_{2}  {\textbf {\textit{n}}}_{2})_{G_7}  \cdot {\textbf {\textit{u}}}_{G_7} .\\
\end{array}
\end{equation}

\noindent Here, four quadrature points will overlap the vertices, the velocity of the edge ${\textbf {\textit{u}}}_{G}$ is equal to the Riemann velocity at the vertices ${\textbf {\textit{u}}}^*_{v}$, while for the non-vertex quadrature points, the velocity of the edge ${\textbf {\textit{u}}}_{G}$ has been specified in \sref{RiemannProblem}. Taking the area normal vector $(a_{2}  {\textbf {\textit{n}}}_{2})_{G_7}$ as an example, the subscript $G_7$ denotes the corner associated with the quadrature point $G_7$.
And for the edge vertex $G_3$ and $G_5$, the superscript $l$ and $r$ denote the left and right corner associated with the edge vertex $G_3$ and $G_5$.
The weights used in this 7-point Lobatto quadrature rule have been included in the weighted area $a_i$. 
The surface area normal vector $s{\textbf {\textit{n}}}$ along the surface $V_3V_4$ can be expressed as, 
\begin{equation}
\label{Gauss-face}
\begin{array}{l}
\int_{V_3V_4} d{\textbf {\textit{s}}}=  \sum_{g}wt_g(s{\textbf {\textit{n}}})_g,\\
\end{array}
\end{equation}
\noindent where $s_g$ represent the determinant of the Jacobian matrix at the corresponding quadrature point along the surface $V_3V_4$.
\noindent Combining  \eref{Gauss-face} with \ref {Face-integral}, we can conclude  that 

\begin{equation}
\label{Simpsonrule}
\begin{array}{l}
(a_{1}  {\textbf {\textit{n}}}_{1})_{G_1}  =  wt_{G_1}(s {\textbf {\textit{n}}})_{G_1}, \\
(a_{1}  {\textbf {\textit{n}}}_{1})_{G_2}  =  wt_{G_2}(s {\textbf {\textit{n}}})_{G_2}, \\
(a_{2}  {\textbf {\textit{n}}}_{2})_{G_3^l}  =  \frac{1}{2}wt_{G_3}(s {\textbf {\textit{n}}})_{G_3^l}, \\
(a_{1}  {\textbf {\textit{n}}}_{1})_{G_3^r}  =  \frac{1}{2}wt_{G_3}(s {\textbf {\textit{n}}})_{G_3^r}, \\
(a_{1}  {\textbf {\textit{n}}}_{1})_{G_4}  =  wt_{G_4}(s {\textbf {\textit{n}}})_{G_4}, \\
(a_{2}  {\textbf {\textit{n}}}_{2})_{G_5^l}  =  \frac{1}{2}wt_{G_5}(s {\textbf {\textit{n}}})_{G_5^l}, \\
(a_{1}  {\textbf {\textit{n}}}_{1})_{G_5^r}  =  \frac{1}{2}wt_{G_5}(s {\textbf {\textit{n}}})_{G_5^r}, \\
(a_{1}  {\textbf {\textit{n}}}_{1})_{G_6}  =  wt_{G_6}(s {\textbf {\textit{n}}})_{G_6}, \\
(a_{2}  {\textbf {\textit{n}}}_{2})_{G_7}  =  wt_{G_7}(s {\textbf {\textit{n}}})_{G_7}. \\
\end{array}
\end{equation}

\noindent In addition, at the edge vertex $V_9$ and $V_{10}$, we need to pay special attention to segment 2 of the right corner  and segment 1 of the left corner in \fref{Quadraturept}.
Likewise,
\begin{equation}
\label{Simpsonrule2}
\begin{array}{l}
(a_{1}  {\textbf {\textit{n}}}_{1})_{V_9^l}   = -(a_{2}  {\textbf {\textit{n}}}_{2})_{V_9^r}   =  \frac{1}{2}wt_{G_3}(s {\textbf {\textit{n}}})_{9t},\\
(a_{1}  {\textbf {\textit{n}}}_{1})_{V_{10}^l}   = -(a_{2}  {\textbf {\textit{n}}}_{2})_{V_{10}^r}   =  \frac{1}{2}wt_{G_5} (s {\textbf {\textit{n}}})_{10t},\\
\end{array}
\end{equation}
\noindent Here, $s_{9t}$ means the determinant of the Jacobian matrix at the vertex $V_9$ from the inside surface $V_6V_9$. 
Please refer to \cite{LiuDGlagSMS2019} for the choice of the weights (\textit{i.e.}, $ \frac{1}{2}wt_{G_3}$ and $ \frac{1}{2}wt_{G_5}$).

\subsection{Limiting}
\label{Limiters}

The modal fields and the pressure field (a nodal field) are limited toward the cell average value (\textit{i.e.,} piece-wise constant over the cell) near shocks and discontinuities, which is an essential ingredient for robust solutions.  Therefore, the limiting strategy is to deliver high-order (e.g., fourth-order) accuracy in smooth regions of a calculation and to reduce the accuracy towards first-order near discontinuities.
A favorable feature in Lagrangian hydrodynamics is that the mesh clusters near the shock, providing natural $h$ refinement for the shock region. Simultaneously, the discretization order is reduced to DG(P1) to guarantee the monotonicity.
A merit of using new orthogonal basis functions is that each moment in the Pn polynomial are decoupled from all other terms, allowing flexibility in limiting of each coefficient. 
Reducing the Pm polynomial space to the Pn one ($m \ge n$) can be done by truncating the higher-order terms that are higher than the Pn polynomial.
We found this aforementioned truncation in the context of the Taylor-series polynomials that are higher than P2 generated extremely poor results because the moments are coupled through the mass matrix.  Using the orthogonal basis functions was essential for obtaining robust solutions on strong shock problems.

To ensure high-order solutions away from a shock, we use a troubled-cell detector \cite{ShutvbDG41990, MichalakShockjcp2009, PerssonShock2006, DarmofalShock2007, ZJWangShockAAMM2009, KrivodonovaShock2004,ChiravalleMARS} to identify the cells near a discontinuity.  In the present work, the troubled-cell detector from \cite{PerssonShock2006} is adopted; this detector was successfully used in \cite{LiuDGlagSMS2019} with a Lagrangian DG(P2) method.  

If a cell is tagged as a troubled-cell, then the quadratic and cubic terms are set to zero, and the first-order derivatives in the modal fields ($v_h$, $\textbf {\textit{u}}_h$, and $\tau_h$) are limited to ensure the reconstructions at the vertices are bounded by the vertex-neighboring cell-averages just like Lagrangian FV CCH methods \cite{Maire1,Maire2,BurtonCCH,MorganContact}. The first-order derivatives in the velocity polynomial reconstructions are limited in either the principle strain direction \cite{Mairestrain2014, BurtonCCH} or the local flow direction \cite{Maire3} to ensure the preservation of symmetry on equal-angle polar meshes.  For a troubled-cell, the pressure value in each corner was limited to be within the bounds of the cell-averages of all adjacent cells to the vertex.  Additional details on the limiting of the first derivatives for scalar and vector fields with the Lagrangian DG(P1) and DG(P2) methods are provided in \cite{LiuDGlagcaf2017, LiuDGlagSMS2019} respectively.  

\section{Test problems}
\label{TestCases}
In this section, a diverse suite of test problems covering gas and solid dynamics are calculated to demonstrate the accuracy and robustness of this Lagrangian DG hydrodynamic method with a hierarchical orthogonal basis on curvilinear meshes.  
Both smooth flows and shock driven flows are calculated.  We will now provide a brief introduction of each test problem.

The Taylor-Green vortex test case (\sref{TGV}) and the Gresho vortex test case (\sref{gresho}) are smooth, vortical flow problems.  We use these test cases to quantify the convergence rate (\textit{i.e}, order of accuracy) of the new Lagrangian DG method using the $L_2$ error norm. We will briefly describe the error analysis used on these test problems.
Taking pressure $p$ as an example, the $L_2$ error norm is defined by,

\begin{equation} \label{l2error}
||p - p_{e}||_{L_2} = \sqrt{\sum \limits_{i=1}^{Num}\int_w (p - p_e)^2 d{w}},\\
\end{equation} 

\noindent where $p_e$ denotes the exact pressure, and $ncell$ is the number of cells in the problem.  For all convergence studies, we use $l1$, $l2$, $l3$ and $l4$ to denote the various levels of mesh resolution where $l1$ is the coarsest resolution. 

The Sedov blast wave test case (\sref{sedov}) is used to demonstrate the accuracy and stability of the Lagrangian DG method with an exceptionally strong shock. We then present results for an elastic vibration of a beryllium plate test (\sref{sec:EVBePlate}) problem and a planar Taylor Anvil impact test (\sref{Taylor-bar}) to show the ability of this new scheme to simulate challenging solid dynamic problems. 

Curvilinear box meshes are used for all calculations, unless stated otherwise. 
A few linear mesh results are included for comparison purposes.
For a box grid, $40 \times 20$ denotes the number of cells along the $x$ and $y$ direction. 
The initial box grid has been specified in \sref{RHS-cal}.

\subsection{ 2D Taylor-Green vortex problem}
\label{TGV}
The Taylor-Green vortex problem is a shockless, smooth flow problem \cite{BurtonCGR, debrovfem2012, VilarDGlagcaf2012, MorganPCHLag}, 
which is of particular interest because it is a vortical flow problem with an analytical solution. 
The material is a gamma-law gas with $\gamma = 7/5$. The initial flow field is

\begin{equation*}
\label{tgvinitial}
\begin{array}{l l l}
\rho^0 &=& 1, \\
 u_x^0   &=& sin(\pi x)cos(\pi y), \\
 u_y^0   &=&-cos(\pi x)sin(\pi y), \\
 p^0   &=& \frac{1}{4}[cos(2\pi x) + cos(2\pi y)] + 1,
\end{array}
\end{equation*}

\noindent where $u_x$ and $u_y$ denote the $x$ and $y$ components of velocity, the superscript $0$ represents the initial state. The specific total energy of the field is 
\begin{equation*} \label{tgvtao}
\tau = \frac{p}{\rho(\gamma-1)}+ \frac{1}{2}(u_x^2 + u_y^2). \\
\end{equation*}

\noindent An energy source term is included in the total energy equation to maintain a steady-state solution for the compressible inviscid case,

\begin{equation*} \label{tgvsrc}
s_\tau= \frac{\pi}{4(\gamma-1)}[cos(3\pi x)cos(\pi y) - cos(3\pi y)cos(\pi x)] .\\
\end{equation*}

\noindent The computational domain for this test case is defined by $(x, y) = [0, 1]  \times [0, 1]$.  
All calculations use a set of uniformly refined quadrilateral meshes, 
namely $5 \times 5$, $10 \times 10$, $20 \times 20$, and $40 \times 40$ respectively. 

\begin{figure*}[h!]
\centering
\subfloat[The $l1$ mesh]{
\includegraphics[width=3.2in]{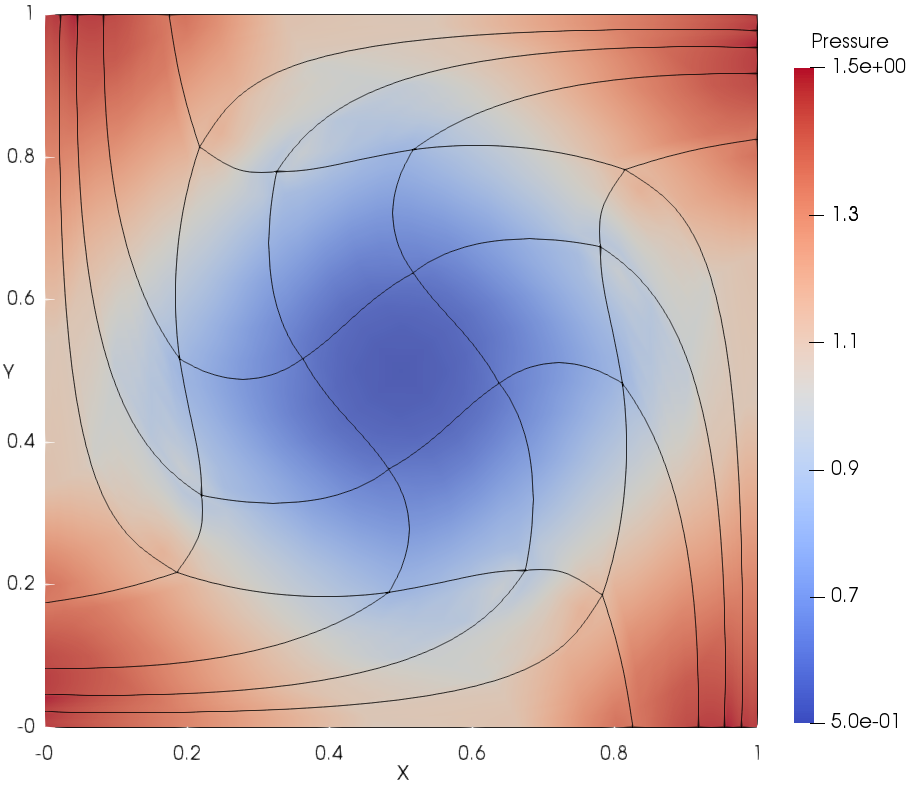}
\label{fig:tgvquadmesht075-p3-l1}}
\subfloat[The $l2$ mesh]{
\includegraphics[width=3.2in]{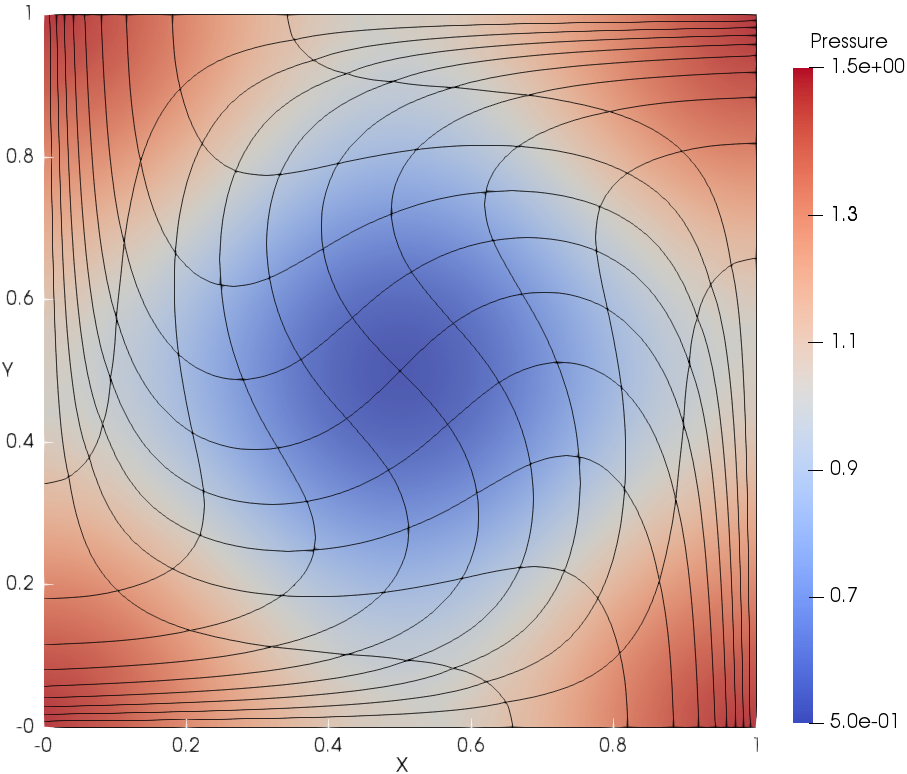}
\label{fig:tgvquadmesht075-p3-l2}}\\
\subfloat[The $l3$ mesh]{
\includegraphics[width=3.2in]{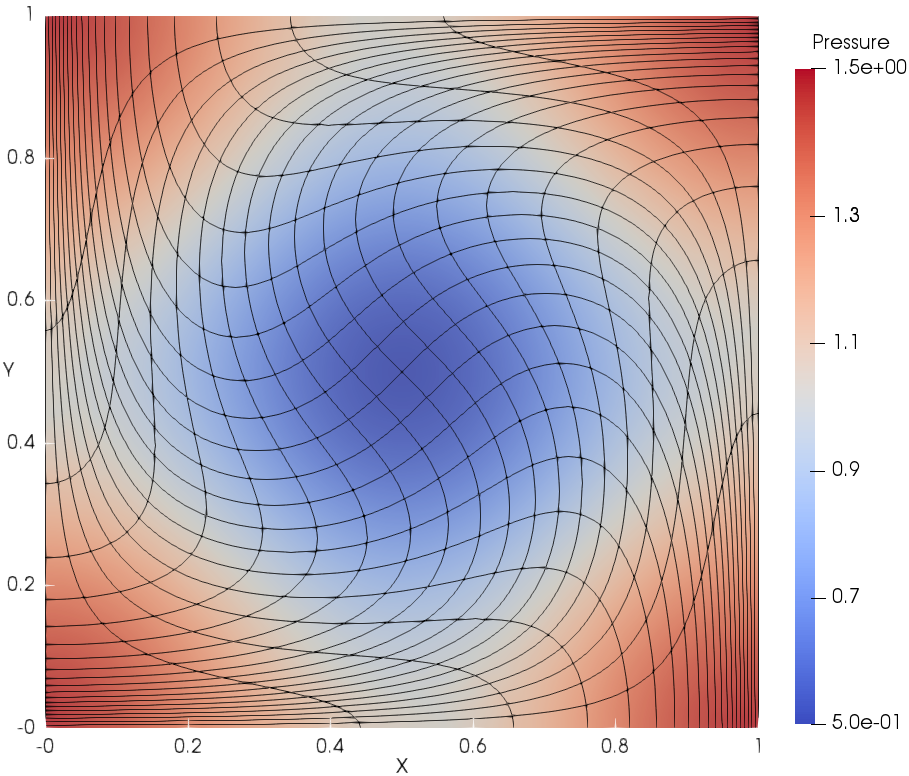}
\label{fig:tgvquadmesht075-p3-l3}}
\subfloat[The $l4$ mesh]{
\includegraphics[width=3.2in]{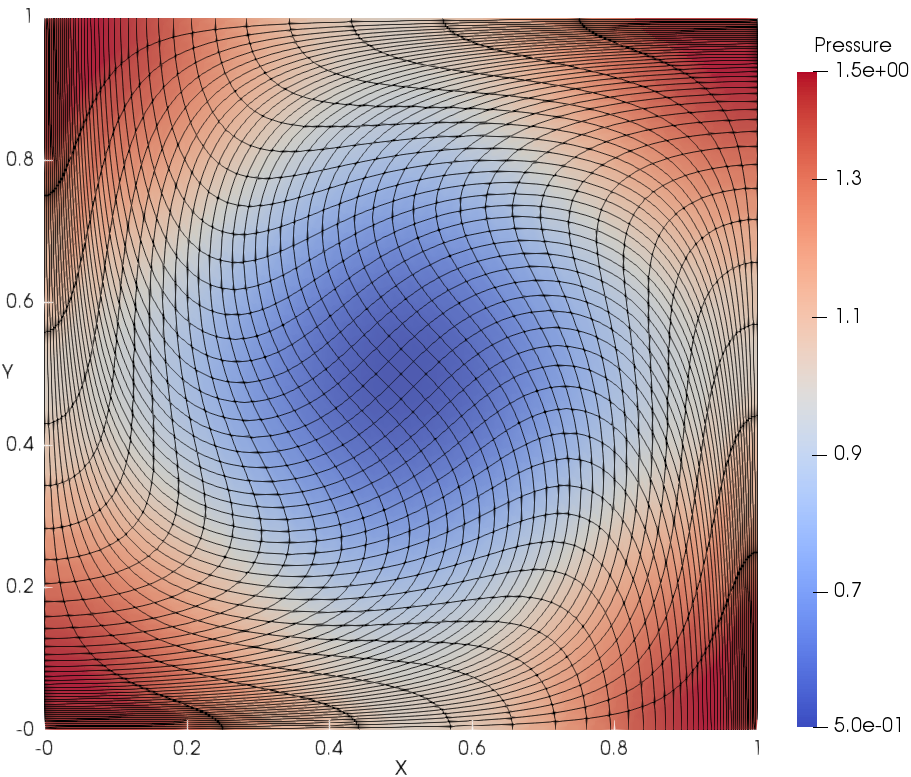}
\label{fig:tgvquadmesht075-p3-l4}}
\caption{ 
A set of cubic meshes and pressure fields at $t=0.75$ are shown for the 2D Taylor-Green vortex problem using DG(P3). 
The meshes are deforming in a smooth and curvilinear way. 
The high-order DG(P3) method is able to capture the details of the flow with very coarse mesh resolutions (\fref{fig:tgvquadmesht075-p3-l1} and \fref{fig:tgvquadmesht075-p3-l2}).}
\label{fig:tgvquadt075}
\end{figure*}

The meshes and pressure fields at $t=0.75$ using DG(P3) with cubic cells are shown in \fref{fig:tgvquadt075}. 
DG(P3) delivers very accurate flow details for the pressure field even with a very coarse mesh resolution ($5 \times 5$).
It can also be observed that the mesh is able to deform with the flow.

\begin{figure*}[h!]
\centering
\subfloat[$L_2$ error of density]{
\includegraphics[width=2.5in]{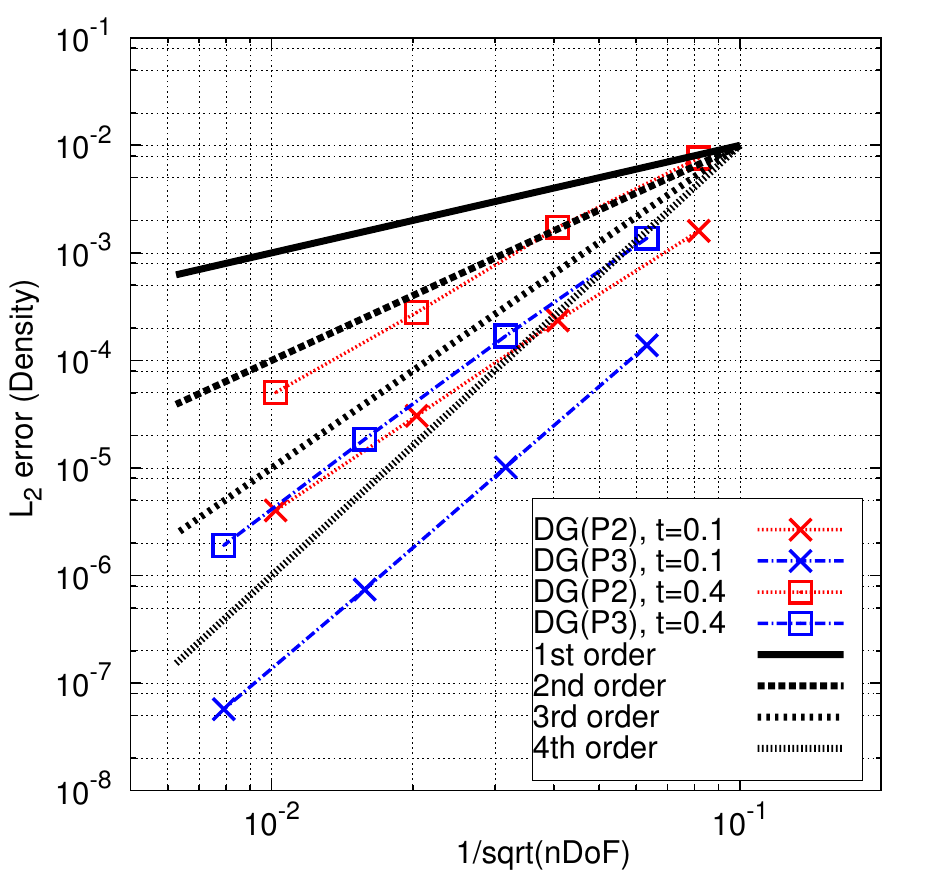}
\label{fig:tgvl2rho}}
\subfloat[$L_2$ error of velocity component]{
\includegraphics[width=2.5in]{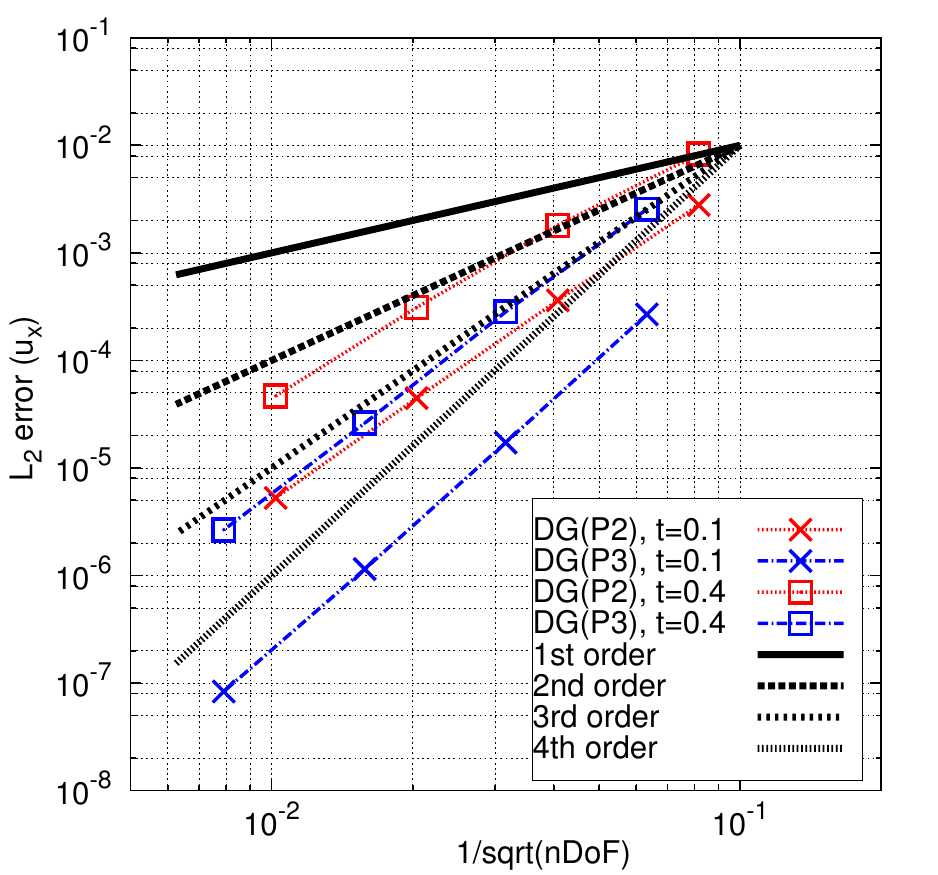}
\label{fig:tgvl2velo}}\\
\subfloat[$L_2$ error of pressure]{
\includegraphics[width=2.5in]{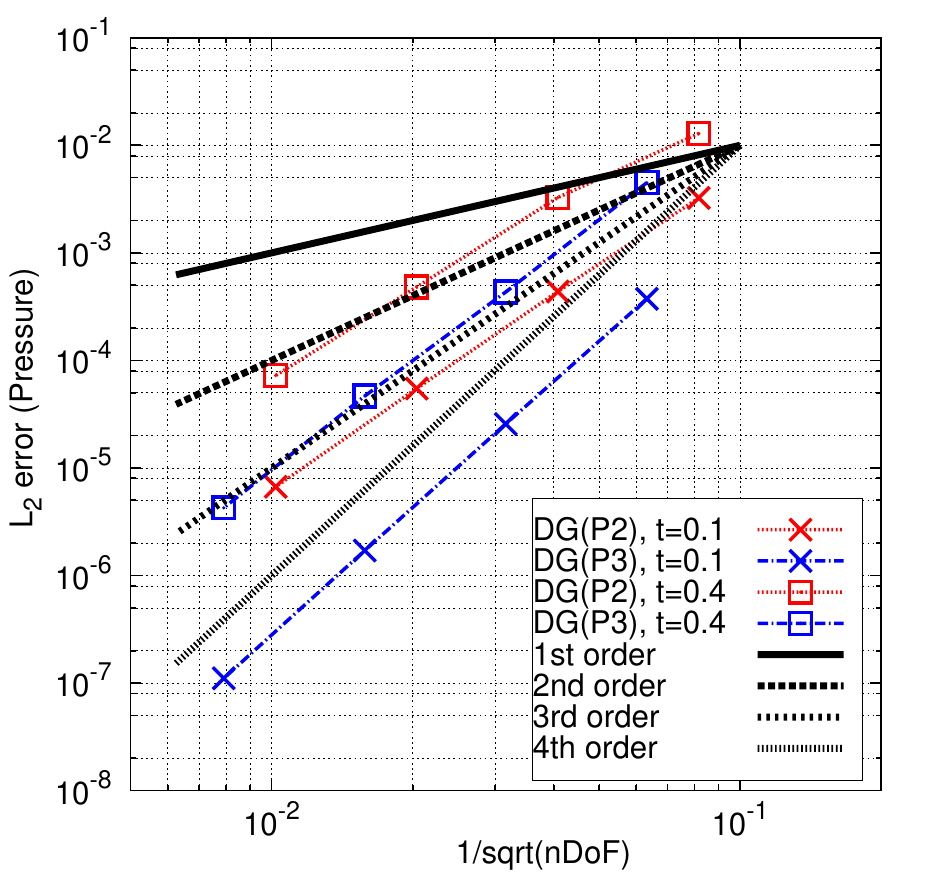}
\label{fig:tgvl2pres}}
\subfloat[$L_2$ error of total energy]{
\includegraphics[width=2.5in]{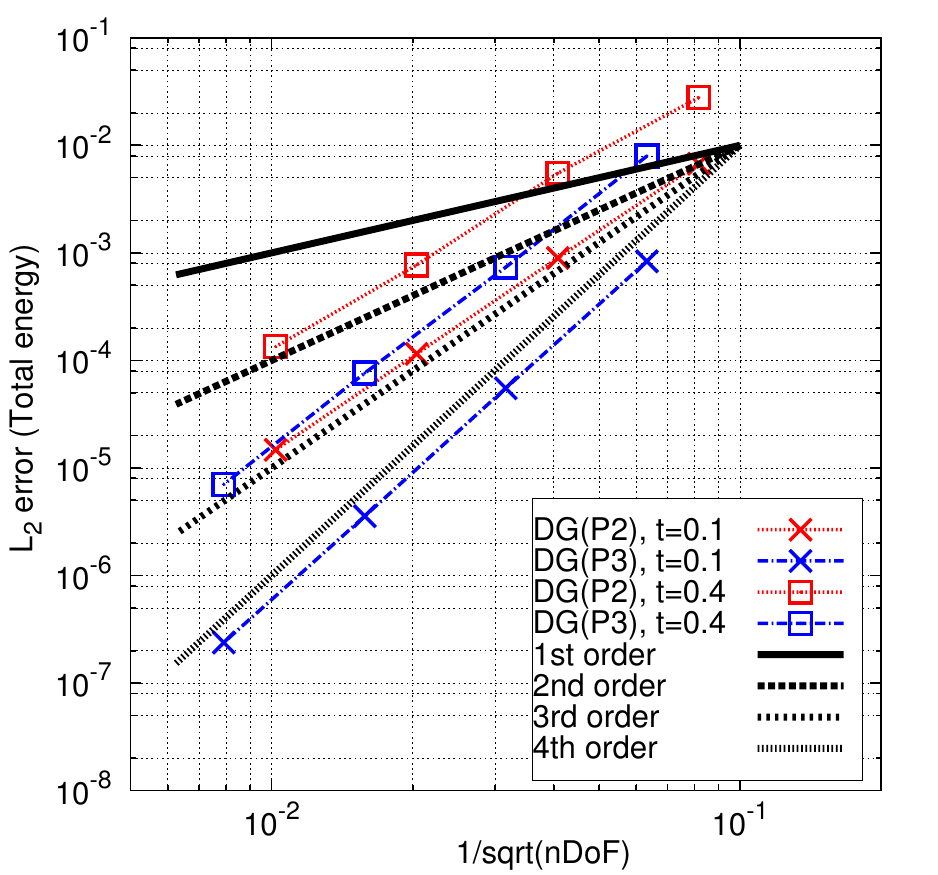}
\label{fig:tgvl2total}}
\caption{ 
Convergence rate (\textit{i.e.,} order of accuracy) is shown for the 2D Taylor-Green vortex problem at $t=0.1$ and $t=0.4$ using DG(P2) and DG(P3).
As shown here, with time marching, the numerical error accrues. 
DG(P2) and DG(P3) method can yield nearly perfect third-order and fourth-order accuracy at $t = 0.1$ for all the variables, respectively. 
However, at a later time $t =0.4$, the convergence rate for all the variables is slightly reduced for both DG(P2) and DG(P3).
In addition,  with the same No. of degree of freedom (nDoF), DG(P3) with cubic meshes delivers less numerical error than DG(P2) with quadratic meshes. }
\label{fig:tgvl2-cmprsn}
\end{figure*}

Using the analytical solution, the Taylor-Green vortex problem is used to assess the order of accuracy of this Lagrangian DG method with curvilinear meshes.
The $L_2$ norm of the error at $t=0.1$ and $t=0.4$ is calculated for the density ($\rho$), velocity component ($u_x$),  pressure ($p$) and total energy field ($\tau$).
The associated results are shown in  \tref{tab:order-quadtgv}.
The DG(P2) and DG(P3) methods can yield nearly perfect third-order and fourth-order accuracy at $t = 0.1$ for all the variables, respectively. 
With time marching, the mesh has greater deformation and the numerical error accrues. 
Accordingly, at $t =0.4$, the convergence rate (\textit{i.e.,} order of accuracy) for all the variables is slightly less than the designed order of accuracy for both DG(P2) and DG(P3).
\fref{fig:tgvl2-cmprsn} presents the numerical errors for all the variables at different times vs. the computational costs (\textit{i.e.}, number of degrees of freedom, \textit{i.e.}, nDoF)
using DG(P2) and DG(P3).
This example demonstrates that higher-order methods can be computationally more efficient to achieve a given level of accuracy compared with lower-order methods.

Finally, we also calculated this problem with DG(P2) using linear and cubic meshes respectively. 
The numerical error for all the variables are shown in \tref{tab:order-quadtgv-p2}.
 DG(P2) with linear meshes delivers second-order accuracy at most for all the variables while DG(P2) with cubic meshes can keep third-order accuracy.
 In addition, in the context of the DG(P2) method, the numerical error and convergence rate on the cubic meshes is close to that on quadratic meshes. 
 This demonstrates that achieving the $(n+1)^{th}$ order of accuracy requires using quadrilaterals meshes with edges defined by a polynomial of degree $n$ with a DG(Pn) method.

\begin{table*}[t]
\begin{center}
\caption
{Numerical error and convergence rate are shown for the Taylor-Green vortex problem at different times using DG(P2) and DG(P3) on curvilinear meshes.
At the early time $t=0.1$, for all the variables (\textit{i.e.}, the density $\rho$,  velocity component $u_x$, pressure $p$ and total energy $\tau$),  DG(P2) and DG(P3) can deliver approximately the designed third- and fourth-order accuracy, respectively. While at a later time $t=0.4$, for both DG(P2) and DG(P3), all the variables converge slightly less than the designed order of accuracy due to the 
acquirement of numerical errors with time marching.}
\label{tab:order-quadtgv}
\mbox{
\begin{tabular}{l c c c c c c c c}
\hline
\multirow{2}{*}{Mesh}& \multicolumn{2}{c} {$\rho$}& \multicolumn{2}{c} {$u_x$} & \multicolumn{2}{c} {$p$} & \multicolumn{2}{c} {$\tau$} \\
                      \cmidrule(lr){2-3}
                      \cmidrule(lr){4-5}
                      \cmidrule(lr){6-7}
                      \cmidrule(lr){8-9}
                     & $L_2$ error & order & $L_2$ error & order & $L_2$ error & order & $L_2$ error & order\\
\hline
DG(P2), t=0.1\\
\hline
$5\times5$         & 1.5995e-3  &           & 2.7796e-3  &         &  3.2363e-3 &        &  6.7452e-3 &       \\
$10\times10$     & 2.3499e-4  &  2.77  & 3.6141e-4  & 2.95 &  4.3700e-4 &2.90 &  8.9561e-4 & 2.92\\
$20\times20$     & 3.0535e-5  &  2.96  & 4.4599e-5  & 3.03 &  5.4800e-5 &3.00 &  1.1491e-4 & 2.97 \\
$40\times40$     & 4.0233e-6 &   2.93  & 5.2717e-6  & 3.09 &  6.6834e-6 &3.04 &  1.4758e-5 & 2.97 \\
\hline
DG(P2), t=0.4\\
\hline
$5\times5$         & 7.6054e-3  &           & 8.2759e-3  &         &  1.2875e-2 &        &  2.7769e-2 &       \\
$10\times10$     & 1.7285e-3  &  2.14  & 1.7996e-3  & 2.21&  3.2562e-3 &1.99  &  5.4817e-3 & 2.35\\
$20\times20$     & 2.7914e-4  &  2.63  & 3.1000e-4  & 2.55 &  4.8153e-4 &2.77 &  7.7198e-4 & 2.84 \\
$40\times40$     & 5.0151e-5 &   2.48  & 4.6679e-5  & 2.74 &  7.2271e-5 &2.75 &  1.3387e-4 & 2.54 \\
\hline
DG(P3), t=0.1\\
\hline
$5\times5$         & 1.3801e-4  &           & 2.6790e-4  &         &  3.7398e-4 &         &8.3408e-4 &    \\
$10\times10$     & 1.0177e-5  & 3.77   & 1.7242e-5  & 3.97 &  2.5624e-5 & 3.88 &5.5205e-5 & 3.93\\
$20\times20$     & 7.3830e-7  & 3.80   & 1.1475e-6  & 3.93 &  1.7080e-6 & 3.92 &3.5667e-6 & 3.97\\
$40\times40$     & 5.7127e-8  & 3.71   & 8.3370e-7  & 3.80 &  1.1058e-7 & 3.96 &2.3785e-7 & 3.92\\
\hline
DG(P3), t=0.4\\
\hline
$5\times5$         & 1.3688e-3  &           & 2.5387e-3  &         &  4.4911e-3 &         &7.9309e-3 &    \\
$10\times10$     & 1.6847e-4  & 3.03   & 2.8333e-4  & 3.17 &  4.3188e-4 & 3.39 &7.3365e-4 & 3.45\\
$20\times20$     & 1.8453e-5  & 3.20   & 2.6109e-5  & 3.45 &  4.6647e-5 & 3.22 &7.6164e-5 & 3.28\\
$40\times40$     & 1.9066e-6  & 3.28   & 2.6566e-6  & 3.31 &  4.2518e-6 & 3.47 &7.0218e-6 & 3.45\\
\hline
\end{tabular}}
\end{center}
\end{table*}

\begin{table*}[t]
\begin{center}
\caption
{Numerical error and convergence rate are shown for the Taylor-Green vortex problem at $t=0.1$ using DG(P2) on both linear and cubic meshes.
For all the variables (\textit{i.e.}, the density $\rho$,  velocity component $u_x$, pressure $p$ and total energy $\tau$),  DG(P2) delivers second-order accuracy at most on linear meshes.
DG(P2) is able to achieve third-order accuracy for all the variables on cubic meshes, and the difference from DG(P2) on quadratic meshes is very small.}
\label{tab:order-quadtgv-p2}
\mbox{
\begin{tabular}{l c c c c c c c c}
\hline
\multirow{2}{*}{Mesh}& \multicolumn{2}{c} {$\rho$}& \multicolumn{2}{c} {$u_x$} & \multicolumn{2}{c} {$p$} & \multicolumn{2}{c} {$\tau$} \\
                      \cmidrule(lr){2-3}
                      \cmidrule(lr){4-5}
                      \cmidrule(lr){6-7}
                      \cmidrule(lr){8-9}
                     & $L_2$ error & order & $L_2$ error & order & $L_2$ error & order & $L_2$ error & order\\
\hline
DG(P2), \\
linear, t=0.1\\
\hline
$5\times5$         & 5.4851e-3  &           & 9.0759e-3  &         &  7.5490e-3 &        &  1.1949e-2 &       \\
$10\times10$     & 1.0247e-3  &  2.43  & 2.2022e-3  & 2.05 &  1.1707e-3 &2.70 &  2.7164e-3 & 2.14\\
$20\times20$     & 2.7111e-4  &  1.93  & 5.4123e-4   & 2.03 &  2.3301e-4 &2.34 &  8.0907e-4 & 1.75 \\
$40\times40$     & 1.1079e-4 &  1.30  & 1.3557e-4   & 2.00 &  5.5829e-5 &2.07 & 3.1164e-4  &  1.38 \\
\hline
DG(P2), \\
cubic, t=0.1\\
\hline
$5\times5$         & 1.7066e-3  &           & 2.8481e-3  &         &  3.2726e-3 &         &6.8003e-3 &    \\
$10\times10$     & 2.4786e-4  & 2.79   & 3.8802e-4  & 2.89 &  4.4973e-4 & 2.87 &8.8884e-4 & 2.95\\
$20\times20$     & 3.2821e-5  & 2.93   & 2.3186e-5  & 2.88 &  5.9167e-5 & 2.94 &1.1354e-4 & 2.98\\
$40\times40$     & 4.3246e-6  & 2.93   & 7.7855e-6  & 2.78 &  7.6519e-6 & 2.96 &1.4637e-5 & 2.97\\
\hline
\end{tabular}}
\end{center}
\end{table*}

\subsection{ 2D Gresho vortex }
\label{gresho}
The Gresho vortex problem is a smooth vortical flow problem.  The velocity in the angular direction $u_\theta$ is solely a function of the radius $r$.  
The pressure gradient balances the centrifugal forces. The material is a gamma-law gas with $\gamma = 7/5$. The initial flow field is

{\small
\begin{equation*}
\label{greshoinitial}
\begin{array}{l l l}
\rho^0            &=& 1, \\
 u_r^0            &=& 0, \\
 u_{\theta}^0  &=& \left \{
\begin{array}{lll}
5r    & \textup{if}  &  0\le r< 0.2,\\
2-5r &  \textup{if} &  0.2\le r< 0.4 , \\
0     &  \textup{if} &  0.4\le r. \\
\end{array} \right. \\
 p^0   &=&  \left \{
\begin{array}{lll}
5+\frac{25}{2}r^2& \textup{if} &  0\le r< 0.2 ,\\
9-4ln(0.2+)5+\frac{25}{2}r^2-20r+4lnr &  \textup{if} &  0.2\le r< 0.4, \\
3+4ln2 &  \textup{if} & 0.4\le r. \\
\end{array} \right. \\
\end{array}
\end{equation*}
}
\noindent Here, $u_r$ and $u_\theta$ denote the radial and tangential components of the velocity. 
Unlike the Taylor-Green vortex case, the source term is not needed to maintain a steady-state solution for this compressible inviscid case.
The density field is uniform and has a value of 1.  The Gresho vortex is calculated to a time of 0.62.  
The computational domain for this test case is defined by $(x, y) = [-0.5, 0.5]\times[-0.5, 0.5]$.
A set of uniformly refined meshes is used to perform the grid convergence study.
In our calculations, the inner and outer boundary are set to be stationary. 
The mesh resolutions are $16 \times 16$, $32 \times 32$ and $64 \times 64$ respectively.

\begin{figure*}[h!]
\centering
\subfloat[The $l1$ mesh at $t=0.4$]{
\includegraphics[width=3in]{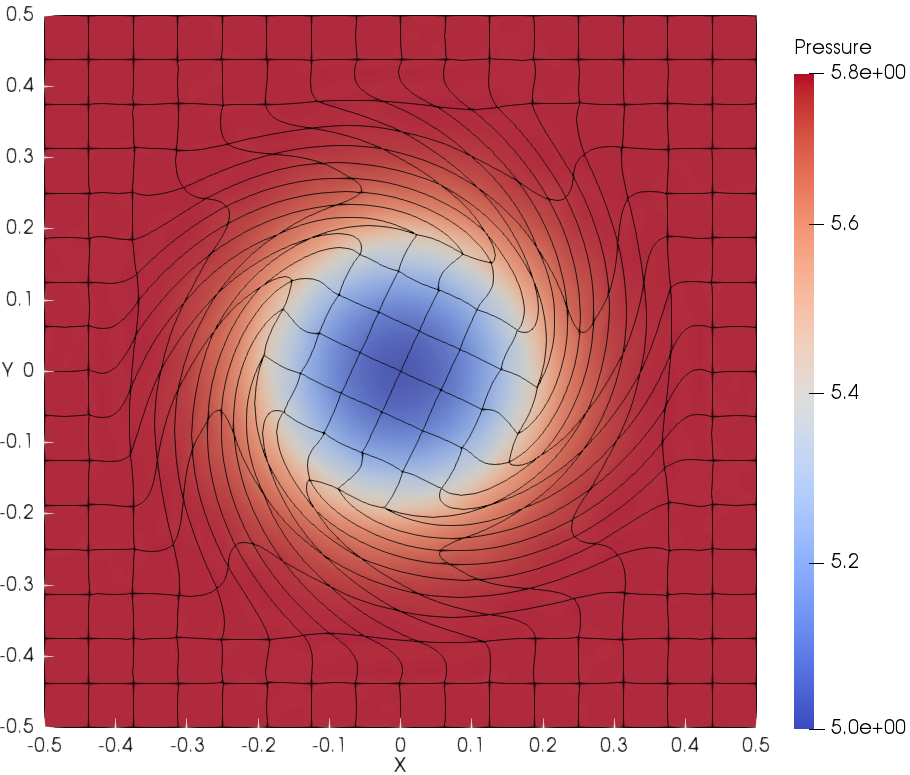} \label{fig:greshoquadl1-p3-t04}
}
\subfloat[The $l2$ mesh at $t=0.4$]{
\includegraphics[width=3in]{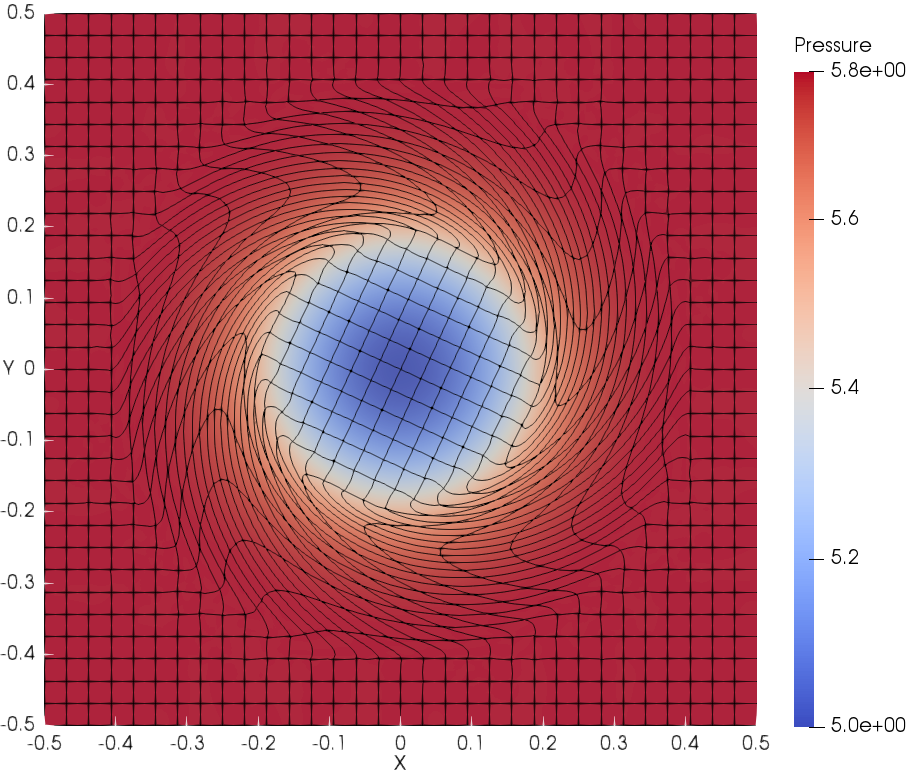} \label{fig:greshoquadl2-p3-t04}
}\\
\subfloat[The $l1$ mesh at $t=0.62$]{
\includegraphics[width=3in]{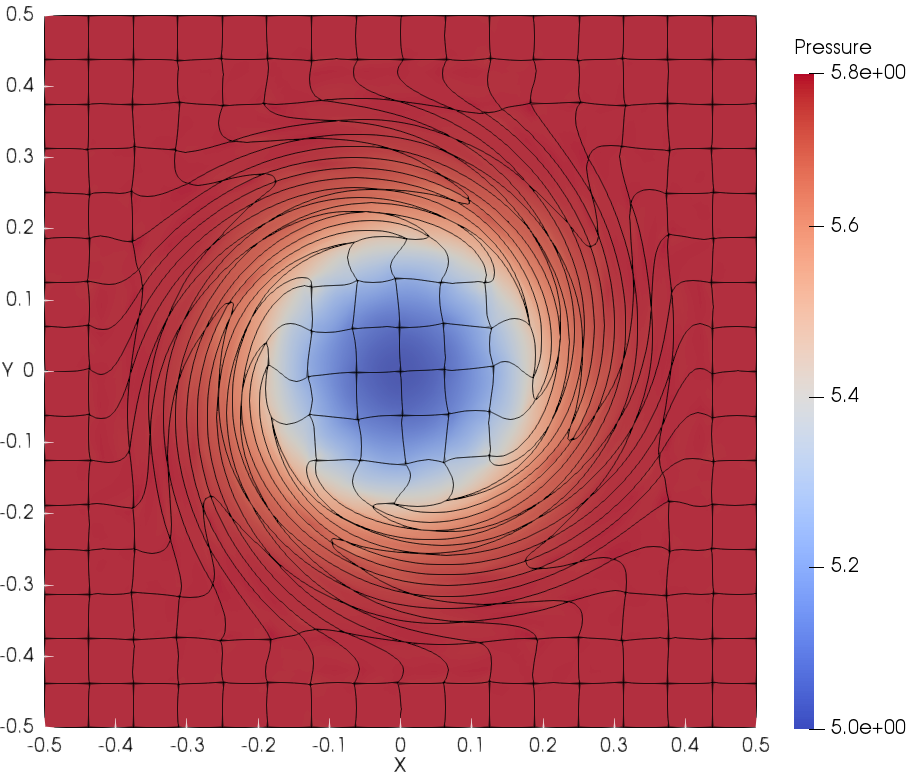} \label{fig:greshoquadl1-p3-t062}
}
\subfloat[The $l2$ mesh at $t=0.62$]{
\includegraphics[width=3in]{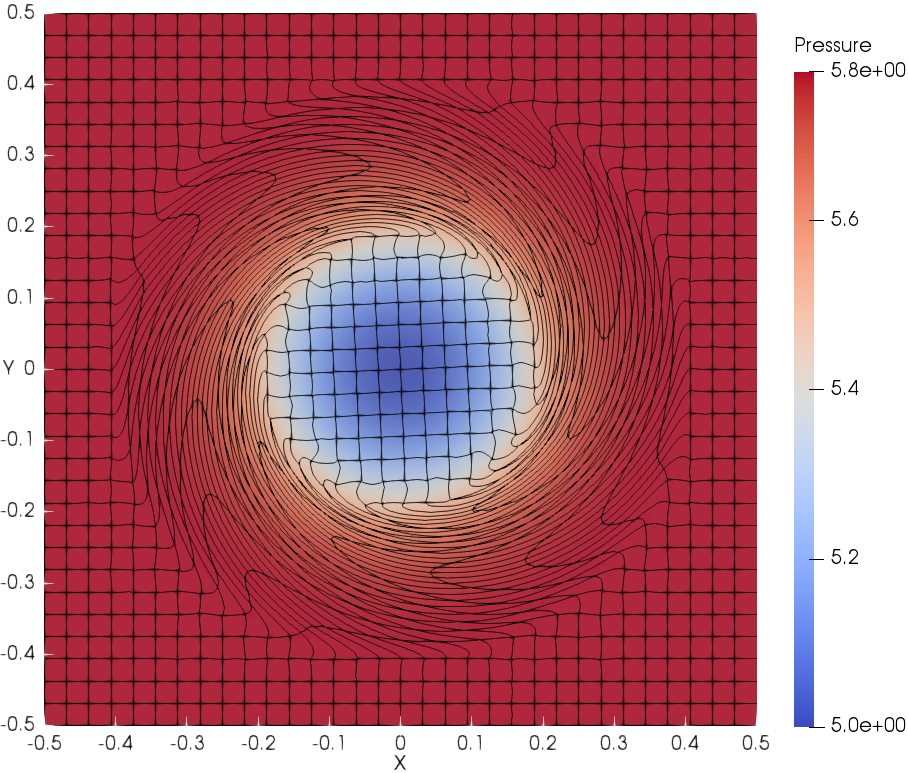} \label{fig:greshoquadl2-p3-t062}
}
\caption{A set of cubic meshes and pressure fields at different times are shown for the Gresho vortex problem using DG(P3). 
From \fref{fig:greshoquadl1-p3-t04} and \ref{fig:greshoquadl1-p3-t062}, 
the meshes are moving in a smooth and curvilinear manner. 
The high-order DG(P3) method is able to capture many flow details with very coarse mesh resolutions (\fref{fig:greshoquadl1-p3-t04} and \ref{fig:greshoquadl1-p3-t062}).
}
\label{fig:greshomesh}
\end{figure*}

\begin{figure*}[h!]
\centering
\subfloat[Density for DG(P2)]{
\includegraphics[width=2.3in]{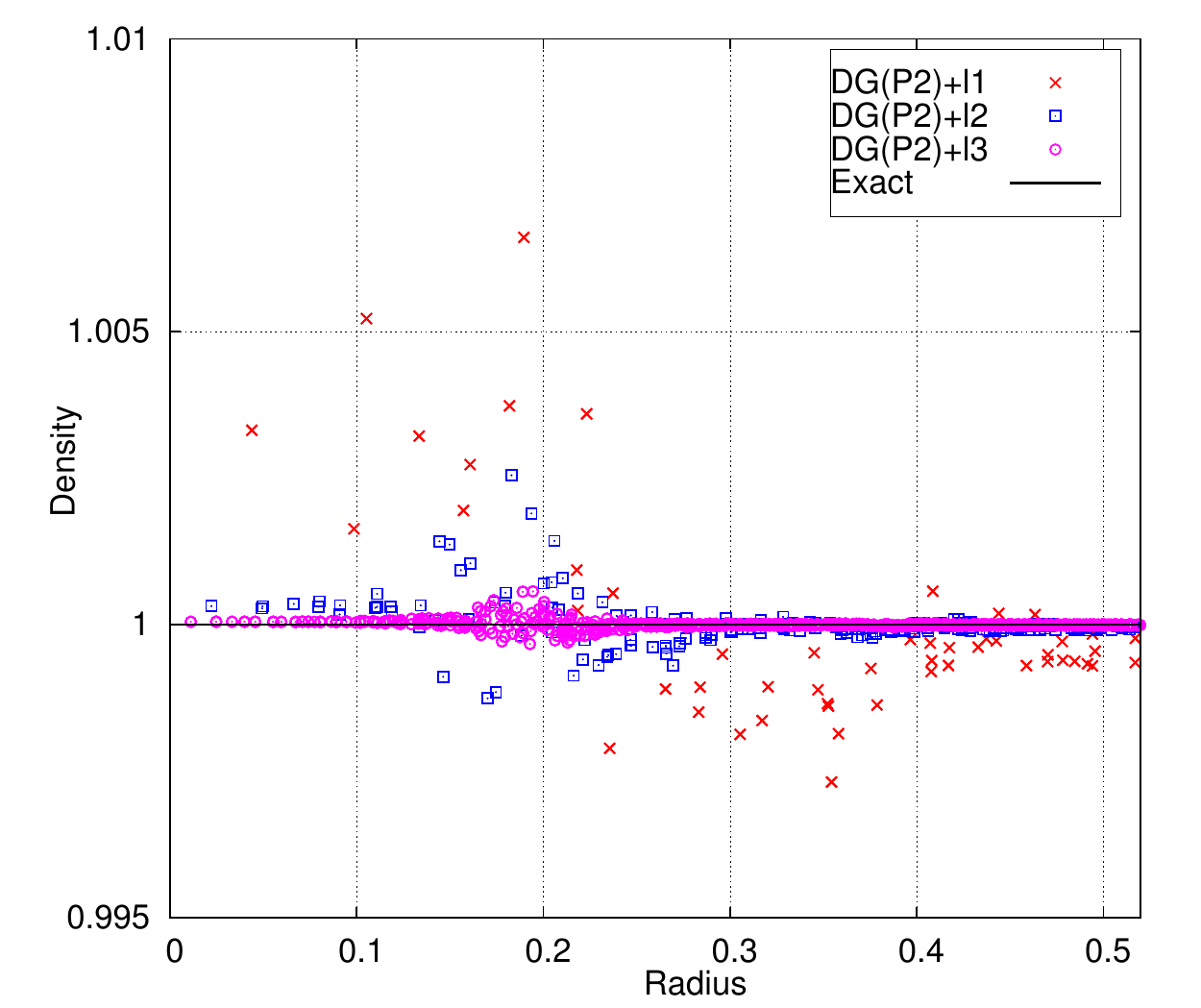}
}
\subfloat[Velocity magnitude for DG(P2)]{
\includegraphics[width=2.3in]{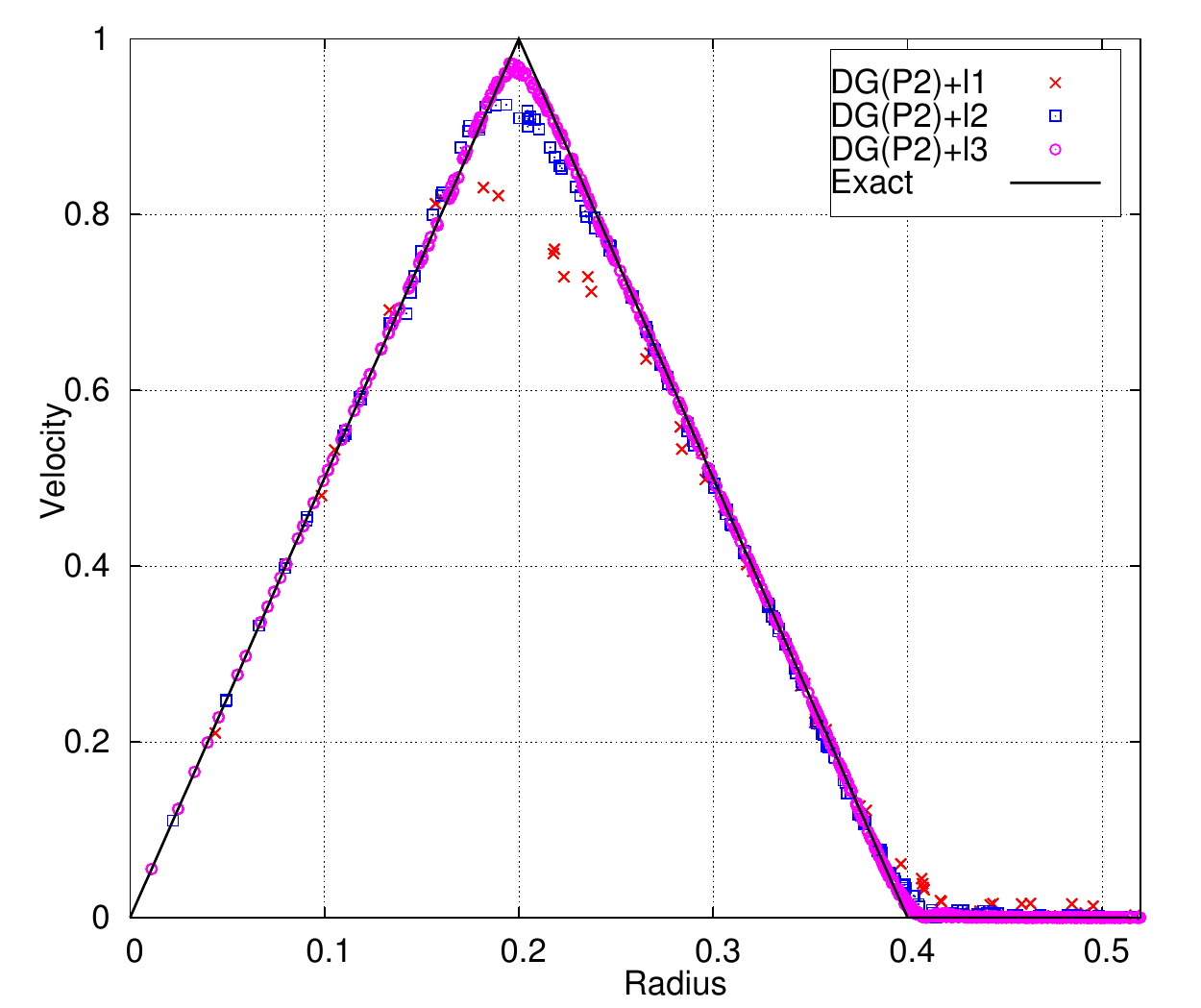}
}
\subfloat[Pressure for DG(P2)]{
\includegraphics[width=2.3in]{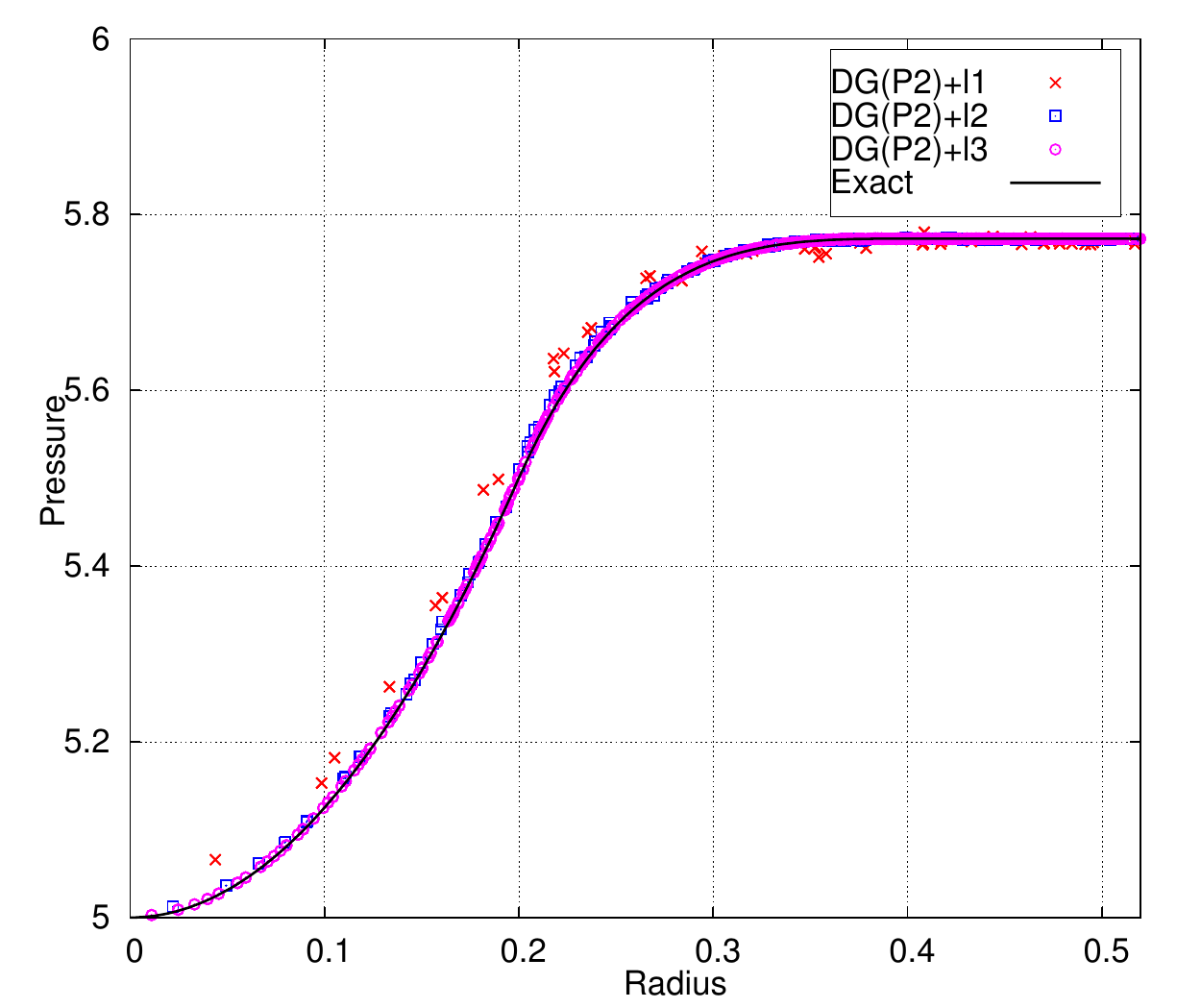}
}\\
\subfloat[Density for DG(P3)]{
\includegraphics[width=2.3in]{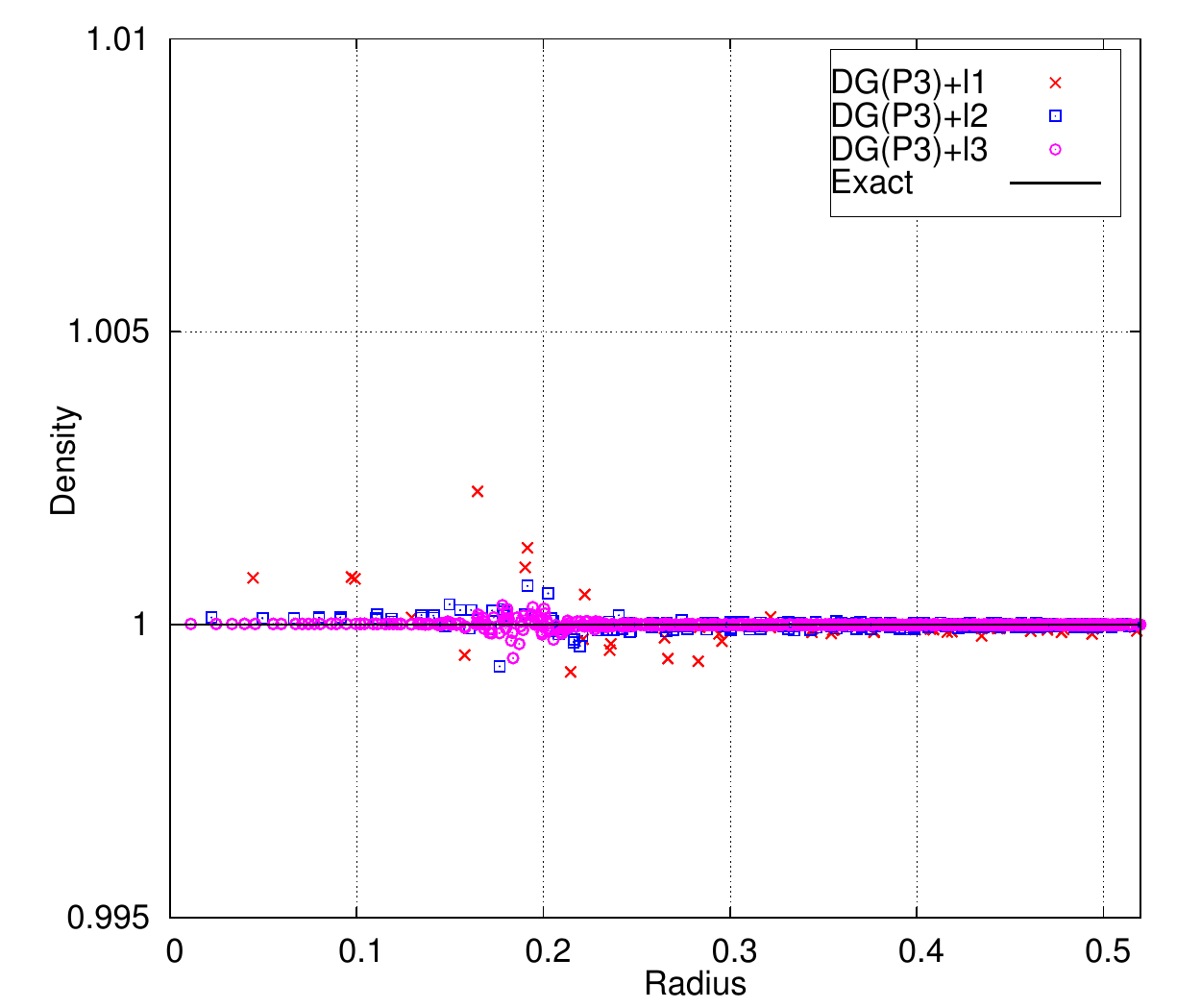}
}
\subfloat[Velocity magnitude for DG(P3)]{
\includegraphics[width=2.3in]{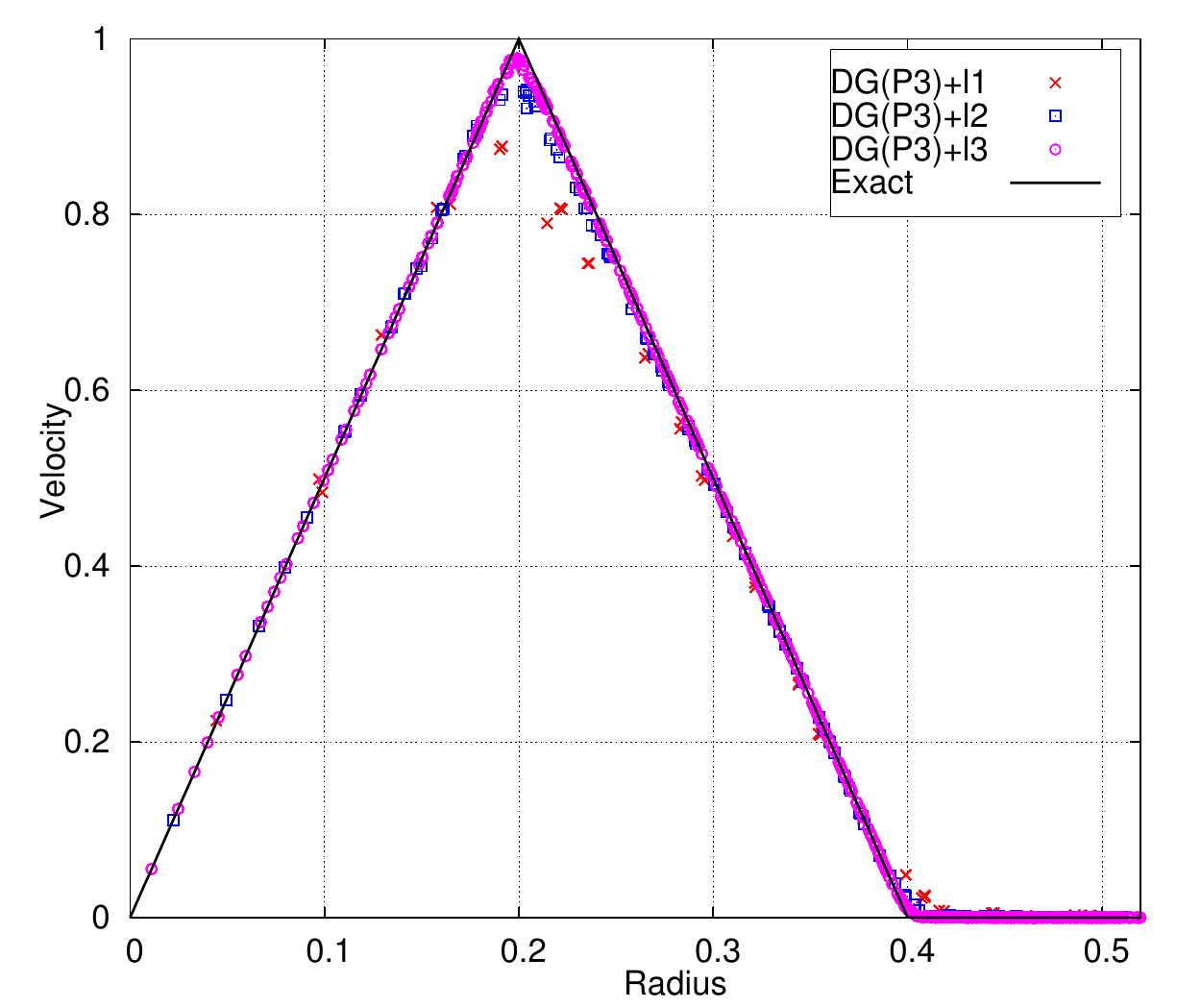}
}
\subfloat[Pressure for DG(P3)]{
\includegraphics[width=2.3in]{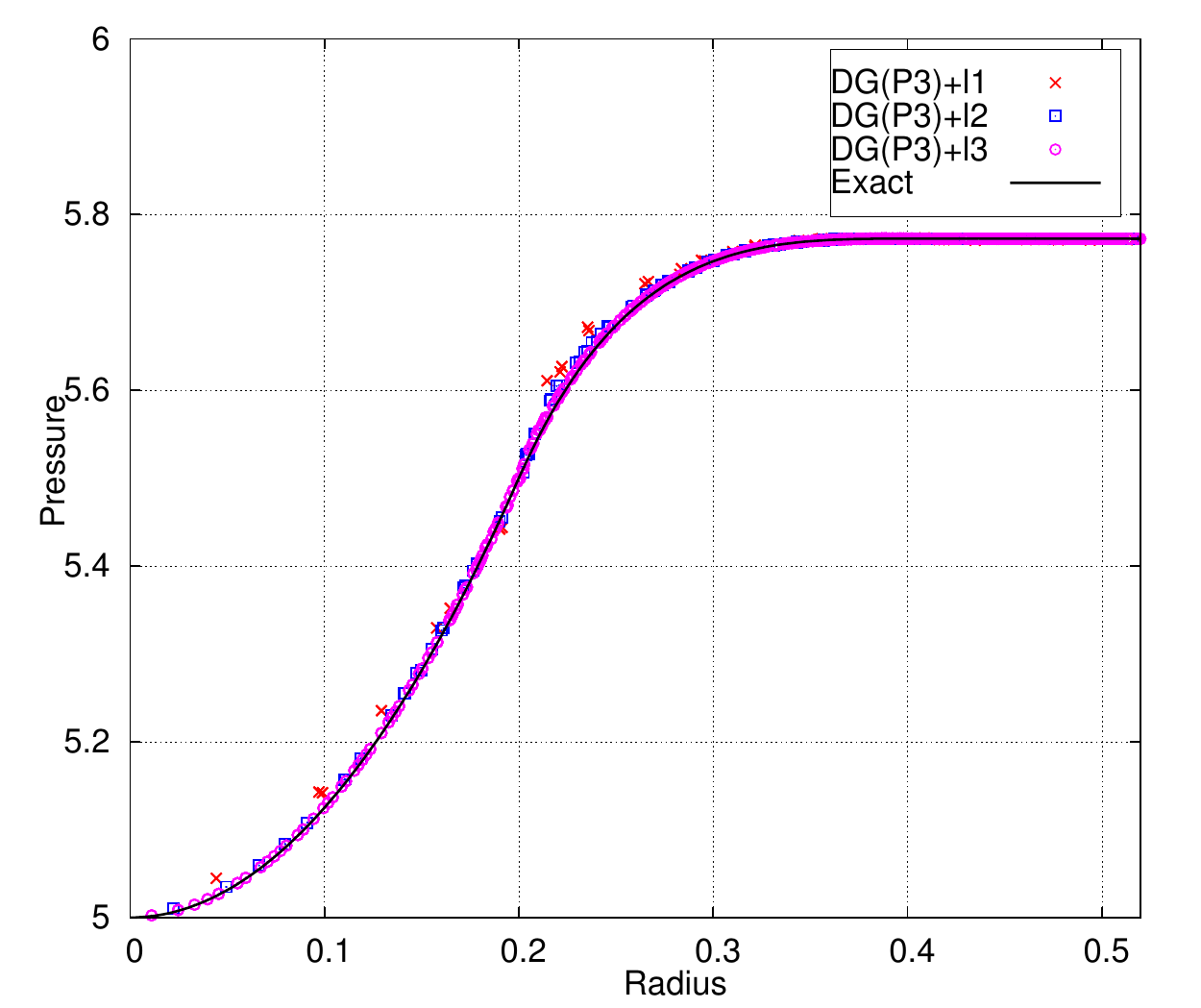}
}
\caption{The scatter plots of density, velocity magnitude, and pressure at $t=0.62$ are shown for the Gresho vortex problem using DG(P2) and DG(P3) on a set of curvilinear meshes.  
For both DG(P2) and DG(P3), the accuracy of the results improves with mesh refinement. 
The DG(P3) method gives more accurate solutions than the DG(P2) with the same mesh resolution.}
\label{fig:gresho-varib}
\end{figure*}

The deformed cubic meshes and pressure field at time $t=0.4$ and $t=0.62$
are shown in \fref{fig:greshomesh} with DG(P3) using two mesh resolutions.  
Indeed the meshes are deforming in a robust and curvilinear manner.  
In addition, \fref{fig:gresho-varib} presents the scatter plots for variables of interest, \textit{i.e.}, density ($\rho$), velocity magnitude ($v$) and pressure ($p$), with 
DG(P2) and DG(P3). With mesh refinement, the numerical results agree better with the analytical solution for both DG(P2) and DG(P3). 
Obviously, DG(P3) provides much better agreement with the analytical solution.
The time evolution until $t=0.62$ demonstrates the robustness of the Lagrangian DG hydrodynamic method with a hierarchical orthogonal basis.

\begin{figure*}[h!]
\centering
\subfloat[$L_2$ error of density]{
\includegraphics[width=3in]{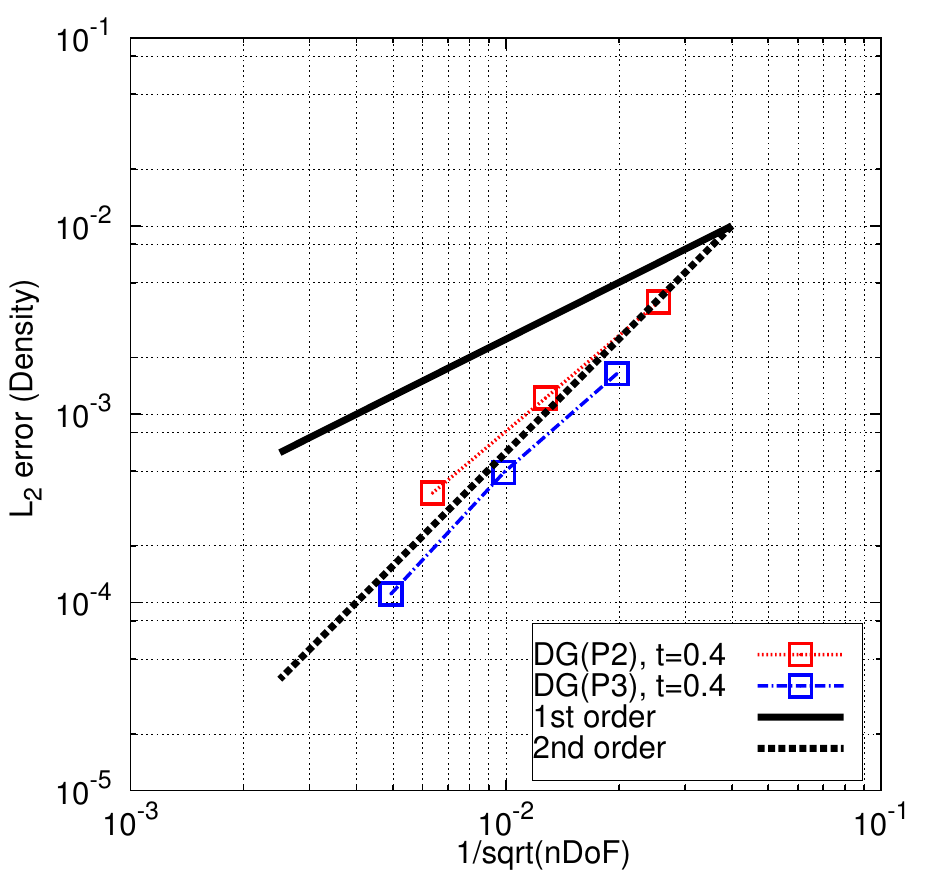}
\label{fig:greshol2rho}}
\subfloat[$L_2$ error of velocity component]{
\includegraphics[width=3in]{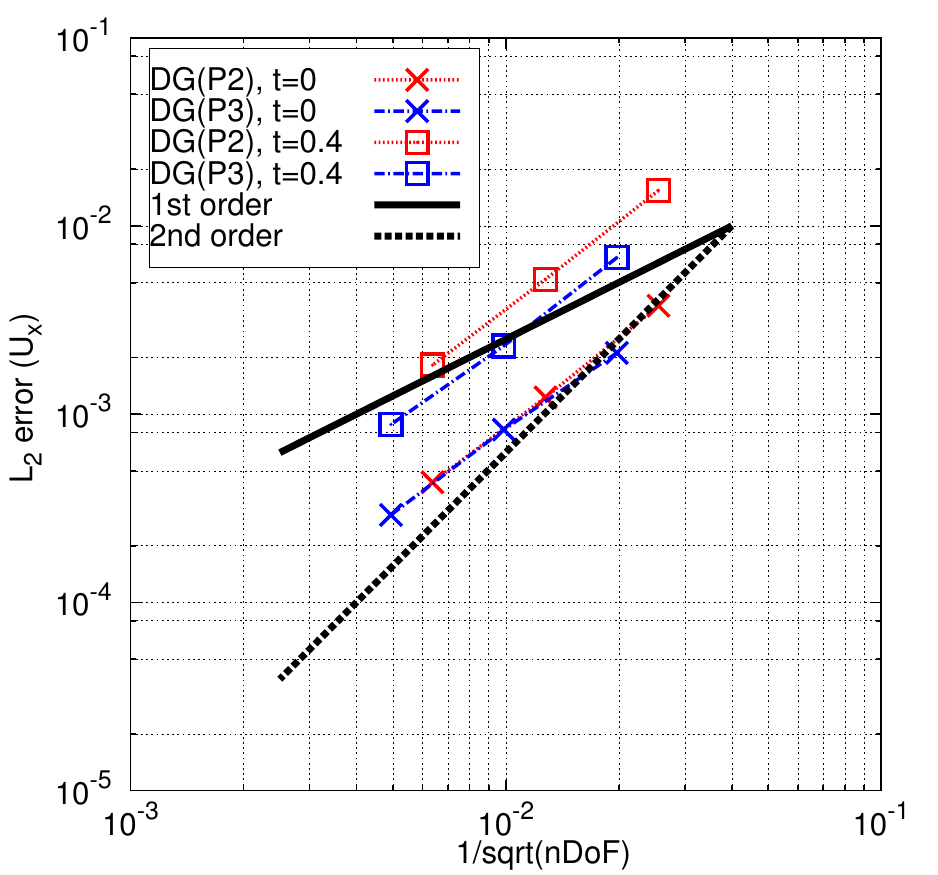}
\label{fig:greshol2velo}}\\
\subfloat[$L_2$ error of pressure]{
\includegraphics[width=3in]{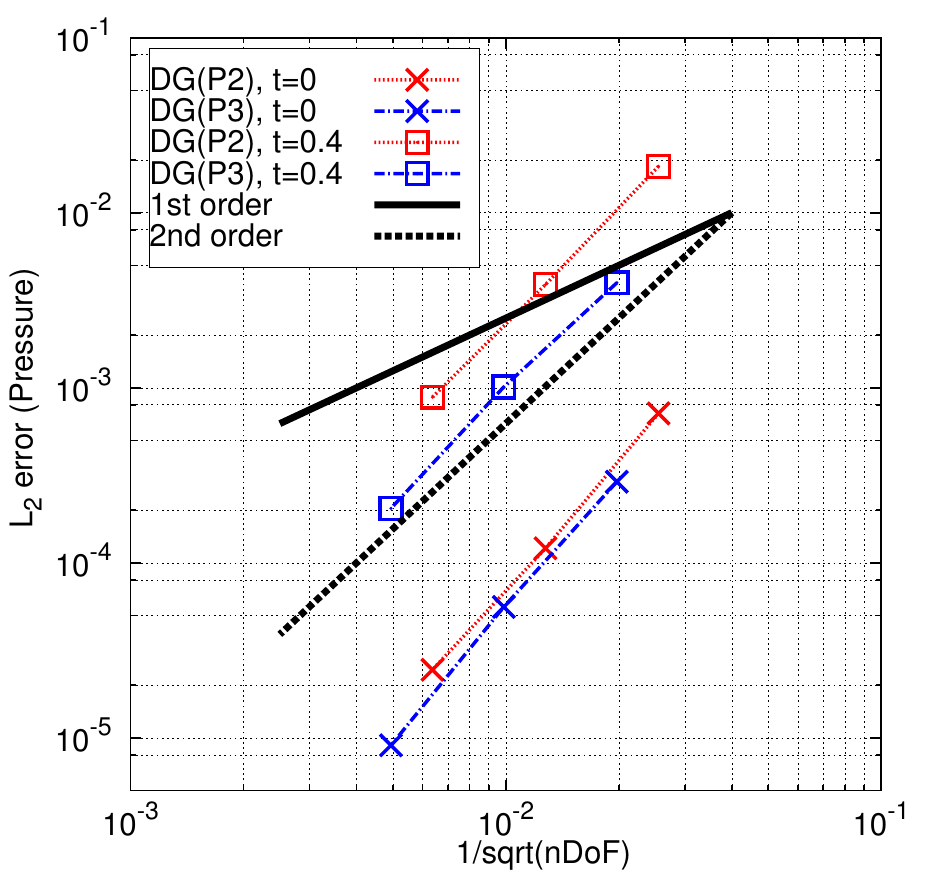}
\label{fig:greshol2pres}}
\subfloat[$L_2$ error of total energy]{
\includegraphics[width=3in]{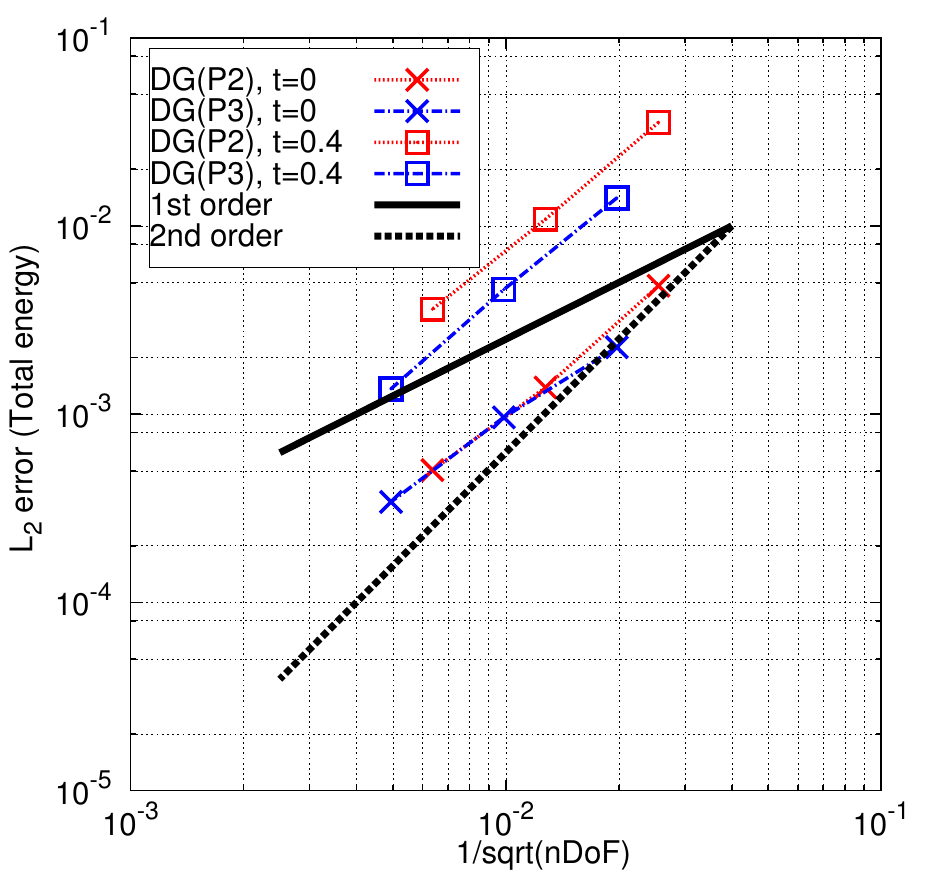}
\label{fig:greshol2total}}
\caption{ 
Convergence rate is shown for the 2D Gresho vortex problem at $t=0$ and $t=0.4$ using DG(P2) and DG(P3).
For any time instance, all the variables converge at a rate close to second-order for both DG(P2) and DG(P3) as expected. 
In addition, the numerical error of DG(P3) with cubic meshes is much less than that of DG(P2) with quadratic meshes even if both are only second-order accurate at most on this test problem. }
\label{fig:greshol2-cmprsn}
\end{figure*}

With the analytical solution, the variant of Gresho vortex problem is used to assess the order of accuracy of this new method.
We present the comparison for the numerical error and order of accuracy between DG(P2) and DG(P3). 
The associated results are shown in \tref{tab:order-gresho}.
For this test problem, it has been shown that the velocity is only $C_1$ continuous at $r=0.2$ and $r=0.4$, indicating that numerical methods can deliver at most second-order accuracy.
The results in \tref{tab:order-gresho} show that at $t=0$ both DG(P2) and DG(P3) deliver approximately second-order accuracy for all the variables except pressure, which is slightly higher. 
Likewise, at time $t=0.4$, the order of accuracy for both DG(P2) and DG(P3) is still second-order at most, which is expected. 
Also \fref{fig:greshol2-cmprsn} compares the efficiency between DG(P2) and DG(P3). 
The numerical error for variables  vs. the computational costs (\textit{i.e.}, nDoF) are presented.
It again shows that the higher-order DG(P3) method is more efficient since it can deliver superior accuracy at the same computational cost as the DG(P2) method, even on a problem that is second-order accurate at most.

\begin{table*}[t]
\begin{center}
\caption
{Numerical error and convergence rate are shown for the 2D Gresho vortex problem at different times using the DG method on curvilinear meshes.
Even for the initial $L_2$ projection, DG(P2) and DG(P3) deliver second-order of accuracy at most for velocity component $u_x$ and pressure $p$.
For a later time $t=0.4$, all the variables converge at approximately second-order for both DG(P2) and DG(P3). 
This is due to the fact that the exact solutions (\textit{i.e.}, velocity and total energy) are only $C^1$ continuous. 
}
\label{tab:order-gresho}
\mbox{
\begin{tabular}{l c c c c c c c c}
\hline
\multirow{2}{*}{Mesh}& \multicolumn{2}{c} {$\rho$}& \multicolumn{2}{c} {$u_x$}& \multicolumn{2}{c} {$p$} & \multicolumn{2}{c} {$\tau$}\\
                      \cmidrule(lr){2-3}
                      \cmidrule(lr){4-5}
                      \cmidrule(lr){6-7}
                      \cmidrule(lr){8-9}
                  & $L_2$ error & order & $L_2$ error & order & $L_2$ error & order & $L_2$ error & order\\
 \hline
DG(P2),\\
t=0\\
\hline
$ 16\times16$   & 4.8991e-16  &         & 3.7666e-3  &          & 7.1518e-4  &         & 4.8080e-3  &\\
$ 32\times32$   & 4.6949e-16  & -       & 1.2268e-3  &  1.62 & 1.2125e-4  & 2.57 & 1.3856e-3  &1.80\\
$ 64\times64$   & 4.6486e-16  & -       & 4.3483e-4  &  1.50 & 2.4526e-5  & 2.31 & 5.0456e-4  &1.46\\
\hline
DG(P2),\\
t=0.4\\
\hline
$ 16\times16$   & 3.9356e-3  &           & 1.5438e-2  &          & 1.8418e-2  &         & 3.5438e-2  &\\
$ 32\times32$   & 1.2165e-3  & 1.70  & 5.2000e-3  &  1.58 & 3.8883e-3  & 2.25 & 1.0854e-2  &1.71\\
$ 64\times64$   & 3.8063e-4  & 1.68   & 1.8206e-3  & 1.52 & 8.8336e-4  & 2.14 & 3.6134e-3  &1.59\\
\hline
DG(P3),\\
t=0\\
\hline
$ 16\times16$   & 5.7498e-16  &         & 2.1138e-3  &          & 2.9084e-4  &         & 2.2668e-3  &\\
$ 32\times32$   & 5.6856e-16  & -       & 8.2622e-4  &  1.37 & 5.5992e-5  & 2.39 & 9.6197e-4  &1.24\\
$ 64\times64$   & 5.8615e-16  & -       & 2.9127e-4  &  1.51 & 9.0702e-6  & 2.64 & 3.4154e-4  &1.50\\
\hline
DG(P3),\\
t=0.4\\
\hline
$ 16\times16$   & 1.6372e-3  &           & 6.7794e-3  &          & 3.9795e-3  &         & 1.4093e-2  &\\
$ 32\times32$   & 4.8989e-4  & 1.75   & 2.3066e-3  &  1.56 & 1.0115e-3  & 1.98 & 4.5845e-3  &1.63\\
$ 64\times64$   & 1.1083e-4  & 2.15   & 8.8086e-4  &  1.39 & 2.0478e-4  & 2.31 & 1.3614e-3  &1.76\\
\hline
\end{tabular}}
\end{center}
\end{table*}

\subsection{Sedov blast problem}
\label{sedov}
The Sedov test problem \cite{Sedov,MorganSedov} is an outward traveling blast wave in a gamma-law gas that is initiated by an energy source at the origin.
In this work, $\gamma=7/5$. The initial conditions are given by $(\rho^{0}, u_x^{0}, u_y^{0},p^{0}) = (1.0, 0, 0, 10^{-6})$ and there is an energy source at the origin.
The pressure in the cell containing the origin is given by $p_o = (\gamma-1){\rho_o}\frac{E_o}{w_o}$, where $w_o$ denotes the volume of the cell containing
the origin and $E_o$ is the total amount of released internal energy. The energy source is selected to be equal to
$0.244816$ so that the shock front is located at a radius equal to 1 at $t=1$. 
The computational domain is defined in 2D Cartesian coordinates $[0, 1.2]\times[0, 1.2]$. 
The mesh resolution is $30 \times 30$ and initially uniform. 
Symmetry boundary conditions are imposed on the left and bottom boundaries.  The outer boundaries are fixed.
The limiting is performed in the local flow direction.

\begin{figure*}[h!]
\centering
\subfloat[DG(P2)]{
\includegraphics[width=3in]{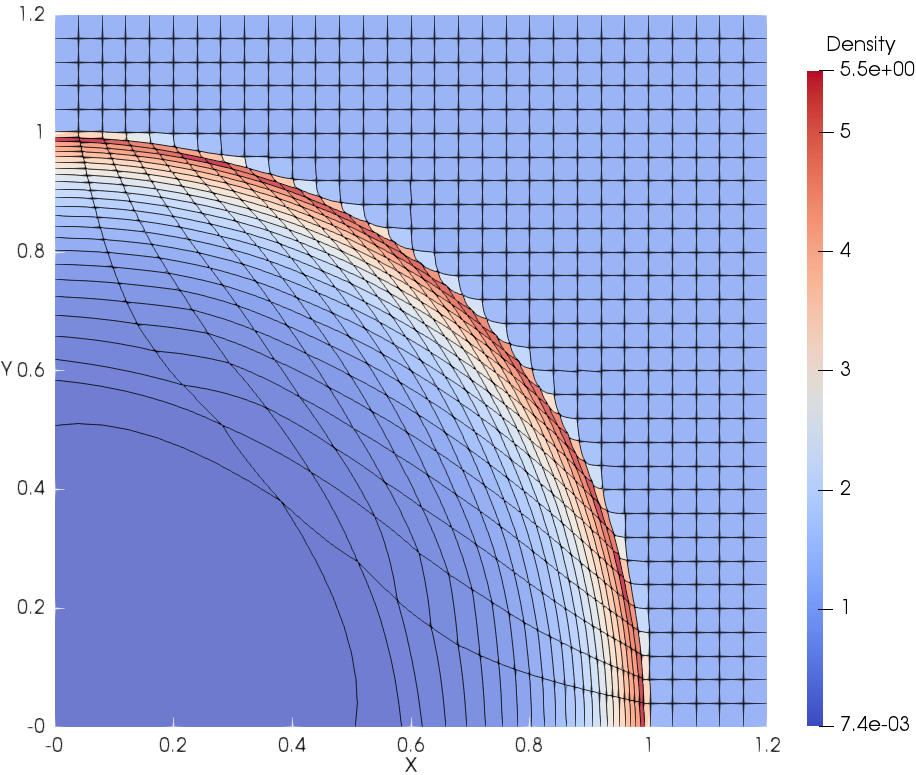}
\label{fig:sedovmesh-p2}}
\subfloat[DG(P3)]{
\includegraphics[width=3in]{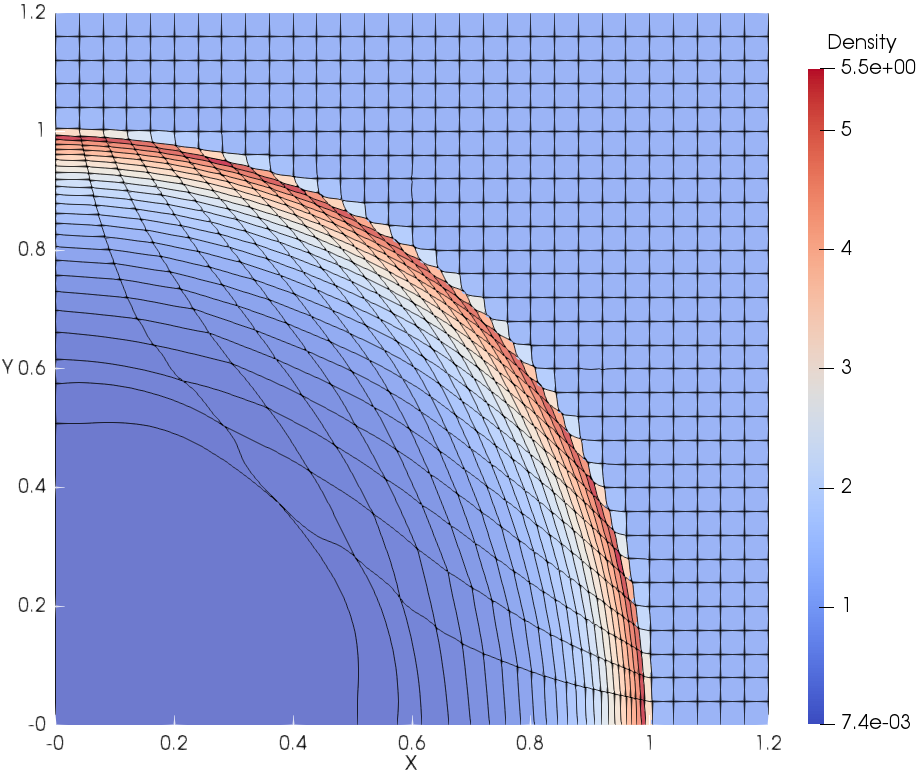}
\label{fig:sedovmesh-p3}}
\caption{ 
The meshes and density fields for the Sedov blast problem at $t=1.0$ are shown using DG(P2) and DG(P3) method on curvilinear meshes with a resolution of $30 \times 30$ cells.  
The limiting is performed in the local flow direction. The meshes move in a stable and curvilinear manner. }
\label{fig:sedovmeshrho}
\end{figure*}

\begin{figure*}[h!]
\centering
\subfloat[Scatter plots of density]{
\includegraphics[width=3in]{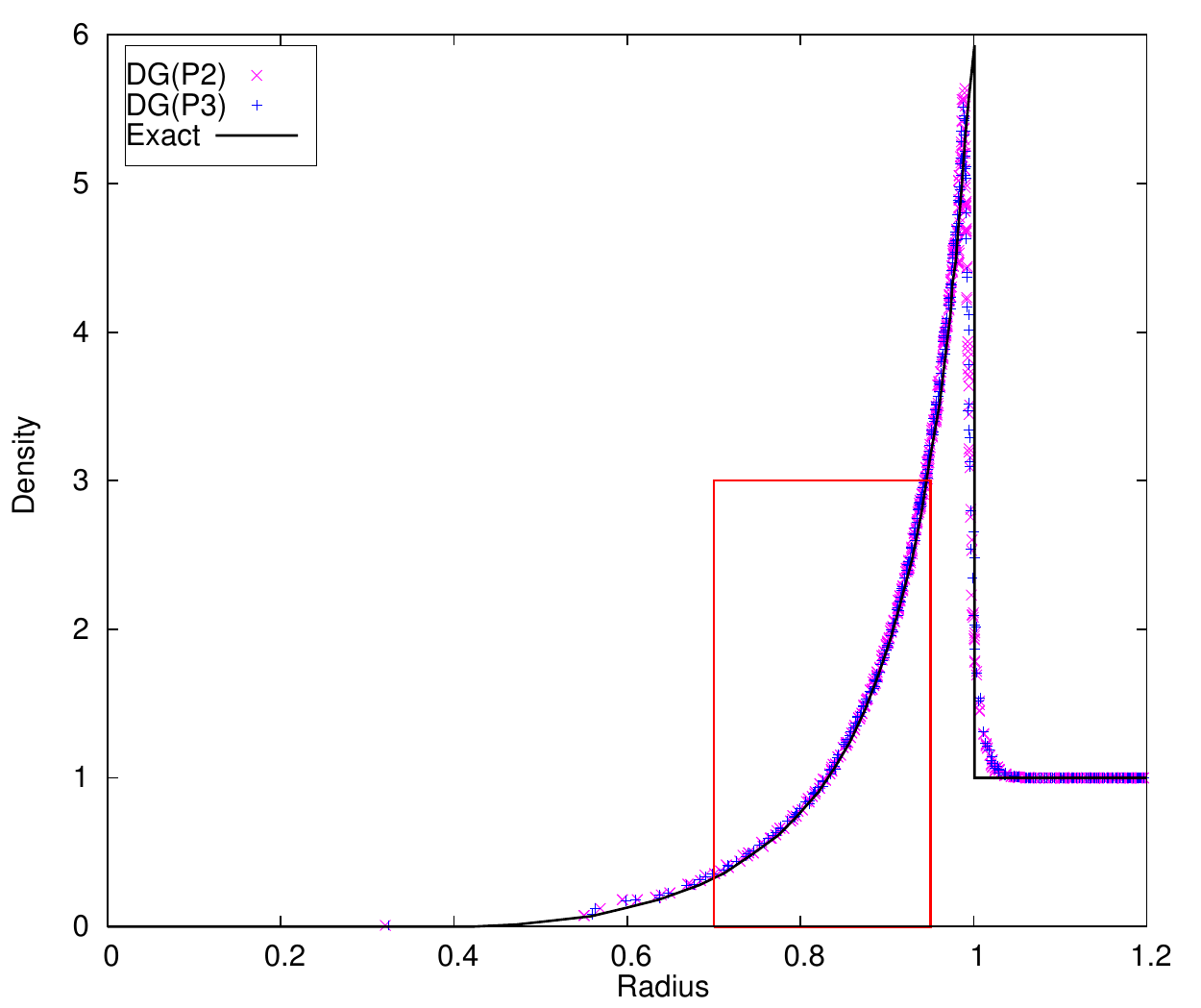}
\label{fig:sedovrho}}
\subfloat[Close-up of the scatter plots]{
\includegraphics[width=3in]{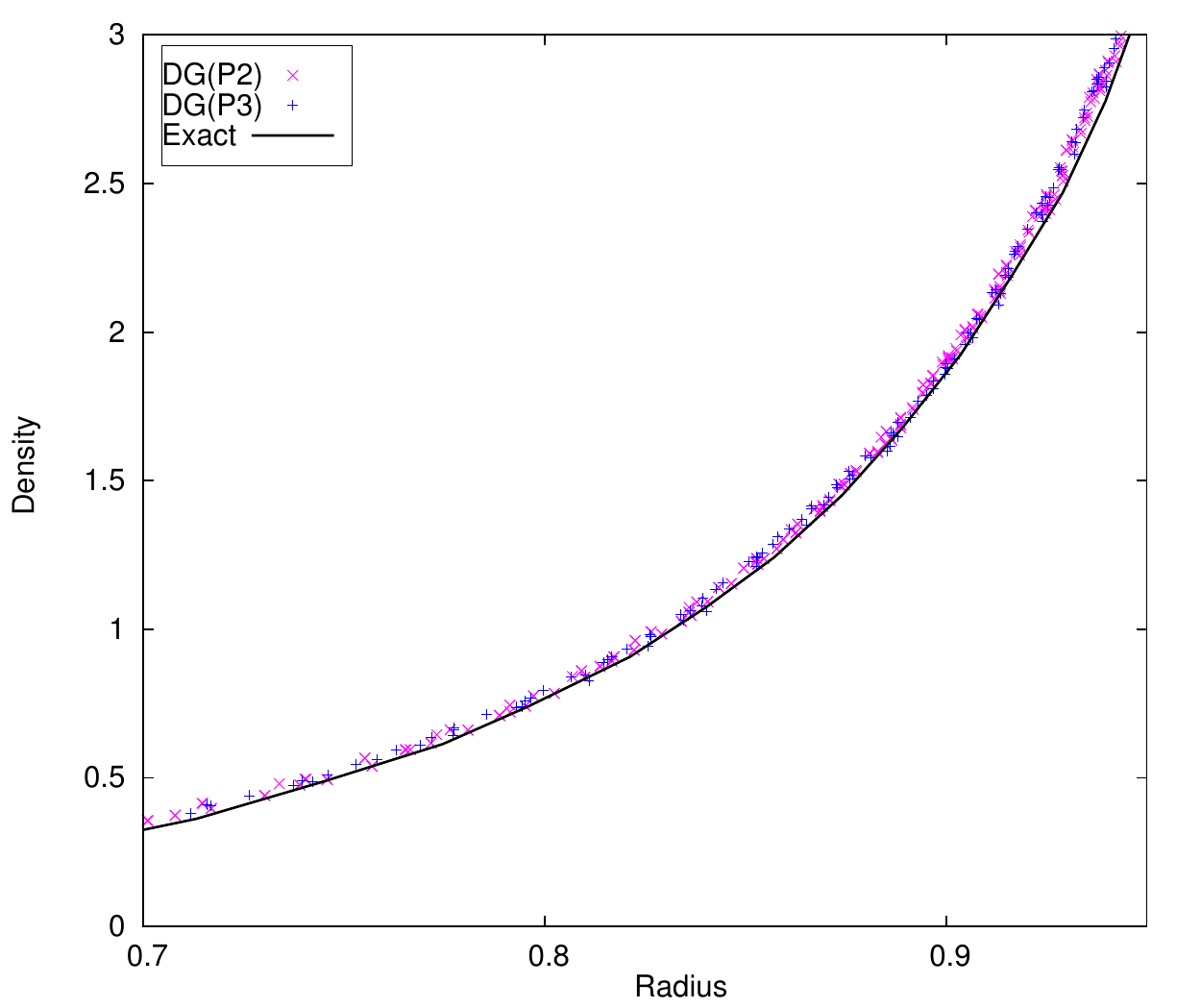}
\label{fig:sedovrho-cu}}
\caption{ 
The scatter plots of density at $t=1.0$ are shown for the Sedov blast problem with the new DG method using a $30 \times 30$ curvilinear mesh.
The numerical results agree very well with the exact solution. In addition, from \fref{fig:sedovrho-cu}, the density scatter for DG(P3) is a little less than that of DG(P2).}
\label{fig:sedovrho-cmprsn}
\end{figure*}

The final meshes and density contours are shown in \fref{fig:sedovmeshrho} using this new Lagrangian DG hydrodynamic method.
The meshes move in a stable and curvilinear manner. 
There is no spurious mesh motion \cite{MorganVeloFilter2017} due to using this new hierarchical orthogonal basis and SMS method.
The scatter plots of density at the final time are shown in \fref{fig:sedovrho-cmprsn}.  
The results are in good agreement with the exact solution. 
From the close-up in \fref{fig:sedovrho-cu}, the density scatter of DG(P3) is a little less than DG(P2).
It is necessary to reiterate that the hierarchical orthogonal basis is crucial in the context of limiting 
for shock problems. 

\subsection{Elastic vibration of a Beryllium plate }
\label{sec:EVBePlate}
Thin beams or plates have been the subject of many test problems, see for example \cite{FB}. The problem considered in this paper is a thick plate with free boundaries. 
The problem has been previously analyzed using PAGOSA \cite{PAGOSAcase}, a three-dimensional Eulerian hydrodynamic code. Here the problem is solved in a Lagrangian framework as described in \cite{BurtonCGR}. This is a two dimensional test problem comprised of an elastic beryllium plate with no supports or constraints. 
The computational domain is defined by $(x, y) \in [-3$ $cm, 3$ $cm] \times [-0.5$ $cm, 0.5$ $cm]$. 
The centerline of the plate initially coincides with the axis $x$. 
There is no analytic solution for a thick plate as specified for this problem. 
However, for a thin unconstrained plate, the solution for the first flexural mode is given by  \cite{BurtonCGR},
\begin{equation*}
y(x, t) = A sin(\omega t)\big(g_1[sinh(\Omega (x+3))+sin(\Omega (x+3))]-g_2[cosh(\Omega (x+3))+cos(\Omega (x+3))] \big),
\end{equation*}
where $\Omega=0.7883401241$ $cm^{-1}$, $\omega=0.2359739922$ ${\mu s}^{-1}$, $A=0.004336850425$  $cm$, $g_1=56.63685154$ ${\mu s}$ and $g_2=57.64552048$ ${\mu s} $.
Then the plate is prescribed with an initial velocity distribution as given by
\begin{equation*}
\begin{array}{lll}
u(x,  t=0) &=&  0 ,\\
v(x,  t=0) &=& A \omega \big(g_1[sinh(\Omega (x+3))+sin(\Omega (x+3))]-g_2[cosh(\Omega (x+3))+cos(\Omega (x+3))] \big).
\end{array}
\end{equation*}

The equation of state is given by the Gr{\"u}neisen model characterized by the following parameters: 
$\rho_0=1.845$ $kg/cm^3$, $c_0=1.287$ $cm/{\mu s}$, $\Gamma_0=1.11$ and $s=1.124$. In addition, the shear modulus $G=1.1519$ $Mbar$ and the yield strength $\sigma_0=1$ $Mbar$.
The first flexural mode has a period of $ 2\pi/\omega = 26.6266$ $\mu s$. 
All the boundaries are set to have free boundary conditions.
A set of uniformly refined meshes (\textit{i.e.}, $10 \times 4$, $20 \times 8$ and $40 \times 16$) are calculated. This case is run to $t=100$ $\mu s$, demonstrating the robustness and accuracy of the DG(P3) method with cubic meshes.  Dissipation errors will artificially dampen the oscillations so the goal is to maintain the peak amplitudes.

\begin{figure}[h!]
\centering
\subfloat[The l1 mesh at $t=6.65$ $\mu s $]{
\includegraphics[width=3.5in]{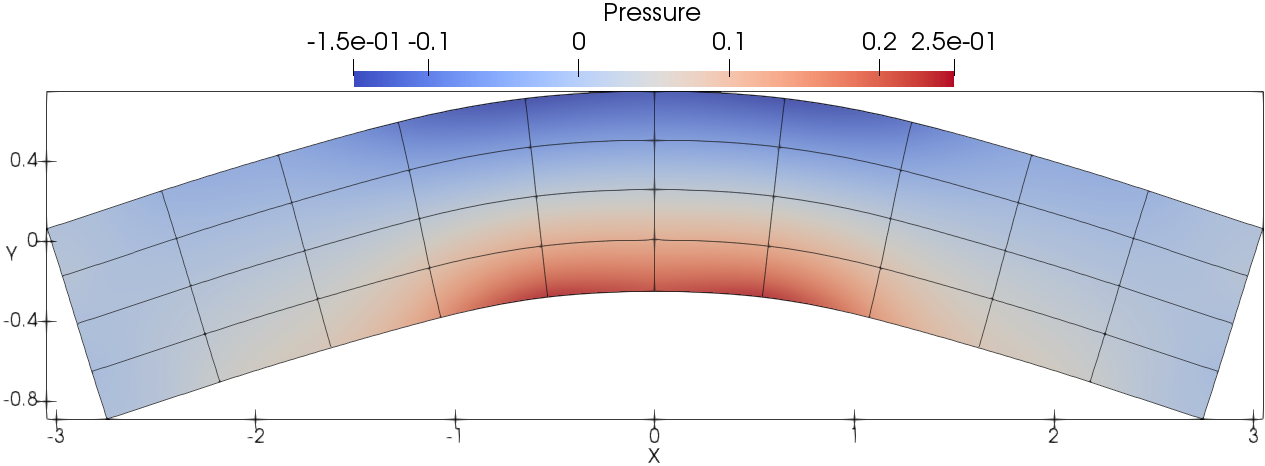}}
\subfloat[The l2 mesh at $t=6.65$ $\mu s$]{
\includegraphics[width=3.5in]{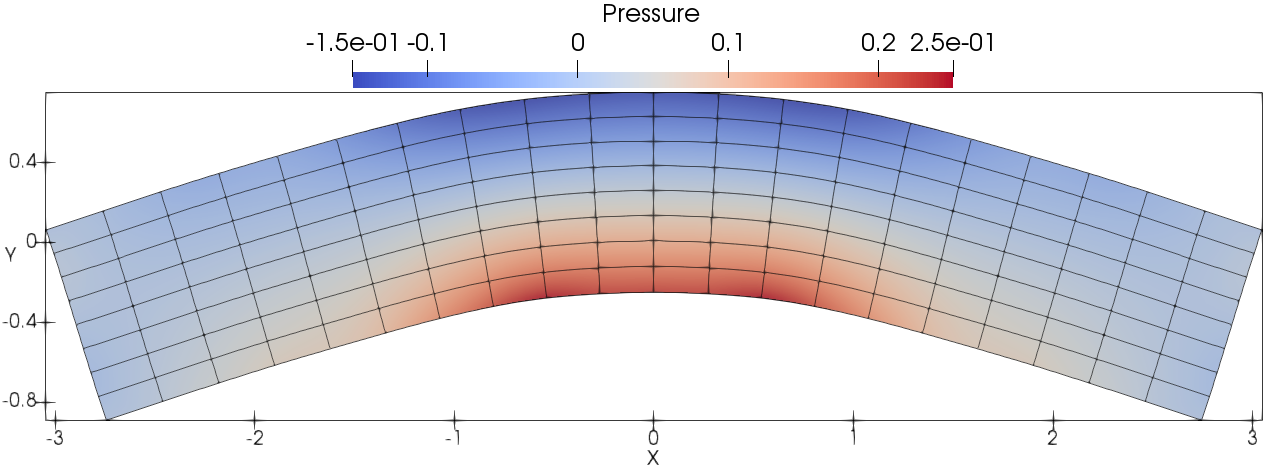}}\\
\vskip 0.2in
\subfloat[The l1 mesh at $t=19.96$ $\mu s$]{
\includegraphics[width=3.5in]{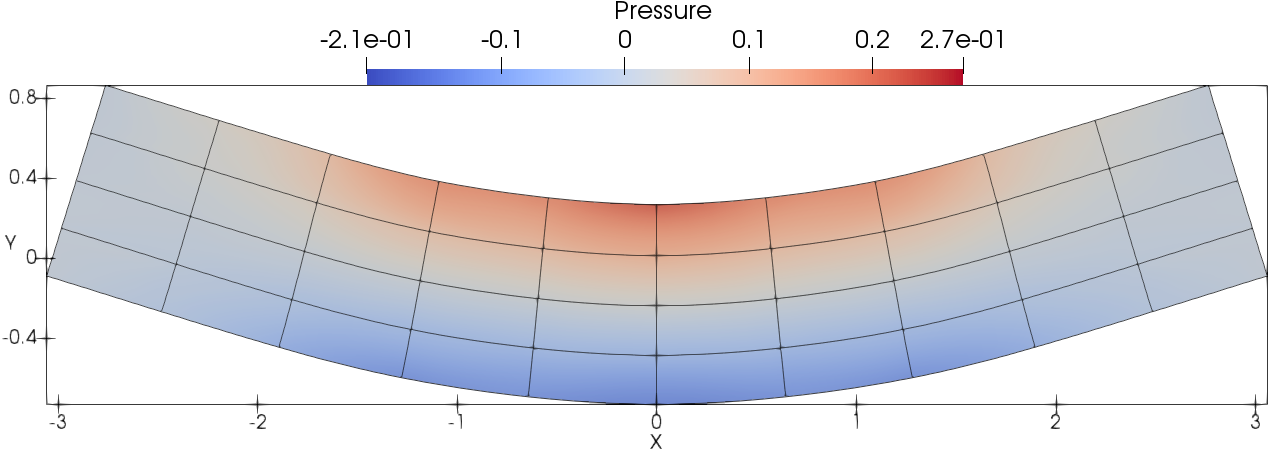}}
\subfloat[The l2 mesh at $t=19.96$ $\mu s$]{
\includegraphics[width=3.5in]{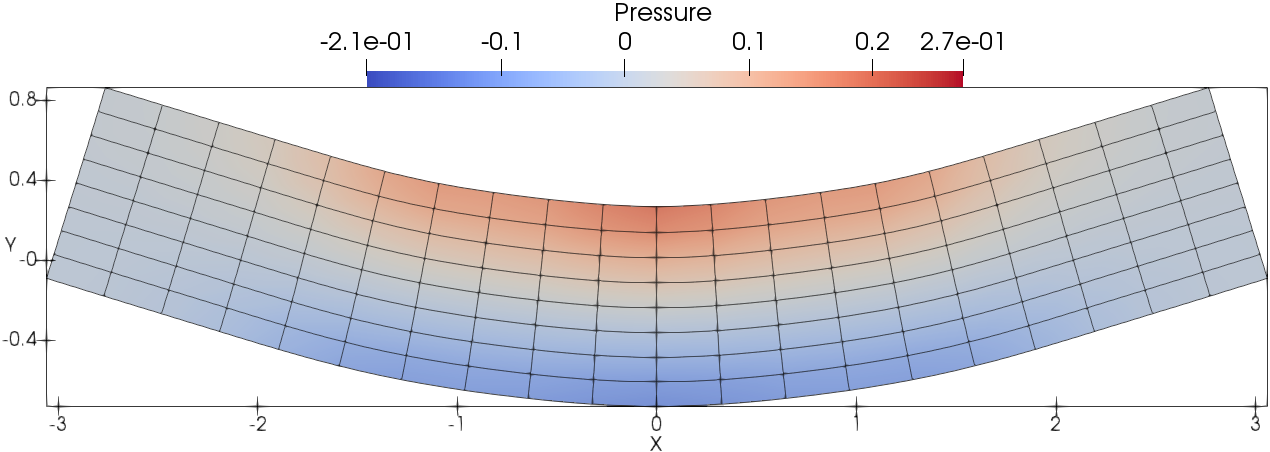}}\\
\vskip 0.2in
\subfloat[The l1 mesh at $t=33.25$ $\mu s$]{
\includegraphics[width=3.5in]{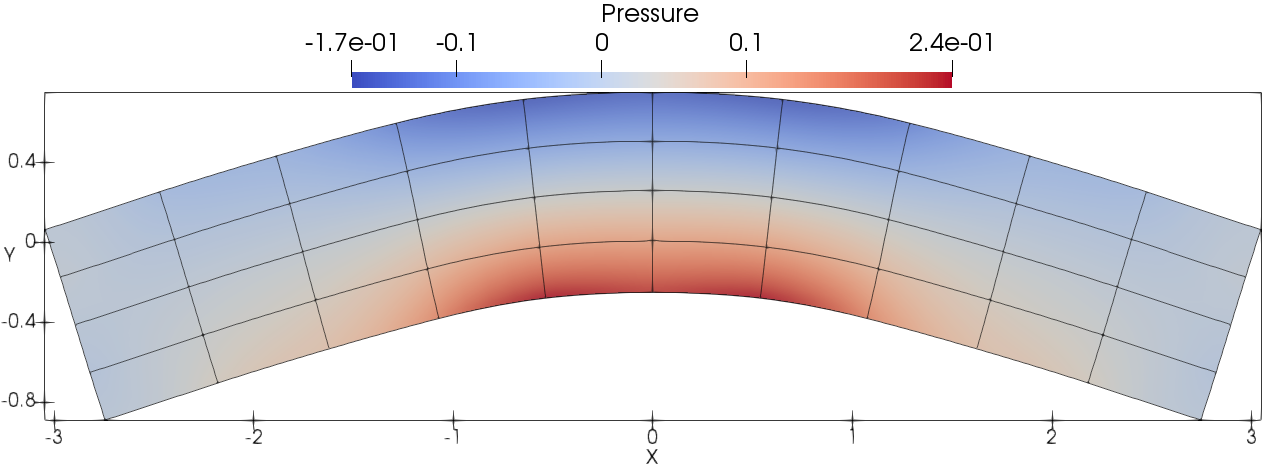}}
\subfloat[The l2 mesh at $t=33.25$ $\mu s$]{
\includegraphics[width=3.5in]{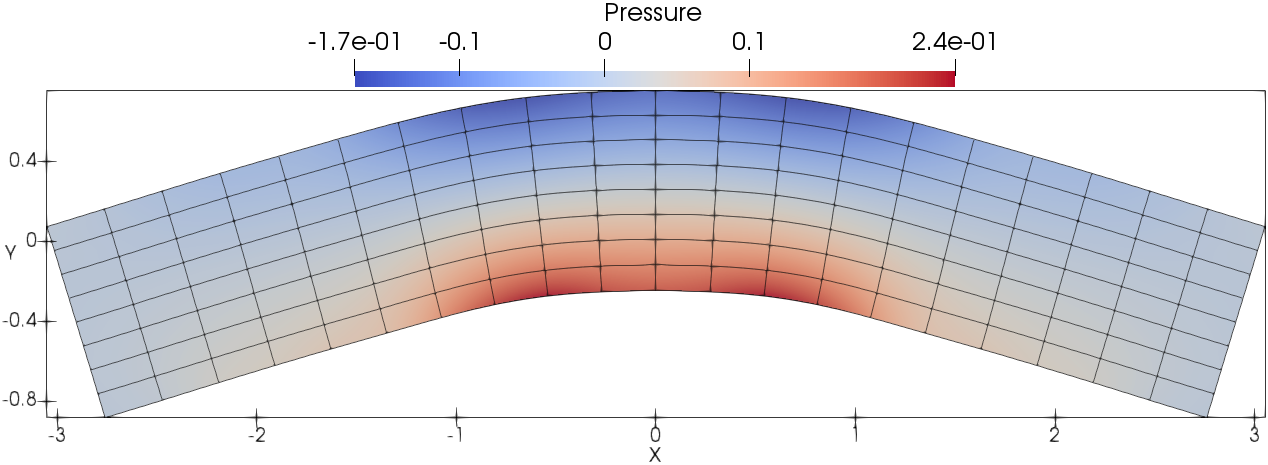}}\\
\vskip 0.2in
\subfloat[The l1 mesh at $t=100$ $\mu s$]{
\includegraphics[width=3.5in]{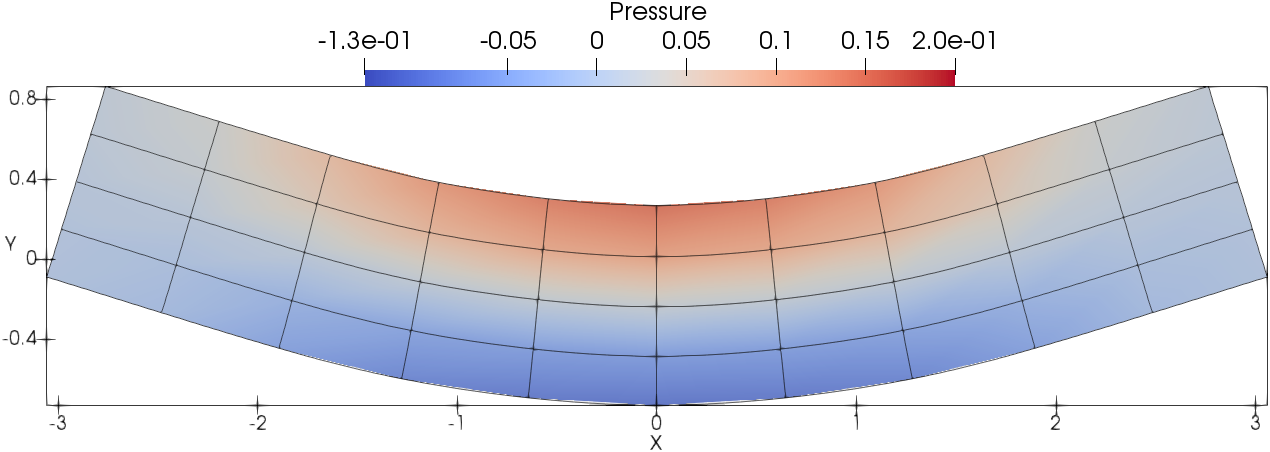}}
\subfloat[The l2 mesh at $t=100$ $\mu s$]{
\includegraphics[width=3.5in]{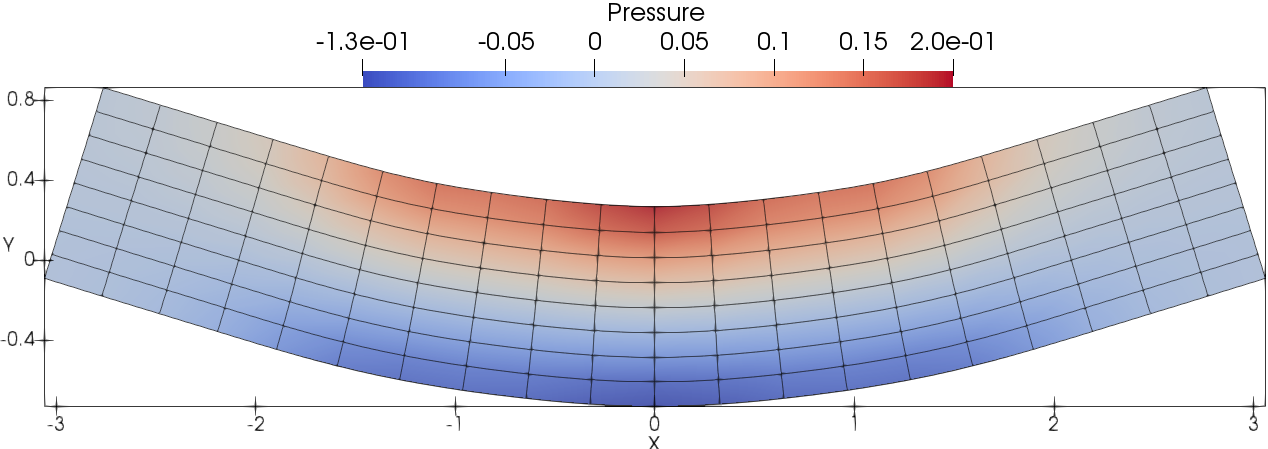}} \\
\caption{ 
A set of cubic meshes and pressure fields for the elastic vibrations of a beryllium plate at different times using DG(P3).
The time instance is selected to be one quarter period ($t=6.65$ $\mu s $), three quarters period ($t=19.96$ $\mu s $), five quarters period ($t=33.25$ $\mu s $), and $t=100$ $\mu s$.
The meshes are moving in a robust and curvilinear manner. DG(P3) captures the flow details for the pressure fields with very coarse mesh resolution. 
This case has been run to a very late time $t=100$ $\mu s$, demonstrating the robustness and accuracy of the new Lagrangian DG hydrodynamic method.}
\label{fig:BBeammesh}
\end{figure}

\begin{figure}[h!]
\centering
\includegraphics[width=3.5in]{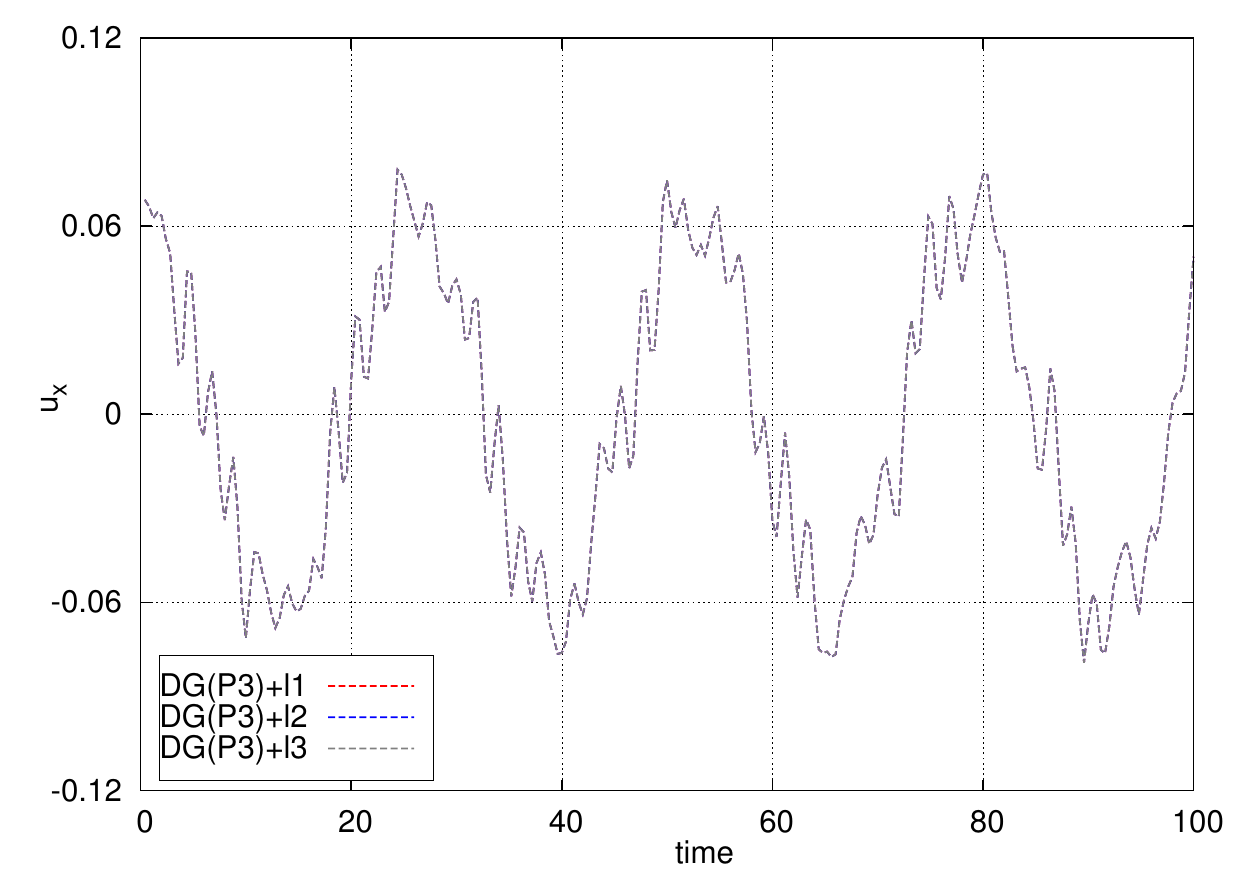}
\caption{ 
This figure presents the convergence histories for the vertical velocity $u_x$ at the central point for the beryllium plate using DG(P3) on a set of cubic meshes.
Here, $l1$ , $l2$ and $l3$ denote the coarse, medium, and fine mesh resolutions respectively. 
Due to the very low dissipation of DG(P3), the convergence histories from the coarse mesh almost overlaps with that of finer mesh resolutions. 
In addition, DG(P3) can maintain a uniform amplitude for the vertical velocity with time marching.}
\label{fig:BBeamConvergence}
\end{figure}

\fref{fig:BBeammesh} presents the mesh and pressure contours at different times respectively. 
The legend scales are set differently for each time.
The meshes and pressure contours at $t=6.65$ $\mu s$ (one quarter period ) and  $t=33.25$ $\mu s$ (three quarters period) (the difference is one period approximately), look very similar qualitatively due to the low numerical dissipation of fourth-order accurate DG(P3) method.
In addition, the vertical velocity of the central point for the plate is recorded for assessing the grid convergence that is shown in \fref{fig:BBeamConvergence}. 
There is no visible difference about the evolution of the vertical velocity profile on uniformly refined meshes, showing the accurate solution with DG(P3). Due to its low dissipation, DG(P3) also maintains a uniform amplitude for the vertical velocity.
And it can be observed that the period based on the vertical velocity is approximately equal to the analytical one.

\subsection{Taylor bar impact on a wall}
\label{Taylor-bar}

This  problem  consists  of  a 2D  aluminum  bar  impacting a  rigid  wall. The kinetic energy in the rod is entirely converted into internal energy through plastic dissipation. This type of test was initially proposed by Taylor \cite{Taylor1948} with a cylindrical rod. 
In this paper, we considered the 2D planar  bar impact test that was simulated in \cite{Mairesolid2013}.  
The equation of state is given by the Gr{\"u}neisen model  characterized by the following parameters: 
$\rho_0=2.785$ $kg/cm^3$, $c_0=0.533$ $cm/{\mu s}$, $\Gamma_0=1.5111$ and $s=1.338$. 
 In addition, the shear modulus $G=0.276$ $Mbar$ and the yield strength $\sigma_0=0.003$ $Mbar$.
The  computational  domain  is  the  initial  projectile $(x,y)\in[0$ $cm, 5$ $cm]\times[0$ $cm,1$ $cm]$.
A  set of uniformly refined meshes (\textit{i.e.}, $10 \times 4$, $20 \times 8$ and $40 \times 16$) are used to perform a grid convergence study. This test case is run to $t=50$  $\mu s$.
  The initial velocity is set to $\textit{v}=(-0.015$  $cm/{\mu s}, 0)$.
There exists no exact solution for this problem, and this case is used to show the robustness of the method and for accuracy bench-marking.

\begin{figure}[h!]
\centering
\subfloat[The $l1$ mesh with DG(P2)]{
\includegraphics[width=3.2in]{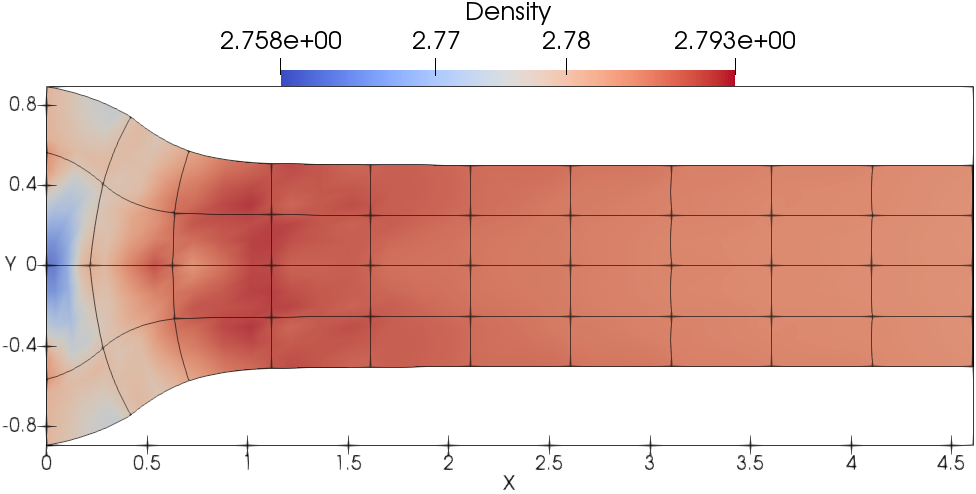}\label{fig:Taylormeshl1p2} }
\subfloat[The $l1$ mesh with DG(P3)]{
\includegraphics[width=3.2in]{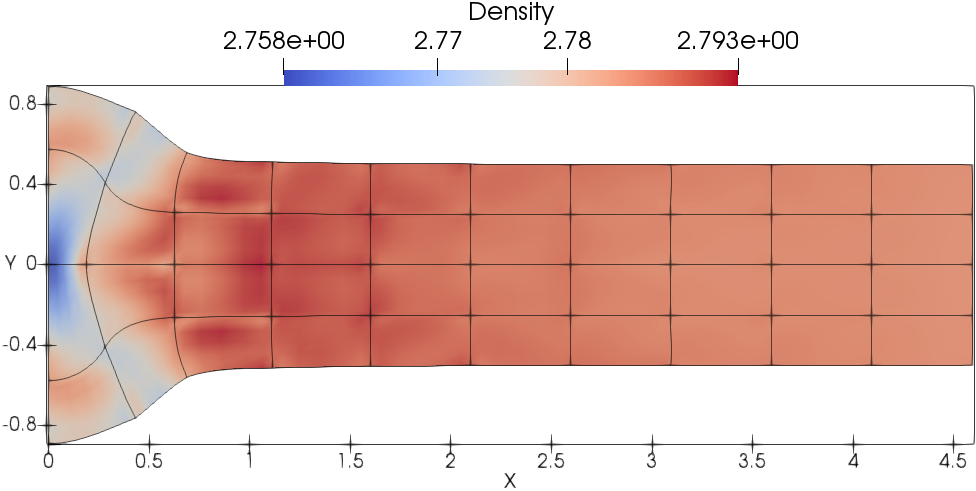}\label{fig:Taylormeshl1p3} }\\
\subfloat[The $l2$ mesh with DG(P2)]{
\includegraphics[width=3.2in]{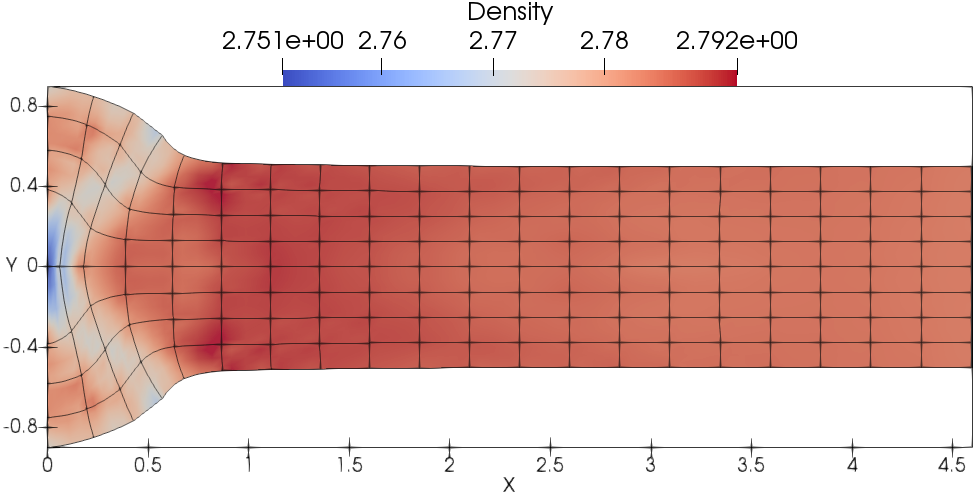}} 
\subfloat[The $l2$ mesh with DG(P3)]{
\includegraphics[width=3.2in]{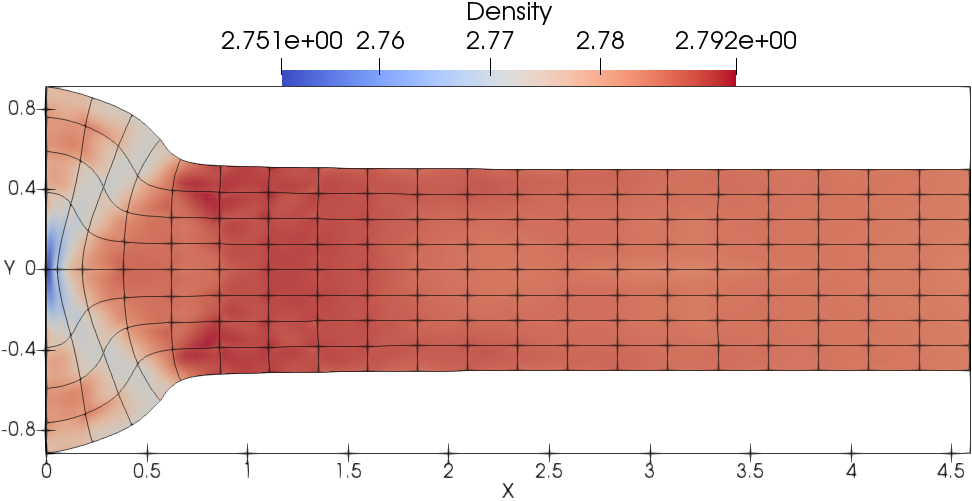}} \\
\subfloat[The $l3$ mesh with DG(P2)]{
\includegraphics[width=3.2in]{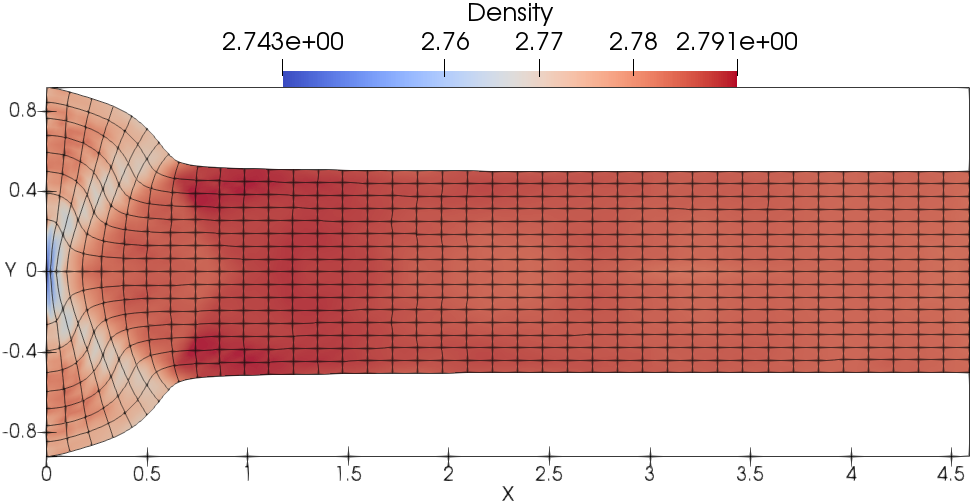}} 
\subfloat[The $l3$ mesh with DG(P3)]{
\includegraphics[width=3.2in]{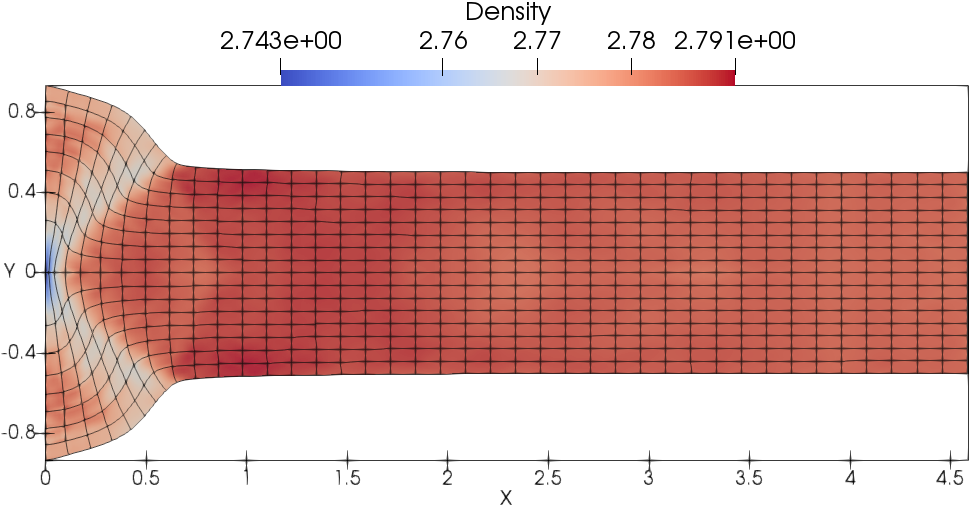}} 
\caption{ 
A set of meshes and density fields shown for a rod impact a wall after $t=50$ $\mu s$ using the DG(P2) and DG(P3) methods. 
By comparing \fref {fig:Taylormeshl1p2} with \ref{fig:Taylormeshl1p3}, DG(P3) delivers more accurate mesh motion and density fields than DG(P2) as expected. 
The wave structure for the density fields is captured in more detail as the mesh is refined.}
\label{fig:Taylormesh}
\end{figure}

\fref {fig:Taylormesh} presents the meshes and density fields.
As shown, the mesh is moving in a curvilinear manner where DG(P2) results are shown in \fref{fig:Taylormeshl1p2} and DG(P3) results are shown in \ref{fig:Taylormeshl1p3}. 
The wave structure is captured in more details by the DG(P3) method compared with the DG(P2) method.

\begin{figure}[h!]
\centering
\subfloat[Energy balance with DG(P2)]{
\includegraphics[width=3in]{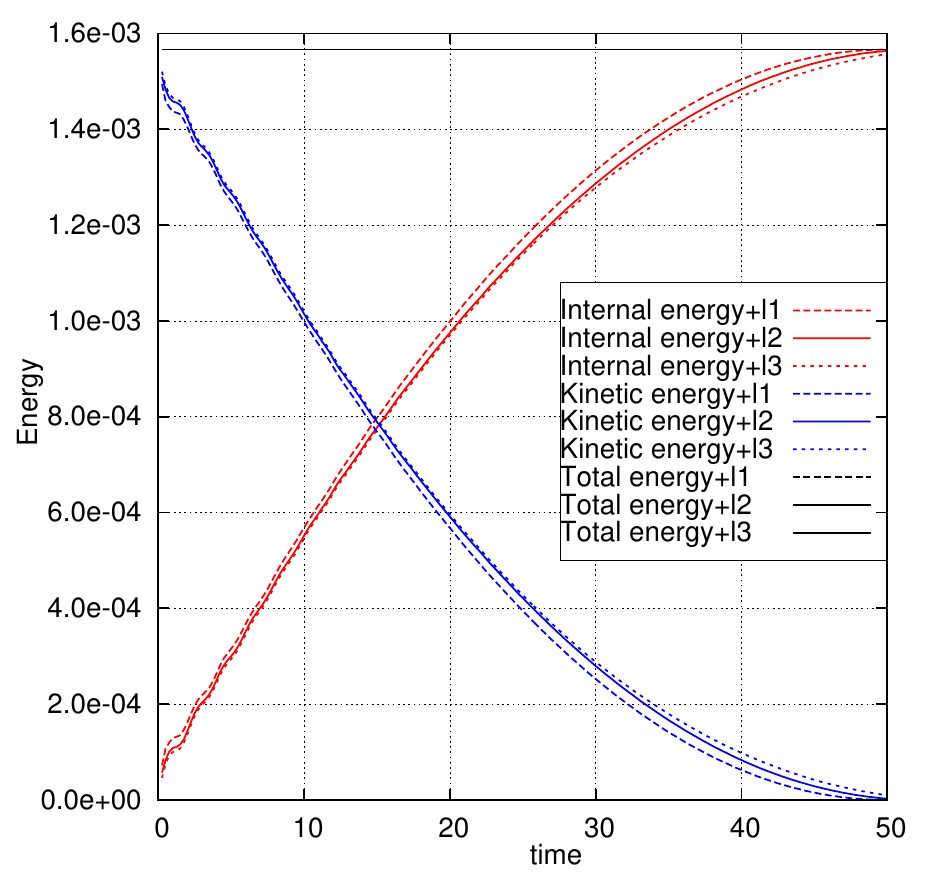}\label{fig:TaylorEnergyp2}}
\subfloat[Energy balance with DG(P3)]{
\includegraphics[width=3in]{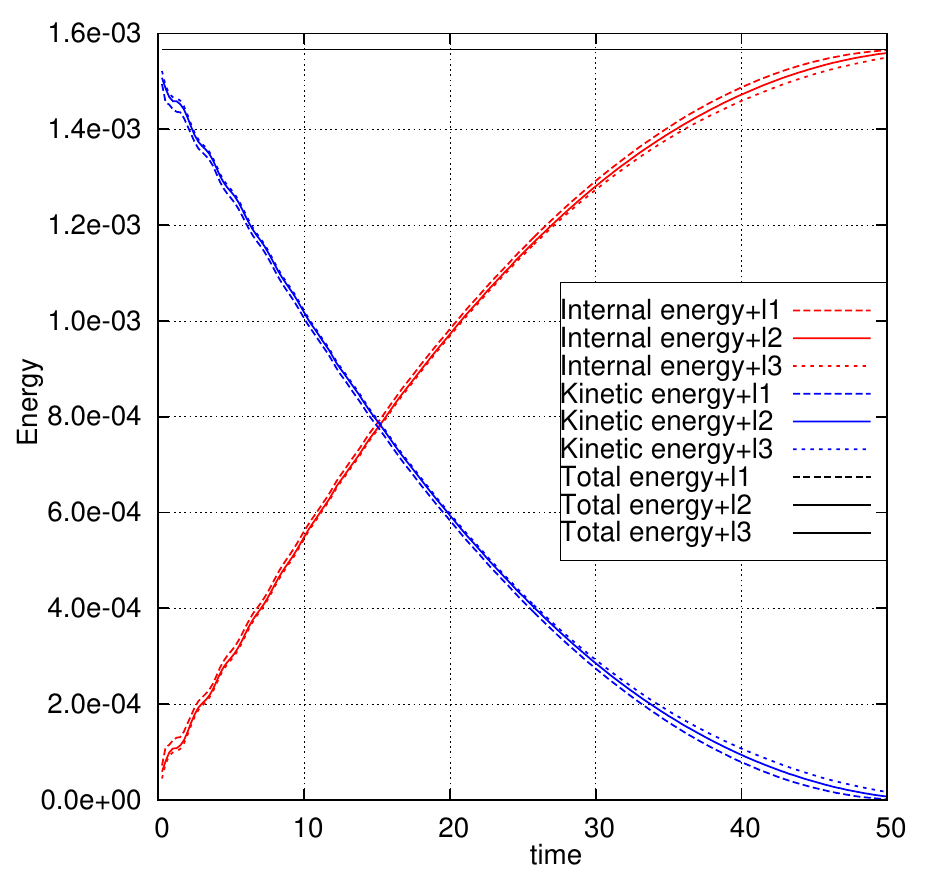}\label{fig:TaylorEnergyp3}}\\
\subfloat[The length evolution with DG(P2)]{
\includegraphics[width=3in]{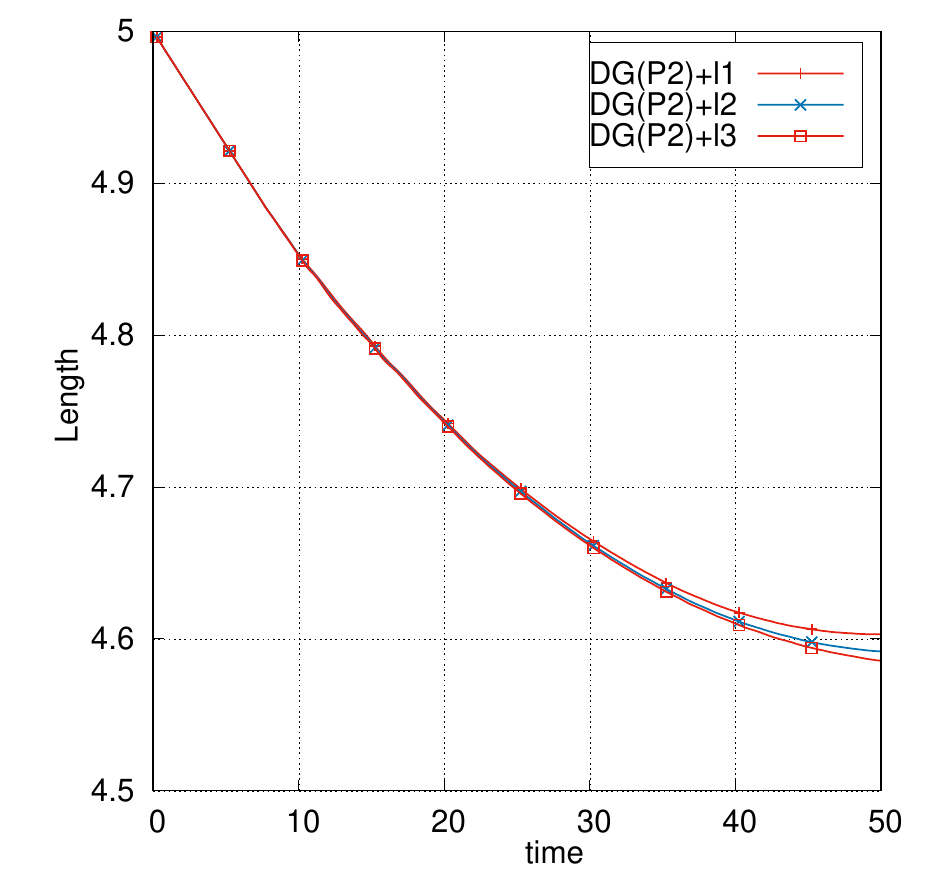}\label{fig:TaylorLengthp2}} 
\subfloat[The length evolution with DG(P3)]{
\includegraphics[width=3in]{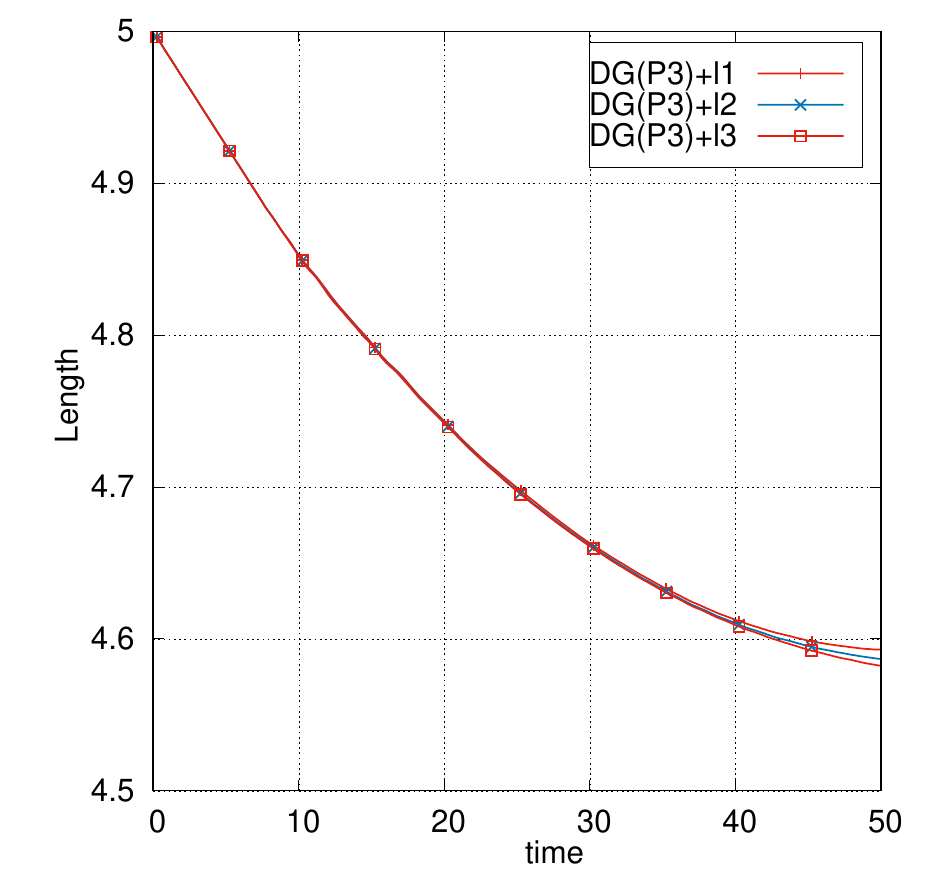}\label{fig:TaylorLengthp3}} 
\caption{ 
This figure presents the convergence histories for the energy conversion (\fref{fig:TaylorEnergyp2} and \ref{fig:TaylorEnergyp3}) 
and length evolution (\fref{fig:TaylorLengthp2} and \ref{fig:TaylorLengthp3}) of the Taylor rod (that impacts a wall) at $t=50$ $\mu s$ using the DG(P2) and DG(P3) methods. 
Here, kinetic energy is converted into internal energy.
The difference between the coarse and fine mesh is very small, showing the high-order accuracy with the DG(P3) method.}
\label{fig:TaylorConvergence}
\end{figure}

In this test problem, the kinetic energy is entirely converted into internal energy through the plastic dissipation shown in \fref{fig:TaylorEnergyp2} and \ref{fig:TaylorEnergyp3}.
The temporal evolution of the length of the Taylor bar is also shown in \fref{fig:TaylorLengthp2} and \ref{fig:TaylorLengthp3}. 
Taking the solution on the finest mesh as a reference, the convergence study shows that the DG(P3) method on cubic meshes converges quickly to the reference solution (\textit{i.e.}, the evolution of energy balance and length). 
Furthermore, favorable results can be obtained using a very coarse mesh resolution (\textit{i.e.}, $10 \times 4$).

\section{Conclusions}
\label{Conclusion}

The motivation for the work was to create a robust Lagrangian DG hydrodynamic method with a hierarchical orthogonal basis for material dynamics that is up to fourth-order accurate on cubic cells, \textit{i.e.}, each edge is defined by a cubic polynomial.  The subcell mesh stabilization (SMS) method was extended to work with cubic cells.  The SMS method is an important part of the new scheme to ensure robust solutions on problems with strong shocks.  
Another key part to ensuring robust solutions on strong shock problems is using hierarchical orthogonal basis functions for the polynomials of higher degrees of freedom (P3 or higher).  We found that the limiting of the polynomials constructed from Taylor basis functions with the DG(P3) method gave very poor results on strong shock problems like the Sedov blast wave.  The poor results were attributed to the moments in the mass matrix being coupled.  This motivated us to create a new DG method based on a hierarchical orthogonal basis about the center of mass,
resulting in a diagonal mass matrix (\textit{i.e.}, no matrix inversion is required) that are constant in time, which ensures accurate and robust solutions on strong shock problems with limiting. According to the operational cost of matrix multiplication, this formulation is also computationally more efficient compared with the Lagrangian DG method based on Taylor-basis functions.

A suite of test cases covering gas and solid dynamics were simulated. Likewise, vortical flows and shock driven flows were simulated to demonstrate the robustness of the proposed new high-order Lagrangian DG hydrodynamic method. The designed order of accuracy (up to fourth order) was demonstrated on the Taylor-Green vortex. The ability to simulate large deformation problems was demonstrated on the Taylor-Green and Gresho vortex problems using cubic cells. In addition, with Taylor-Green vortex, DG(Pn) can deliver the $(n+1)^{th}$ order of accuracy on quadrilaterals meshes with edges defined by a polynomial of degree $n$ or higher in Lagrangian hydrodynamics. The Sedov blast wave problem showed that the new method is robust and accurate on very strong shocks.  The results from the Taylor anvil impact test and the vibrating beam problem demonstrate the method can accurately simulate solid dynamics. For the Taylor anvil test case, we also showed that the fourth-order accurate method was superior to the third-order accurate method. 

This work fills an existing algorithmic gap by delivering a computationally efficient high-order hydrodynamic method to robustly and accurately evolve cubic meshes.
Using cubic meshes is key to improving the fidelity of engineering and physics simulations, because a cubic cell can reproduce the as-built part dimensions in a simulation so that the exact volume, shape, mass, and density can be used. With linear meshes, the part contours do not reproduce the correct volume, so to match the mass, the part contours must be artificially shifted. Mesh refinement with linear meshes will only recover the exact part contours in the limit of a zero mesh size.  As such, the impact of this work covers accurate simulations of large deformation flows and shock driven material dynamics to enabling the correct initial conditions to be used in simulations with manageable resolutions.

\section{Acknowledgments}

We gratefully acknowledge the support of the NNSA through the Laboratory Directed Research and Development (LDRD) program at Los Alamos National Laboratory.  The Los Alamos unlimited release number is LA-UR-21-21718. Los Alamos National Laboratory is managed by Triad National Security, LLC for the U.S. Department of Energy's NNSA .

\appendix

\renewcommand{\thesubsection}{\Alph{section}.\arabic{subsection}} 

\setcounter{figure}{0}
\renewcommand{\thefigure}{\Alph{section}\arabic{figure}} 

\setcounter{table}{0}
\renewcommand{\thetable}{\Alph{section}\arabic{table}} 

\section{Study of the hierarchical basis functions $\boldsymbol{\phi}$}
\label{BFproof}
For this section, we aim at proving the material derivatives for the basis functions $\boldsymbol{\phi}$ are equal to 0,
consisting of the following three steps.
From the previous work \cite{LiuDGlagSMS2019}, 

  \begin{equation}
    \label{drhojdt}
    \frac{d}{dt} (\rho j) =0.
   \end{equation}
   
   \begin{equation}
    \label{dxidt}
    \begin{split}
    \frac{d{\boldsymbol \xi}}{dt} =0,
    \frac{d{\boldsymbol \xi}_{cm}}{dt} =0,
    \end{split}
   \end{equation}
   
\noindent
where, ${\boldsymbol \xi} = (\xi, \eta)$ and ${\boldsymbol \xi}_{cm} = (\xi_{cm}, \eta_{cm})$ . 
   Based on \eref{drhojdt} and \ref{dxidt}, it is possible to get 
   
   \begin{equation}
    \label{dpsidt}
    \frac{d \boldsymbol \psi}{dt}  =0.
   \end{equation} 

\noindent
Let's assume 
   \begin{equation}
    \label{coefmat}
    A_{mn} = -\frac{ \langle \psi_m , \phi_n \rangle }{ \langle \phi_n , \phi_n \rangle },  \quad \quad m = 2,...,N(P), \quad 1 \le n < m.\\
   \end{equation} 
\noindent
For DG(P3), 
 \begin{equation}
 \begin{array}{l}
   \begin{bmatrix}
   \phi_1\\
   \phi_2\\
   \phi_3\\
   \phi_4\\
   \phi_5\\
   \phi_6\\
   \phi_7\\
   \phi_8\\
   \phi_9\\
   \phi_{10}\\
   \end{bmatrix} 
   =
   \begin{bmatrix}
    1 & \\
    A_{21} & 1\\
    A_{31} & A_{32}& 1\\
    A_{41} & A_{42}& A_{43} &1\\
    A_{51} & A_{52}& A_{53} & A_{54} &1\\
    A_{61} & A_{62}& A_{63} & A_{64} & A_{65} &1\\
    A_{71} & A_{72}& A_{73} & A_{74} & A_{75} & A_{76}&1\\
    A_{81} & A_{82}& A_{83} & A_{84} & A_{85} & A_{86}& A_{87} & 1\\
    A_{91} & A_{92}& A_{93} & A_{94} & A_{95} & A_{96}& A_{97} & A_{98} &1\\
    A_{X1} & A_{X2}& A_{X3} & A_{X4} & A_{X5} & A_{X6}& A_{X7} & A_{X8} & A_{X9} & 1\\
   \end{bmatrix}   
   \begin{bmatrix}
   \psi_1\\
   \psi_2\\
   \psi_3\\
   \psi_4\\
   \psi_5\\
   \psi_6\\
   \psi_7\\
   \psi_8\\
   \psi_9\\
   \psi_{10}\\
   \end{bmatrix} 
   \end{array}
  \end{equation}

\noindent
Therefore,  
    \begin{equation}
    \label{dphi1dt}
     \frac{d \phi_1}{dt}=0.\\
   \end{equation} 

\noindent
Then it is possible to derive  
    \begin{equation}
    \label{da21d-pre}
    \begin{split}
     \frac{d}{dt}\left( {\int \rho \psi_2 \phi_1dw}\right)=\int \frac{d}{dt} (\rho \psi_2 \phi_1 j)d\Omega = \int \Big [ \frac{d}{dt} (\rho j)\psi_2 \phi_1 + \frac{d}{dt}(\psi_2 \phi_1) \rho j \Big ] d\Omega = 0,\\
      \frac{d}{dt}\left( {\int \rho \phi_1 \phi_1dw}\right)=\int \frac{d}{dt} (\rho \phi_1 \phi_1 j)d\Omega = \int \Big [ \frac{d}{dt}  (\rho j)\phi_1 \phi_1 + \frac{d}{dt}(\phi_1 \phi_1) \rho j \Big ] d\Omega = 0.\\
      \end{split}
   \end{equation} 
   
\noindent
Therefore, 
   \begin{equation}
    \label{da21dt}
     \frac{d A_{21}}{dt}=\frac{d}{dt}\left(- \frac{\int \rho \psi_2 \phi_1dw}{\int \rho \phi_1 \phi_1dw}\right)=0.\\
   \end{equation} 
   
\noindent
As such, 
    \begin{equation}
    \label{dphi2dt}
     \frac{d \phi_2}{dt}=\frac{d}{dt}(\psi_2+A_{21}\psi_1)=0.\\
   \end{equation} 
   
\noindent
Likewise, it is also possible to derive, 
    \begin{equation}
    \label{da31d-pre}
    \begin{split}
     &\frac{d}{dt}\left( {\int \rho \psi_3 \phi_1dw}\right)=\int \frac{d}{dt} (\rho \psi_3 \phi_1 j)d\Omega = \int \Big [ \frac{d}{dt} (\rho j)\psi_3 \phi_1 + \frac{d}{dt}(\psi_3 \phi_1) \rho j \Big ] d\Omega = 0,\\
     & \frac{d}{dt}\left( {\int \rho \phi_1 \phi_1dw}\right) = 0.\\
      \end{split}
   \end{equation}  
   
\noindent   
and 
   \begin{equation}
    \label{da31d-pre2}
    \begin{split}
     &\frac{d}{dt}\left( {\int \rho \psi_3 \phi_2dw}\right)=\int \frac{d}{dt} (\rho \psi_3 \phi_2 j)d\Omega = \int  \Big [\frac{d}{dt}(\rho j)\psi_3 \phi_2 + \frac{d}{dt}(\psi_3 \phi_2) \rho j \Big ] d\Omega = 0,\\
     &\frac{d}{dt}\left( {\int \rho \phi_2 \phi_2dw}\right) = 0.\\
      \end{split}
   \end{equation} 

\noindent   
 Therefore, 
   \begin{equation}
    \label{da31dt}
     \frac{d A_{31}}{dt}=\frac{d}{dt}\left(-\frac{\int \rho \psi_3 \phi_1dw}{\int \rho \phi_1 \phi_1dw}\right)=0,\\
     \frac{d A_{32}}{dt}=\frac{d}{dt}\left(-\frac{\int \rho \psi_3 \phi_2dw}{\int \rho \phi_2 \phi_2dw}\right)=0.\\      
  \end{equation} 

\noindent
Using a similar way, we can get, 
   \begin{equation}
    \label{dphidt}
     \begin{split}
      \frac{d \boldsymbol \phi}{dt}=0, \\
      \frac{d A_{mn}}{dt}=0. \\      
     \end{split}  
  \end{equation} 
  
\noindent
It is easy to get the orthogonal basis is still hierarchical as long as the initial basis is ordered in a hierarchical manner.

\begin{equation}
\label{OrthoP3-hier}
 {\mathbb U}_h({\boldsymbol \xi},t) =\underbrace{\underbrace{ \underbrace{\underbrace{{\mathbb U}_1}_{P0} + {\mathbb U}_2\phi_2 + {\mathbb U}_3\phi_3}_{P1} + {\mathbb U}_4\phi_4 + {\mathbb U}_5\phi_5 + {\mathbb U}_6\phi_6}_{P2} + {\mathbb U}_7\phi_7
 + {\mathbb U}_8\phi_8+ {\mathbb U}_9\phi_9+ {\mathbb U}_{10}\phi_{10}}_{P3}
\end{equation}

\noindent
But we need to mention that, although 
the hierarchy is the same as the original Taylor bases, the unknown solution is not the same as the original unknown individually. 
For example, ${\mathbb U}_3$ is the linear combination of $\frac{\partial \mathbb U}{\partial \xi}$ and $\frac{\partial \mathbb U}{\partial \eta}$.

\section{Gauss-Lobatto quadrature rules}
Some Gauss-Lobatto quadrature rules are tabulated for the interval $[-1, 1]$ in Table \ref{tab:Lobatto-rule}. 
\begin{table*}[t]
\begin{center}
\caption
{The Gauss-Lobatto point positions and weights for the  Gauss-Lobatto quadrature rule. $\xi$ and $wt$ denote the position and weight of the quadrature points. }
\label{tab:Lobatto-rule}
\scriptsize
\mbox{
\begin{tabular}{l c c c c c c c}
\hline
$G$ & 1 & 2 & 3 & 4 & 5 & 6 & 7\\    
\hline
2 points      \\
\hline
 $wt$    &  1.0	& 1.0   &       &        & 	& \\
$\xi$     & -1.0	& 1.0   &       &        & 	& \\   
\hline
3 points      \\
\hline
 $wt$    &  $\frac{1}{3}$   &  $\frac{4}{3}$  &  $\frac{1}{3}$  &       &        & 	 \\
$\xi$     & -1.0	& 0   &  1.0      &        & 	& \\  
\hline
4 points      \\
\hline
 $wt$    &  $\frac{1}{6}$   &  $\frac{5}{6}$  &  $\frac{5}{6}$  &  $\frac{1}{6}$  &       &        & 	 \\
$\xi$     & -1.0	& $-\frac{\sqrt{5}}{5}$ & $\frac{\sqrt{5}}{5}$  &  1.0      &        &  \\   
\hline
5 points      \\
\hline
 $wt$    &  $\frac{1}{10}$   &  $\frac{49}{90}$  &  $\frac{32}{45}$  &  $\frac{49}{90}$  & $\frac{1}{10}$  &        & 	 \\
$\xi$     & -1.0	& $\frac{-\sqrt{21}}{7}$  & 0 & $\frac{-\sqrt{21}}{7}$  &  1.0      &         \\   
\hline
6 points      \\
\hline
 $wt$    & $\frac{1}{15}$  	& $ \frac{\left(14 - \sqrt{7}\right)}{30}$  		& $ \frac{\left(14 +\sqrt{7}\right)}{30}$	& $ \frac{\left(14 +\sqrt{7}\right)}{30}$ 		& $ \frac{\left(14 - \sqrt{7}\right)}{30}$ 		& $\frac{1}{15}$   \\
$\xi$     & $-1$      		& $-\sqrt{\frac{1}{21}\left( 7 + 2\sqrt{7} \right) }$ 	& $-\sqrt{\frac{1}{21}\left( 7 - 2\sqrt{7} \right) }$  & $\sqrt{\frac{1}{21}\left( 7 - 2\sqrt{7} \right) }$ 	& $\sqrt{\frac{1}{21}\left( 7 + 2\sqrt{7} \right) }$  & 1\\
\hline
7 points      \\
\hline
 $wt$    & $\frac{1}{21}$  	& $ \frac{124-7\sqrt{15}}{350}$  		& $ \frac{124+7\sqrt{15}}{350}$  		& $\frac{256}{525}$ & $ \frac{124+7\sqrt{15}}{350}$ 	& $ \frac{124-7\sqrt{15}}{350}$  		& $\frac{1}{21}$   \\
$\xi$     & $-1$      		& $-\sqrt{\frac{5}{11}+\frac{2}{11}\sqrt{\frac{5}{3}} }$ 	& $-\sqrt{\frac{5}{11}-\frac{2}{11}\sqrt{\frac{5}{3}} }$  & 0 & $\sqrt{\frac{5}{11}-\frac{2}{11}\sqrt{\frac{5}{3}} }$ 	& $\sqrt{\frac{5}{11}+\frac{2}{11}\sqrt{\frac{5}{3}} }$   & 1\\
\hline
\end{tabular}}
\end{center}
\end{table*}

%
\bibliography{Lag-DGsubg} 
\bibliographystyle{unsrt}

\end{document}